# General Bayesian inference schemes in infinite mixture models

## María Dolores Lomelí-García

Gatsby Computational Neuroscience Unit

University College London

THESIS

Submitted for the degree of
**Doctor of Philosophy**

# Declaration

I declare that this thesis was composed by myself, the work contained herein is my own except where explicitly stated otherwise in the text. This work has not been submitted for any other degree or professional qualification.

(Maria Lomeli)

# Abstract

Bayesian statistical models allow us to formalise our knowledge about the world and reason about our uncertainty, but there is a need for better procedures to accurately encode its inherent complexity. One way to do so is through compositional models, which are formed by combining blocks or components consisting of simpler models. One can increase the complexity of the compositional model by either stacking more blocks or by using a not-so-simple model as a building block. This thesis is an example of the latter. One first aim is to expand the choice of Bayesian nonparametric (BNP) blocks for constructing tractable compositional models. So far, most of the models that have a Bayesian nonparametric component use either a Dirichlet Process or a Pitman–Yor process because of the availability of tractable and compact representations. This thesis shows how to overcome certain intractabilities in order to obtain analogous compact representations for the very wide class of Poisson–Kingman priors which includes the Dirichlet and Pitman–Yor processes.

A major impediment to the widespread use of Bayesian nonparametric building blocks is that inference is often costly, intractable or difficult to carry out. This is an active research area since dealing with the model's infinite dimensional component forbids the direct use of standard simulation-based methods. The main contribution of this thesis is a variety of inference schemes that tackle this problem: Markov chain Monte Carlo and Sequential Monte Carlo methods, which are exact inference methods since they target the true posterior.

The contributions of this thesis, in a larger context, provide general purpose exact inference schemes in the flavour or probabilistic programming: the user is able to choose from a variety of models, focusing only on the modelling part. We show how if the wide enough class of Poisson–Kingman priors is used as one of our blocks, this objective is achieved.

# Acknowledgements

I thank my supervisor Yee Whye Teh, who taught me how to strive for clarity, for his useful critiques and recommendations, and for teaching me how to formulate interesting research questions. He let me spend some time in Oxford and my PhD was greatly enriched with this visit.

The advice given by Stefano Favaro, my collaborator from the start, has been of great asistance in my development as a researcher. I thank him for his time, patience and invaluable teachings. Also, I thank Pierre Jacob, for the stimulating discussions and for providing invaluable feedback for my research talks and for this thesis.

My special thanks to my examiners Dario Spano and Maria De Iorio, for their time and for helping me improve this manuscript greatly.

I am particularly grateful to my Gatsby, greater Gatsby and UCL colleagues: Charles Blundell, Balaji Lakshminarayanan and Jan Gausthaus, who guided me through the beginning of my PhD and were always there to answer questions or for insightful discussions. Ioanna Manolopoulou, provided invaluable feedback for my research talks. Joana Soldado-Magraner and Arthur Gretton have been amazing colleagues and friends. Peter Dayan and Zoubin Ghahramani, who have provided me with ongoing support and encouragement.

I would also like to extend my thanks to my professors in Mexico: Eduardo Gutiérrez-Peña, who helped me improve this manuscript and has always been a guiding light; Ernesto Barrios, who taught me some essential tools to be a researcher, and Ramsés Mena, for his ongoing support.

I wish to acknowledge my excellent proofreaders: Laurence Aitchison, Tamara Fernández, Frauke Harms, Lawrence Murray, Konstantina Palla, Christian Steinruecken, and Mark Rowland. Despite all this invaluable assistance, some errors surely still remain and the author would be grateful to be e-mailed about these at maria.lomeli@eng.cam.ac.uk.

I thank my mother Alicia, who has been incredibly supportive and for her encouragement through all these years. My aunt Catalina and cousin Yankel have also been there for me always.

Finally, I am very grateful to Adam Walker, who has been there for me during this chaotic period, offering a friendly ear and making me laugh out loud.

# Contents

# Chapter 1

# Introduction

## 1.1 Motivation

The world is a complex and uncertain place. In principle, we could use Bayesian statistical models to formalise our knowledge about the world and reason about our uncertainty. The adjective "Bayesian" refers to a statistical paradigm or philosophy based on a subjective interpretation of probability. Indeed, it consists of a subjective quantification of our state of knowledge rather than anything that can be measured directly in an experiment. All statistical approaches recognise the need to take into account the uncertainty about all quantities of interest in our models. The Bayesian approach incorporates the uncertainty through a prior probability distribution for the model parameters and allows us to deal with unobserved quantities in a principled way. However, the need remains for better procedures that accurately encode the world's inherent complexity. There are several modelling strategies that we could envisage. One possibility — discussed here — is to use compositional models, which are built from blocks or components of simple models stacked together (Orbanz, 2014). We can effectively increase the compositional model's complexity by either stacking more blocks together or using a not-so-simple model as a building block. For instance, a hierarchical model with more than one level of latent variables (Gelman et al., 1995) is an example of the former; the nested Dirichlet process (Rodríguez et al., 2008), the hierarchical Dirichlet process (Teh et al., 2006) and the infinite hidden Markov model (Beal et al., 2002), which use Dirichlet process blocks, are examples of the latter.

According to Ghahramani (2015), models that have a Bayesian nonparametric component provide more flexibility that can lead to better predictive performance. This is so because their capacity to learn does not saturate, hence their predictions should continue to improve as we get more and more data. Furthermore, we are able to better describe our uncertainty about predictions thanks to the Bayesian paradigm.

Bayesian modelling has developed steadily during the last two or three decades. So far, however, most of the compositional models that have a Bayesian nonparametric (BNP) component use either a Dirichlet (Ferguson, 1973; Sethuraman, 1994) or a Pitman–Yor process (Pitman and Yor, 1997) because of the availability of tractable representations in these cases. One aim of this thesis is to expand the choice of BNP blocks for constructing tractable compositional models. Specifically, the class of Poisson-Kingman processes is used as a building block. This is a very large class of Bayesian nonparametric priors which contains the Pitman–Yor and Dirichlet processes, and most random discrete distributions used in the literature.

A major impediment to the widespread use of Bayesian nonparametric models is the problem of inference. Given the complexity of these models, posterior inference is only possible by means of sophisticated computational methods. This is an active research area since dealing with the model's infinite dimensional component forbids the direct use of standard simulation-based methods. This thesis shows how to overcome certain intractabilities for the class of Poisson–Kingman priors. We provide compact representations that make inference possible based on simplifying properties of the class of priors just as inference algorithms for graphical models (Lauritzen, 1996; Lauritzen and Spiegelhalter, 1988; Pearl, 1988) rely on the conditional independence relationships encoded by the graph. The main focus is given to exact inference schemes that target the true posterior: Markov chain Monte Carlo (MCMC) (Metropolis et al., 1953; Hastings, 1969; Geman and Geman, 1984) and Sequential Monte Carlo (SMC) (Doucet et al., 2001) methods fall under this umbrella. Indeed, according to Wagner (1987), an advantage of exact methods, where no systematic bias remains, is that the error is entirely due to the variation in the output of the Monte Carlo algorithms and, thus, it is straightforward to interpret and quantify.

The contributions of this thesis, in a larger context, provide a step towards wider usage of flexible Bayesian nonparametric models. The methods developed herein allow automated inference in probabilistic programs built out of a wide variety of Bayesian nonparametric building blocks. Hence, the user should be able to choose from a variety of models, focusing only on the modelling part.

## 1.2 Outline of the Thesis

This thesis is split into five parts. In Chapter 2, the required background material is presented. A detailed review of Bayesian nonparametrics is given: from the mathematical spaces of interest to the interplay between random probability measures and exchangeable random partitions. Some useful characterisations of the family of Poisson–Kingman priors are described, which lay the foundations for the algorithms presented in later chapters. In Chapter 3, an MCMC scheme for a subclass of infinite mixture models is described in detail. This subclass is called $\sigma$-Stable Poisson–Kingman, it is interesting because it encompasses most of the Bayesian nonparametric priors previously explored in the literature. Some of its properties, which are later used in deriving the algorithm, are reviewed. In Chapter 4, a new MCMC scheme, based on a novel compact way of representing the infinite dimensional component, is introduced. Each step of the algorithm is derived in detail, and some of the difficulties that arise when constructing it are discussed. Also, some experiments are carried out to show that it outperforms other inference schemes. In Chapter 5, various SMC schemes are presented. These samplers are useful in a sequential scenario: they allow us to repeatedly estimate the marginal likelihood and hence, compare models in a principled way. In Chapter 6, some discussion is given for the previous chapters, including critical remarks about the current limitations of each contribution. Finally, some possible directions of future research that could benefit from the proposed algorithms are suggested.

# Chapter 2

# Background

The aim of this chapter is to describe the main class of models that are dealt with in this thesis, namely, mixture models and its infinite dimensional analogue, infinite mixture models (Rasmussen, 2000). These models are specific examples of compositional models (Orbanz, 2014) or Bayesian hierarchical models (Box and Tiao, 1973; Gelman et al., 1995). Firstly, the notions of exchangeable partitions and their corresponding laws are stated. These objects are then formally surveyed starting with some definitions and terminology. The underlying spaces for these probabilistic objects have the useful property of compactness with respect to a metric. This guarantees that the space of all measures defined on any of these base spaces inherits the compactness property (Parthasarathy, 1967, Chapter 6, Theorem 6.4). Hence, once we endow these base spaces with any probability measure, this measure is tight, no probability mass escapes to infinity (Billingsley, 1999). Secondly, the theoretical justification for the interplay between random probability measures and random partitions of $\mathbb{N}$, known in the literature as Kingman's paintbox construction (Kingman, 1982), is discussed. Thirdly, the main type of block used in infinite mixture models is reviewed: a family of Bayesian nonparametric priors or random probability measures called Poisson–Kingman priors. This class, which has some useful characterisations that lead us to derive tractable representations, is then described in detail herein. Some fundamental proofs are given for the interested reader. Finally, a brief overview is provided of exact inference schemes for Bayesian nonparametric mixture models built with Pitman–Yor process blocks. Sections 2.4, 2.2 and 2.3 are mostly based on Bertoin (2006, Chapter 2); Section 2.5, on Perman et al. (1992), and Section 2.3.2, in Gnedin and Pitman (2006).

## 2.1 Mixture and infinite mixture models

Let $\{Y_i\}_{i=1}^n$ a sample of $n$ independent observations. A mixture model is an example of a compositional model which has a single discrete latent variable per observation that takes $m \in \mathbb{N}$ finitely many values. It can be written as the following hierarchical model

$$\underline{\pi} \in \Delta_m := \left\{ (\pi_1, \ldots, \pi_m) : 0 \leqslant \pi_\ell \leqslant 1 \;\; \forall \ell \in \{1, \ldots, m\} \;\; \text{and} \;\; \sum_{\ell=1}^m \pi_\ell = 1 \right\}$$

$$X_i \sim \text{Discrete}(\underline{\pi})$$
$$Y_i \mid X_i \sim F(\cdot \mid \phi_{X_i}) \tag{2.1}$$

where $\phi_{X_i}$ is the parameter corresponding to the $X_i$-th component of the mixture model. The discrete or categorical probability mass function gives the prior probability that the corresponding latent variable $X_i \in \{1, \ldots, m\}$ takes a specific value

$$P(X_i = \ell) = \pi_\ell, \quad \ell = 1, \ldots, m.$$

The marginal distribution for the $i$-th observation is given by

$$P(Y_i \in dy_i) = \sum_{\ell=1}^m P(Y_i \in dy_i \mid X_i) P(X_i = \ell)$$
$$= \sum_{\ell=1}^m \pi_\ell F(Y_i \in dy_i \mid \phi_\ell).$$

In most cases, the parameters $\{\phi_\ell\}_{\ell=1}^m$ for each mixture component, the total number of components $m$ and the prior probabilities $\underline{\pi}$ are unknown quantities. If we are interested in density estimation or clustering, we could proceed in a number of ways. All the approaches discussed in this thesis render a flat clustering of the data. However, in certain contexts, it is natural to think of superclasses so a hierarchical clustering would be more appropriate. Some examples of hierarchical clustering schemes are: ad hoc approaches such as agglomerative clustering (Medvedovic and Sivaganesan, 2002), Bayesian hierarchical clustering (Heller and Ghahramani, 2005) or a full Bayesian non-parametric model (Teh et al., 2006), among others.

A simple way of selecting the total number of components is to perform a model selection procedure among models with different total number of components. First, for each model, the total number of components $m$ is fixed. Then, the model param-eters $\{\phi_\ell\}_{\ell=1}^m$ and the probabilities $\underline{\pi}$ are learnt using, for instance, the Expectation Maximization algorithm (Dempster et al., 1977), Gibbs sampling (Geman and Geman, 1984), variational Bayes (Beal and Ghahramani, 2003; Blei and Jordan, 2006; Minka and Ghahramani, 2003), or some other inference scheme.

We repeat these steps for different values for the total number of components $m$ and use the output of the learning algorithm to compute certain quantities for a model selection procedure to select the best model. Some popular model selection criteria include the Bayes factor (Jeffreys, 1935) or the Bayesian Information Criterion (Schwarz, 1978), which are parametric likelihood-based procedures. The caveat is that a high computational overhead could be incurred since we need to repetitively learn the model's parameters for every model with different total number of components.

Furthermore, in order to be fully Bayesian, a prior distribution for all unknown quantities must be included in the hierarchical model (2.1) as follows

$$M \sim \mathcal{Q}$$
$$\underline{\pi} \mid M = m \sim \mathcal{P}_m$$
$$(\phi_\ell)_{\ell=1}^m \mid M = m \overset{\text{i.i.d}}{\sim} H_0$$
$$X_1, \ldots, X_n \mid \underline{\pi} \overset{\text{i.i.d.}}{\sim} \text{Categorical}(\underline{\pi})$$
$$Y_i \mid X_i \overset{\text{ind.}}{\sim} F(\cdot \mid \phi_{X_i}). \tag{2.2}$$

$H_0$ corresponds to a non-atomic probability measure, it is usually called a base measure. Ideally, the probability measure $\mathcal{Q}$ should have full support on $\mathbb{N} \cup \{\infty\}$: it should assign, *a priori*, a positive probability $0 < q(m) < 1$ to each value $m \in \mathbb{N} \cup \{\infty\}$ in order to learn any possible total number of components, *a posteriori*. Some common choices for this probability mass function are: Poisson($\lambda$), Negative Binomial $(r, p)$, Geometric$(p)$, among others. The truncated versions of the first two probability mass functions are used such that positive probabilities are assigned to the positive integers only; the last one, is parameterised to count the number of number of Bernoulli trials necessary for one success. See Section 5.5 from Chapter 5 for other common choices of probability mass functions for the total number of components. However, these common choices of probability mass functions have a finite expected total number of components *a priori* and do not assign a positive probability to $\infty$. These can be strong assumptions when choosing a prior for the total number of components. In practice, the hyperparameters have to be chosen as well and, depending on this choice, the prior could be "very sticky" or "hard to get away from" in a misspecified scenario. Gnedin (2010) derived an interesting prior distribution $\mathcal{Q}$ that can have an infinite or finite expected total number of components, depending on its hyperparameter value.

Given $M = m$, a popular choice for $\mathcal{P}_m$ is the Dirichlet distribution due to its conjugacy with respect to the categorical distribution (Bernardo and Smith, 1994). Furthermore, the symmetric Dirichlet distribution with a common parameter $\alpha$ is sometimes chosen since there is no reason, *a priori*, to believe that the weights are distinguishable from each other. The corresponding expressions for the second and fourth levels of the hierarchical specification (2.2) are

$$\Pr(\underline{\pi} \mid M = m) = \frac{\Gamma(m\alpha)}{\Gamma(\alpha)^m} \prod_{\ell=1}^{m-1} \pi_\ell^{\alpha-1} (1 - \pi_1 - \ldots - \pi_{m-1})^{\alpha-1}$$

$$\Pr(X_i \mid \underline{\pi}) = \prod_{\ell=1}^{m} \pi_\ell^{\mathbb{I}\{X_i = \ell\}}$$

$$\Pr(\underline{\pi} \mid M = m)\Pr(X_1, \ldots, X_n \mid \underline{\pi}) = \frac{\Gamma(m\alpha)}{\Gamma(\alpha)^m} \prod_{\ell=1}^{m-1} \pi_\ell^{\alpha + n_\ell - 1}$$
$$\times (1 - \pi_1 - \ldots - \pi_{m-1})^{\alpha - 1 + n - \sum_{\ell < m} n_\ell} \quad (2.3)$$

where $\underline{\pi} \in \Delta_m$, $\mathbb{I}\{X_i = \ell\}$ denotes the indicator function and $n_\ell = \sum_{i=1}^{n} \mathbb{I}\{X_i = \ell\}$. It is possible that there exists one or more than one mixture component $\ell \in \{1, \ldots, m\}$ such that $n_\ell = 0$. Hence, the total number of components $m$ should be separated into $k$ occupied components, for which $n_\ell$ is positive, and, $k_0$ empty components, those for which $n_\ell = 0$. The Dirichlet-Multinomial conjugacy property allows us to integrate out the weights in Equation (2.3), see the Appendix A.1 for details of these derivations. Then,

$$\Pr(X_1, \ldots, X_n \mid M = m) = \frac{\alpha^k \Gamma(m\alpha)}{\Gamma(\alpha m + n)} \prod_{\{\ell: n_\ell > 0\}} \frac{\Gamma(n_\ell + \alpha)}{\Gamma(\alpha + 1)}.$$

There are $\binom{m}{k}$ possible ways in which the $n$ indicator variables $(X_i)_{i=1}^{n}$ take $k$ different values out of $m$, with exactly $n_\ell$ indicators taking the same value $\ell$ ($\ell = 1, \ldots, m$, $k = 1, \ldots, n, m \in \mathbb{N}$, $n_\ell > 0$, $\sum_{\ell=1}^{k} n_\ell = n$). For this reason, we can group together the indicator variables that take the same value $\ell$ in a set. This gives us a family of $k$ sets. Since each indicator has to take one but only one $\ell$ value, every indicator strictly belongs to one and only one set. For this reason, the family of sets has the property of being a mutually exclusive and exhaustive collection of $k$ sets, such a family is called a partition. Furthermore, we can use the index of the $i$-th data point to denote the elements of each set, since the resulting partition is equivalent to a partition of the first $n$ integers, denoted by $\Pi_n$. The corresponding total set is $\{1, \ldots, n\}$, successively denoted by $[n] = \{1, \ldots, n\}$. This concept is formally stated in Definition 4. This collection of sets can be arranged in $k!$ different ways. The combinatorial factor $k!\binom{m}{k}$ takes into account the number of ways we can arrange $n$ data points in $m$ bins such that $k$ are occupied and $m - k$ are empty and the number of ways in which we can arrange the $k$ occupied bins. If we include it, we obtain the following expression

$$\Pr(\Pi_n = \pi \mid M = m) = \alpha^k \frac{m\Gamma(m)}{\Gamma(m-k+1)} \frac{\Gamma(m\alpha)}{\Gamma(m\alpha + n)} \prod_{\ell=1}^{k} \frac{\Gamma(n_\ell + \alpha)}{\Gamma(\alpha + 1)}. \qquad (2.4)$$

Equation (2.4) reflects the fact that we can move to the space of partitions since sampling the indicator random variables induces a partitioning of the dataset due to their discreteness. The combinatorial factor takes into account that all resulting partitions with the same number of blocks and cardinalities per block are assigned the same probability, i.e. they form an equivalence class. Finally, if the total number of components is randomised and marginalised out, we obtain

$$\Pr(\Pi_n = \pi) = \sum_{m=0}^{\infty} \alpha^k \frac{m\Gamma(m)}{\Gamma(m-k+1)} \frac{\Gamma(m\alpha)}{\Gamma(m\alpha + n)} q(m) \prod_{\ell=1}^{k} \frac{\Gamma(n_\ell + \alpha)}{\Gamma(\alpha + 1)}. \qquad (2.5)$$

Equation (2.5) gives the probability of an unordered partition of $[n]$, with k blocks and $n_\ell$ is the corresponding cardinality of the $\ell$-th block, $\ell = 1, \ldots, k$. This probability is for an unordered partition $\Pi_n$ because it takes into account all possible orders of its sets.

There are two ways to construct an MCMC scheme for this model. One possibility is to choose a distribution $\mathcal{Q}$ such that the sum of Equation (2.5) can be computed analytically. Alternatively, an augmentation can be introduced for the random variable for the total number of components $M$, and then, we can sample it explicitly together with the other quantities. In the former case, the Gibbs sampling scheme of Miller and Harrison (2015) can be used for certain choices of $\mathcal{Q}$; for the latter case, the reversible jump MCMC (RJMCMC) of Richardson and Green (1997). The first MCMC scheme is easy to implement but it is restricted to the very few choices of $\mathcal{Q}$ which yield a tractable sum. In contrast, the RJMCMC scheme is more general but is difficult to implement since sometimes it requires the computation of a Jacobian term. Furthermore, in RJMCMC it is hard to design a good proposal which can make the algorithm slow mixing: it is difficult for it to jump between models of different dimension which is reflected in low acceptance rates.

Alternatively, a distribution $\mathcal{Q}$ that places all its mass in $M = \infty$ can be chosen for the top level of the generative model given by Equation (2.2). This is the motivation for a Bayesian nonparametric mixture model, first derived by Rasmussen (2000) as a limit of a finite mixture model when the total number of components tends to infinity. This limit can be derived if we start with Equation (2.4) and set $\alpha = \frac{\theta}{m}$, then,

$$\Pr(\Pi_n = \pi \mid M = m) = \left(\frac{\theta}{m}\right)^k \frac{m\Gamma(m)}{\Gamma(m-k+1)} \frac{\Gamma(\theta)}{\Gamma(\theta+n)} \prod_{\ell=1}^{k} \frac{\Gamma(n_\ell + \frac{\theta}{m})}{\Gamma(\frac{\theta}{m}+1)}.$$

Given that $\Gamma(m+1) = m!$, if the Stirling approximation is used $m! \sim \sqrt{2\pi} m^{m+1/2} \exp(-m)$ for large $m$, then,

$$\Pr(\Pi_n = \pi \mid M = m) \simeq \left(1 - \frac{k}{m}\right)^{k-m-1/2} \exp(-k) \frac{\theta^k \Gamma(\theta)}{\Gamma(\theta+n)} \prod_{\ell=1}^{k} \frac{\Gamma(n_\ell + \frac{\theta}{m})}{\Gamma(\frac{\theta}{m}+1)}.$$

Let $m \to \infty$, then, we obtain

$$\Pr(\Pi_n = \pi) = \frac{\theta^k \Gamma(\theta)}{\Gamma(\theta+n)} \prod_{\ell=1}^{k} \Gamma(n_\ell). \tag{2.6}$$

Equation (2.6) coincides with the finite dimensional distributions of a Chinese restaurant process (Aldous, 1985). The Chinese restaurant process is a distribution over partitions of $\mathbb{N}$. The finite-dimensional distributions of such process have the property that they do not distinguish between those partitions with the same cardinalities per block and number of blocks. For example, $\Pr(\{\{1,2\},\{3\}\}) = \Pr(\{\{1,3\},\{2\}\}) = \Pr(\{\{1\},\{2,3\}\}) = \frac{\theta^2 \Gamma(\theta)}{\Gamma(\theta+3)}$. This property makes the Chinese restaurant process a mathematically convenient object. Since sampling the indicator random variables induces a partitioning of the dataset due to their discreteness, it is also possible to work in the space of partitions.

In the previous derivations, we showed how if we take the limit $m \to \infty$ then we obtain Equation (2.6), this shows how an *infinite mixture model* is derived as a mixture model which assigns all the prior mass to an infinite total number of components. Informally, in the generative model given by Equation (2.2), we are choosing $\mathcal{Q}$ to be a delta function at $\infty$. This encodes the idea that it is desirable that the indicator variables $(X_i)_{i=1}^n$ take countably infinitely many values *a priori* so any possible number of occupied components can be learnt *a posteriori*. Because we only observe a dataset of finite size at any given point, the number of occupied components $k$ is always less than or equal to the size of the dataset and, as a consequence, finite. Even though the number of occupied components is finite for a dataset of fixed size, it is unbounded and grows sublinearly as a function of the number of observations when we observe more data points. See Gnedin et al. (2007); Gnedin and Pitman (2006) for further details about the relevant asymptotic results about the growth rate. For this reason, these models are more flexible than their parametric counterpart and can adapt to the data. They can be formulated in the following hierarchical fashion

$$P := \sum_{\ell=1}^{\infty} \pi_\ell \delta_{\phi_\ell} \sim \mathcal{P}$$
$$X_1, \ldots, X_n \mid P \overset{\text{i.i.d.}}{\sim} P$$
$$Y_i \mid X_i \sim F(\cdot \mid \phi_{X_i}). \qquad (2.7)$$

Here, $\mathcal{P}$ is the law of a random discrete distribution or random probability measure (RPM). The $(\pi_\ell)_{\ell \in \mathbb{N}}$ is an infinite sequence of random probabilities and the $(\phi_\ell)_{\ell \in \mathbb{N}}$ are independent identically distributed random variables distributed according to $H_0$, which refers to the category labels being picked uniformly at random. For convenience, the infinite sequence of random probabilities $(\pi_\ell)_{\ell \in \mathbb{N}}$ is sorted in decreasing order, denoted by $(s_\ell)_{\ell \in \mathbb{N}}$, where $s_1 := \underset{\pi_\ell}{\text{Sup}} \{\pi_1, \ldots\}$, $s_2 := \underset{\pi_\ell}{\text{Sup}} \{\pi_1, \ldots\} \setminus \{s_1\}$, etc. Such space of ranked infinite sequences $(s_\ell)_{\ell \in \mathbb{N}}$ is called a mass-partition space and denoted by $\mathcal{P}_M$ (Bertoin, 2006). Without loss of generality, we refer to a RPM with a ranked infinite sequence of probabilities as a random mass partition. Informally, if we sample repeatedly from (2.7), we can put together those indices $i \in \{1, \ldots, n\}$ for which the corresponding $X_i$ take the same $\phi_\ell$ value. In this way, a partition can be induced (Kingman, 1982). In the Bayesian hierarchical model given by Equation (2.7) $\mathcal{P}$ was first set to be a Dirichlet process (DP) (Lo, 1984). A Dirichlet process is a random discrete distribution such that $\pi_\ell \stackrel{d}{=} Z_\ell \prod_{j<\ell}(1-Z_j)$ where $Z_\ell \sim \text{Beta}(1,\theta)$ for $\ell \in \mathbb{N}$ and $(Z_\ell)_{\ell \in \mathbb{N}}$ is a sequence of i.i.d. random variables (Sethuraman, 1994). The hierarchical model from Equation (2.7) with $\mathcal{P} \sim$ DP was first called *infinite mixture model* by Rasmussen (2000). We refer herein to the hierarchical model of Equation (2.7) as an *infinite mixture model* allowing the top level to be any random discrete distribution. In the next section, the space of mass-partitions $\mathcal{P}_M$ is formally defined. Also, some of its properties and its relationship to both the space of unordered infinite sequences and the space of partitions of $\mathbb{N}$ are stated. The concept of an RPM as a mathematical object is then reviewed, and some metrics on the base space to ensure it has the nice property of compactness are stated.

## 2.2 Partitions

This section goes through some fundamental definitions and proofs required for a better understanding of the theoretical underpinnings of Kingman's paintbox construction (Kingman, 1982). This fundamental result, given in Theorem 2, formally justifies the interplay between exchangeable random partitions and random discrete distributions. Indeed, the law of a partition $\Pi$ of $\mathbb{N}$ is given as an integral representation in terms of the *asymptotic frequencies* of the partition $\Pi$ with a DeFinetti mixing measure based on a *random mass partition*, i.e. a RPM where the weights are ranked in decreasing order and the locations are independent and identically distributed from a Uniform distribution. Hence, all of these concepts need to be formally defined before the Theorem can be stated.

**Definition 1.** *(Mass-partition) A* mass partition *is an infinite numerical sequence* $(s_1, s_2, \ldots)$ *arranged in decreasing order, i.e.*

$$s_1 \geqslant s_2 \geqslant \ldots \geqslant 0$$

*and*

$$\sum_{i=1}^{\infty} s_i \leqslant 1.$$

The space of mass partitions is denoted by $\mathcal{P}_M$. $s_0 := 1 - \sum_{i=1}^{\infty} s_i$ is referred to as dust. A mass partition **s** is proper if it has no dust, i.e. if $s_0 = 0$.

A metric $d_M$ on the space of mass partitions is the following

$$d_M(\mathbf{s}, \mathbf{s}') = \max\left\{|s_i - s'_i|, i \in \mathbb{N}\right\}, \quad \mathbf{s}, \mathbf{s}' \in \mathcal{P}_M \tag{2.8}$$

then, $(\mathcal{P}_M, d_M)$ is a compact metric space (Bertoin, 2006).

For every member of the mass partition space **s**, one can associate a discrete random variable $I^*$, which takes values in $\bar{\mathbb{N}} = \mathbb{N} \cup \{\infty\}$. The probability mass function for $I^*$ is

$$\Pr(I^* = i) = s_i \quad i \in \mathbb{N} \quad \text{and} \quad \Pr(I^* = \infty) = s_0.$$

**Definition 2.** *(Interval-partition) A collection of disjoint open intervals of an arbitrary open set, which is a subset of $(0,1)$, is called an* interval partition. *Let us denote this collection by $\vartheta$ and the space of interval-partitions, by $\mathcal{P}_I$.*

The lengths of the interval components are called *spacings*. One can rank these in decreasing order and complete with an infinite sequence of 0 to obtain a mass partition. The sequence of ranked spacings is denoted by $|\vartheta| \downarrow$. Let $\chi$ be a function such that

$$\chi_\vartheta(x) = \max\left\{|y - x|, y \in \vartheta^c\right\}, \quad x \in [0,1],$$

where $\vartheta^c = [0,1] \setminus \vartheta$ is the complement of the set $\vartheta$. Then, a metric $d_I$ on the space of interval partitions is the following

$$d_I(\vartheta, \vartheta') = \max\left\{|\chi_\vartheta(x) - \chi_{\vartheta'}(x)|, x \in [0,1]\right\}.$$

$(\mathcal{P}_I, d_I)$ is a compact metric space and the map $\vartheta \to |\vartheta| \downarrow$ is continuous from $\mathcal{P}_I$ to $\mathcal{P}_M$. (Bertoin, 2006).

**Definition 3.** *(Size-biased reordering of a mass-partition) Let $\mathbf{s} \in \mathcal{P}_M$ be a proper mass partition. A* size-biased reordering *of a mass partition (based on **s**) is a random map $\sigma : \mathbb{N} \to \mathbb{N}$ whose finite dimensional distributions are given as follows*

- For every $n \in \mathbb{N}$ such that $s_n > 0$ and every n-tuple $k_1, \ldots, k_n$ of distinct integers,

$$Pr(\sigma(1) = k_1, \ldots, \sigma(n) = k_n) = \prod_{i=1}^{n} \frac{s_{k_i}}{1 - (s_{k_1} + s_{k_2} + \ldots + s_{k_{i-1}})} \qquad (2.9)$$

  where by convention, the term in the product above corresponding to $i = 1$ is equal to $s_{k_1}$.

- When $\ell = \inf\{n \in \mathbb{N} : s_n = 0\} < \infty$, we agree that for every $n > \ell$ and every n-tuple $k_1, \ldots, k_n$ of distinct integers with $k_i = i$ for every $i \geqslant \ell$,

$$Pr(\sigma(1) = k_1, \ldots, \sigma(n) = k_n) = \prod_{i=1}^{\ell} \frac{s_{k_i}}{1 - (s_{k_1} + s_{k_2} + \ldots + s_{k_{i-1}})} \qquad (2.10)$$

The random sequence $\mathbf{s}^* = (s_{\sigma(1)}, s_{\sigma(2)}, \ldots)$ is then called a size-biased reordering of $\mathbf{s}$.

Equation (2.9) refers to the joint probability of picking the first $n$ elements of the mass partition in a particular order $(k_1, \ldots, k_n)$, i.e., $s_{k_1}$ was picked first, with probability $s_{k_1}$; $s_{k_2}$ was then picked, with probability $\frac{s_{k_2}}{1-s_{k_1}}$, etc. This is a *sampling without replacement* scheme and the $k$-th time an element of the mass partition is picked, the probabilities are restricted to sum $1 - \sum_{j \leqslant k} s_j$. Equation (2.10) refers to the case when some elements of the mass partition are equal to zero. If you pick a zero element, the joint probability of $n$ picks is equal to zero.

A size-biased reordering $\mathbf{s}^*$ of a mass partition $\mathbf{s}$ is an element of the space of unordered numerical sequences $\mathbf{r} = (r_1, \ldots)$ with values in $[0, 1]$, denoted by $\mathcal{S}_{[0,1]}$. If this space if equipped with the distance $d_{\mathcal{S}_{[0,1]}}$

$$d_{\mathcal{S}_{[0,1]}}(\mathbf{r}, \mathbf{r}') = \sum_{i=1}^{\infty} 2^{-i} |r_i - r_i'|, \quad \mathbf{r} = (r_1, \ldots), \mathbf{r} = (r_1', \ldots)$$

then $(\mathcal{S}_{[0,1]}, d_{\mathcal{S}_{[0,1]}})$ is a compact metric space (Bertoin, 2006).

**Definition 4. (Partition)** *A partition $\Pi_n = \{A_1, \ldots, A_{|\Pi_n|}\}$ of the first $n$ integers' set, denoted by $[n] := \{1, \ldots, n\}$, $n \in \mathbb{N}$ is a finite collection of $|\Pi_n|$ non-empty, non-overlapping and exhaustive subsets of $[n]$ called blocks and denoted by $A_j, j = 1, \ldots, |\Pi_n|$, i.e.*

1. *$\emptyset \subset A_j \subseteq [n], \forall j = 1, \ldots, |\Pi_n|$.*

2. *$A_i \cap A_j = \emptyset, \forall i, j \in [n], i \neq j$.*

3. $\bigcup_{j=1}^{|\Pi_n|} A_j = [n]$,

where $|\Pi_n|$ is the cardinality or number of blocks of the partition.

Definition 4 can be modified to include the case when $n = \infty$ and the corresponding partition of $\mathbb{N}$ has countably infinitely many blocks, for instance, $\{\{1\}, \{2\}, \ldots\}$, but we do not require this case for our purposes.

The space of partitions of $[n]$ is denoted by $\mathcal{P}_{[n]}$. Let $\mathcal{P}_\infty$ denote the space of partitions of $\mathbb{N}$, where the blocks are ordered by increasing order of the least element in each block. This space can be endowed with the following metric

$$d_\infty(\mathbf{\Pi}, \mathbf{\Pi}') = \frac{1}{\max\left\{k \in \mathbb{N} : \mathbf{\Pi}|_{[k]} = \mathbf{\Pi}'|_{[k]}\right\}}, \quad \mathbf{\Pi}, \mathbf{\Pi}' \in \mathcal{P}_\infty \qquad (2.11)$$

where $\frac{1}{\max \mathbb{N}} = 0$. Then, $(\mathcal{P}_\infty, d_\infty)$ is a compact metric space (Bertoin, 2006).

**Definition 5.** *(Restriction of a partition)* Let $m, n \in \mathbb{N}$, for all $m < n$, a partition $\Pi_m$ of $[m]$ is a restriction of a partition $\Pi_n$ of $[n]$, denoted by $\Pi|_{[m]}^{[n]}$ if,

$$\Pi|_{[m]}^{[n]} = \{A \cap [m] : A \in \Pi_n, A \cap [m] \neq \varnothing\}.$$

**Definition 6.** *(Finite permutation)* A permutation is a bijection $\sigma : \mathbb{N} \to \mathbb{N}$ of $\mathbb{N}$; it is finite when there exists an $N_\sigma \in \mathbb{N}$ such that $\forall n > N_\sigma$ we have that $\sigma(n) = n$.

**Definition 7.** *(Permutation of a block/partition)* Let $\sigma : \mathbb{N} \to \mathbb{N}$ be a finite permutation. The permutation applied to a block $A$ is defined as follows

$$\sigma(A) = \{\sigma(n) : n \in A\}.$$

The permutation applied to a partition $\Pi_n$ is defined to be

$$\sigma(\Pi_n) = \{\sigma(A) : A \in \Pi_n\}.$$

## 2.3 Exchangeable random partitions

**Definition 8.** *(Finite exchangeability)* A random partition $\Pi_n$ is finitely exchangeable if $\sigma(\Pi_n) \stackrel{d}{=} \Pi_n$ for every finite permutation $\sigma$ of $n$.

**Definition 9.** *(Infinite exchangeability)* *A random partition $\Pi_n$ is infinitely exchangeable if it is exchangeable in the sense of Definition 8 for all $n \in \mathbb{N}$.* (Bernardo and Smith, 1994).

**Definition 10.** *(Compatibility)* *Let $\mathcal{R}_n(\Pi_m)$ be the set of all partitions of $[n]$ whose restriction to $[m]$ is $\Pi_m$, i.e. $\Pi_m = \Pi|_{[m]}^{[n]}$. The sequence of random partitions $(\Pi_n)_{n \in \mathbb{N}}$ is* compatible in distribution *if, for all $m, n \in \mathbb{N}$ such that $m < n$, we have*

$$Pr(\Pi_m = \pi_m) = \sum_{\Pi_n \in \mathcal{R}_n(\Pi_m)} Pr(\Pi_n = \pi_n). \tag{2.12}$$

*This concept is sometimes referred to as* consistency under marginalisation.

If we start with a sequence of partitions of $(\Pi_n)_{n \in \mathbb{N}}$ which are finitely exchangeable and compatible, we obtain an infinitely exchangeable random partition of $\mathbb{N}$. Conversely, an infinitely exchangeable partition of $\mathbb{N}$ is compatible with all of its restrictions and finitely exchangeable by definition.

We have stated some definitions to verify if a general random partition is finitely exchangeable and compatible. The first property, invariance under permutation, can be verified through the finite dimensional distributions of a partition. Since all exchangeable random partitions with the same cardinalities per block $(\lambda_j)_{j=1}^k$ and number of blocks $k$ are mapped to one single object, called a *composition of $n$*, it is easier to verify this symmetry. A *composition of the first $n$ integers* is a finite family of positive integers $(\lambda_1, \ldots, \lambda_k)$ such that their sum $\lambda_1 + \lambda_2 + \ldots + \lambda_k = n$. In principle, the compatibility property requires a careful enumeration of all the finite partitions which correspond to the restrictions of a partition of $\mathbb{N}$. In Section 2.3.2,, we review a subclass of a random partition whose finite dimensional distributions have a factorised form and are exchangeable. Due to this factorised form, it is easier to check the compatibility requirement.

### 2.3.1 Kingman's paintbox construction

Kingman's paintbox is a constructive way to sample an exchangeable random partition in $\mathcal{P}_\infty$. Indeed, the law of an exchangeable random partition cannot be recovered from a unique mass partition but from a mixture of these. Let $\mathbf{s} \in \mathcal{P}_M$ and $(U_i)_{i=1}^\infty$ be a sequence of independent, identically distributed Uniform random variables on $(0, 1)$. Let $\vartheta$ be an interval representation of the mass partition $\mathbf{s}$, for instance, from left to right: $J_1 = (0, s_1), J_2 = (s_1, s_1 + s_2), \ldots$ and, if the dust component $s_0$ is positive, it can be put in the remainder of the interval. An equivalence relation can then be defined relating the interval representation to the sequence of uniform random variables. It associates the corresponding component of the interval representation to one or more of the random variables in $(U_i)_{i=1}^\infty$ that fall into it, this is denoted by $\overset{\vartheta}{\sim}$, as follows

$$i \overset{\vartheta}{\sim} j \iff i = j \quad \text{or} \quad U_i \text{ and } U_j \text{ belong to the same interval component of } \vartheta. \tag{2.13}$$

This procedure produces a partition of $\mathbb{N}$. Lemma 2.7 of Bertoin (2006, Chapter 2) states that the paintbox's law does not depend on the interval representation of the mass partition $\mathbf{s} \in \mathcal{P}_M$ and that the generated partition is exchangeable.

**Definition 11.** *(Asymptotic frequency)* *Given a partition $\Pi = \{B_1, B_2, \ldots\}$ and a block $B \in \Pi$, the* asymptotic frequency*, denoted by $|B|$, is given by the following quantity (if it exists)*

$$|B| = \underset{n \to \infty}{lim} \frac{1}{n} Card(B \cap [n])$$

*it is a measure of the block's relative size. If each block of $\Pi$ has an asymptotic frequency then we say that $\Pi$ possesses asymptotic frequencies.*

**Theorem 1.** *(de Finetti)* *Let $\xi_1, \xi_2, \ldots$ be an exchangeable sequence of real-valued variables, that is, for every finite permutation $\sigma$ of $\mathbb{N}$, $(\xi_1, \xi_2, \ldots)$ and $(\xi_{\sigma(1)}, \xi_{\sigma(2)}, \ldots)$ have the same finite-dimensional distributions. Then the sequence of empirical distributions*

$$\mu_n(dx) := \frac{1}{n} \sum_{i=1}^n \delta_{\xi_i}(dx)$$

*converges almost-surely when $n \to \infty$, in the sense of weak convergence of probability measures in $\mathbb{R}$, to some random probability measure $\mu(dx)$. Moreover, conditionally on $\mu$, the variables $\xi_1, \xi_2, \ldots$ are independent and identically distributed with joint distribution $\mu$.*

See Durret (1996, Chapter 4, Example 6.4) for a detailed proof of de Finetti's theorem based on a reverse martingale argument and Billingsley (1999) for a comprehensive account of convergence of probability measures.

**Theorem 2.** *(Kingman's paintbox construction) Let $\Pi$ be an exchangeable random partition of $\mathbb{N}$. Then $\Pi$ possesses asymptotic frequencies almost-surely. More precisely, the law of $\Pi$ can be expressed as a mixture of paintboxes*

$$Pr(\Pi \in \cdot) = \int_{\mathcal{P}_M} Pr(|\Pi|^\downarrow \in d\mathbf{s}) \varrho_{\mathbf{s}}(\cdot), \qquad (2.14)$$

where $\varrho_{\mathbf{s}}$ stands for the law of the paintbox based on $\mathbf{s}$.

The following proof is the one given in Bertoin (2006, Chapter 2) and reviewed in Berestycki (2009) with additional details. The original proof by Kingman (1982) is based on a reverse martingale argument.

*Proof.* Let $\Pi_n = \{A_1, \ldots, A_k\}$ be a partition of $[n]$ with $k$ blocks and $b : \mathbb{N} \to \mathbb{N}$, a function that maps all points of each block of the partition to a point in that block, e.g. $b(i) = \min_j (j \in A_\ell : i \in A_\ell)$ and $(U_i)_{i \in \mathbb{N}}$ a sequence of independent, identically distributed uniform random variables on $(0,1)$. Let $\xi_i = U_{b(i)}$, this relates the uniform random variable to the partition via the selection map $b$. If we pick a permutation $\sigma$ of $[n]$, $\xi_{\sigma(i)} = U_{b(\sigma(i))} = U'_{b'(i)}$ where $U'_j = U_{\sigma(j)}$ and $b' = \sigma^{-1} \circ b \circ \sigma$. Thus, it turns out that $b'$ is also a selection map of the partition $\sigma(\Pi_n)$ and $(U'_i)_{i \in \mathbb{N}}$ are i.i.d. independent of $\sigma(\Pi_n)$ and the selection map $b'$.

It follows that the sequence $(\xi_1, \xi_2, \ldots)$ is exchangeable, by de Finetti's theorem, there exists $\mu$ such that the sequence is conditionally i.i.d. Let $(V_i)_{i \in \mathbb{N}}$ be a sequence of i.i.d. uniform random variables in $(0,1)$, $x \in \mathbb{R}$ and $q(x) = \inf(y \in \mathbb{R} : \mu((0,y)) > x)$ is the quantile function. Let $\vartheta = \{x \in (0,1) : \exists \epsilon > 0 \text{ such that } q(x) = q(y) \text{ when } |x-y| < \epsilon\}$ be an interval partition. The interval components of $\vartheta$ coincide with the atoms of the masses of $\mu$. Thus, $q(V_1), q(V_2), \ldots$ has the same law as $\xi_1, \xi_2, \ldots$ conditionally on $\mu$. Then $i$ and $j$ will be in the same block if and only if $V_j$ and $V_i$ belong to the same interval component of $\vartheta$, i.e. $i \overset{\vartheta}{\sim} j$. Hence, conditionally on $\mu$, $\Pi_n$ is distributed as a paintbox based on $\vartheta$. □

Theorem 2 gives an integral representation for the law of an exchangeable partition of $\mathbb{N}$. The finite dimensional distributions inherit this invariance under permutations of its indices. The way in which the finite dimensional distributions of an exchangeable partition satisfy this invariance is that they do not depend on the particular elements that belong to each block.

The first example provided herein of the finite dimensional distributions of an exchangeable partition of $\mathbb{N}$ is the Chinese restaurant process. In that case, the corresponding finite dimensional distribution, given in Equation (2.6), assigns the same probability to all partitions with the same number of blocks and cardinalities per block. This property is shared among the finite dimensional distributions of all exchangeable random partitions. For this reason, it is convenient to define a space where all partitions that share the same number of blocks and cardinalities per block are mapped to one object, called a composition of $n$. Indeed, a composition of $n$ is a many-to-one mapping from the set of partitions of $[n]$: there are $\binom{n}{\lambda_1,\ldots,\lambda_k}$ partitions of size $k$ that have cardinalities given by $(\lambda_1, \ldots, \lambda_k)$. The finite dimensional distributions of all exchangeable random partitions are stated in terms of a composition of $n$, as follows

**Definition 12.** *(Exchangeable partition probability function (EPPF))* Let $\varphi = \{A_1, \ldots, A_k\}$ be a partition of $[n]$ of size $k$ and let $\Pi$ be an exchangeable random partition of the set of natural numbers $\mathbb{N}$. Exchangeability implies that the law of its restriction to $[n]$, denoted by $\Pi_n$, can be expressed in the form

$$Pr(\Pi_n = \varphi) = p(|A_1|, \ldots, |A_k|) \qquad (2.15)$$

*for some non-negative function $p(\underline{\lambda})$ of compositions $\underline{\lambda} := (\lambda_1, \ldots, \lambda_k)$ of $n$ such that $p$ is symmetric in its arguments. Furthermore, suppose that $p$ satisfies the following sum rule*

$$p(\underline{\mu}) = \sum_{\underline{\lambda} \downarrow \underline{\mu}} p(\underline{\lambda}) \qquad (2.16)$$

*where the sum is over all compositions $\underline{\mu}$ derived from $\underline{\lambda}$ by increasing the cardinality of an existing block or by appending a new block of cardinality one. Then, Equation (2.15) is called* exchangeable partition probability function *of $\Pi$.*

In the next subsection, we review a subclass of exchangeable random partitions such that their EPPFs have a specific factorised form.

### 2.3.2 Exchangeable Gibbs partitions

**Definition 13.** *(Gibbs-type EPPF)* An exchangeable random partition $\Pi$ of the set of natural numbers $\mathbb{N}$ is said to be of Gibbs form if, for some nonnegative weights

$\mathbf{W} = (W_j)_{j=1}^{\infty}$ and $V_{n,k}$, the EPPF of $\Pi$ satisfies

$$p(\lambda_1, \ldots, \lambda_k) = V_{n,k} \prod_{j=1}^{k} W_{\lambda_j} \qquad (2.17)$$

$\forall k \in \{1, \ldots, n\}$ and for a composition $(\lambda_1, \ldots, \lambda_k)$ of $n$.

The following normalisation condition makes it a valid probability distribution

$$\sum_{k=1}^{n} V_{n,k} B_{n,k}(\mathbf{W}) = 1, \quad n \in \mathbb{N}, \qquad (2.18)$$

where the partial Bell polynomial is

$$B_{n,k}(\mathbf{W}) = \sum_{\{A_1, \ldots, A_k\}} \prod_{j=1}^{k} W_{|A_j|} = \frac{1}{k!} \sum_{(\lambda_1, \ldots, \lambda_k)} \binom{n}{\lambda_1, \ldots, \lambda_j} \prod_{j=1}^{k} W_{\lambda_j}.$$

The sum on the left hand side is over all partitions with $k$ blocks and cardinalities per block $|A_1|, \ldots, |A_k|$. On the right hand side, the sum is over those compositions of n $\{\lambda_1, \ldots, \lambda_k\}$ multiplied by the number of partitions that are mapped to the specific composition, given by $\binom{n}{\lambda_1, \ldots, \lambda_k}$. Suppose that we start with an EPPF that factorises as stated in Definition 13. The functional form of the infinite sequence of $\mathbf{W}$-weights can be derived if we impose the compatibility requirement of Equation (2.16) on the subset of weights associated to a partition of $[n]$ of size $k$ with cardinalities $(\lambda_1, \ldots, \lambda_k)$. This form is given in the following Lemma and corresponds to Lemma 2 of Gnedin and Pitman (2006).

**Lemma 1.** *The weights $\mathbf{W}$ and $V_{n,k}$ with $W_1 = V_{1,1}$ define a partition of Gibbs form if and only if for some $b \geq 0$ and $a \leq b$ the following two conditions are satisfied*

i) $W_j = (b-a)_{j-1 \uparrow b}$, $j = 1, 2, \ldots$

ii) the $V_{n,k}$ satisfy the recursion

$$V_{n,k} = (bn - ak)V_{n+1,k} + V_{n+1,k+1} \quad 1 \leq k \leq n. \qquad (2.19)$$

Where $(b-a)_{j-1 \uparrow b}$ denotes the ascending factorial, see Apendix A.2 for details about how to compute this quantity.

*Proof.* ⇒)Let $\underline{\mu}$ denote all compositions of $n+1$ whose restriction to $n$ is $\underline{\lambda}$. From the compatibility requirement of Equation (2.16), if the factorised form from Equation (2.17) is assumed and, without loss of generality, that $W_1 = V_{1,1} = 1$ ( the Gibbs-type EPPF can be multiplied and divided by $W_1$ and the corresponding weights can be redefined to include this term) we have

$$p(\underline{\lambda}) = \sum_{\underline{\mu} \downarrow \underline{\lambda}} p(\underline{\mu})$$

$$V_{n,k} \prod_{j=1}^{k} W_{\lambda_j} = V_{n+1,k} \sum_{\ell=1}^{k} \prod_{j \leqslant k, j \neq \ell} W_{\lambda_j} \times W_{\lambda_\ell + 1} + V_{n+1,k+1} \prod_{j=1}^{k} W_{\lambda_j} \times 1$$

$$V_{n,k} = V_{n+1,k} \sum_{\ell=1}^{k} \frac{W_{\lambda_\ell + 1}}{W_{\lambda_\ell}} + V_{n+1,k+1}.$$

Let $r_j = \frac{W_{j+1}}{W_j}$, then we can write

$$\frac{V_{n,k} - V_{n+1,k+1}}{V_{n+1,k}} = \sum_{\ell=1}^{k} r_{\lambda_\ell}.$$

Note that the right hand side does not depend on $n$. Let $k = 2$

$$\frac{V_{n,2} - V_{n+1,3}}{V_{n+1,2}} = r_1 + r_2.$$

Note that for $i \geqslant 0$ with $r_0 = 0$

$$r_i + r_{i+1} \pm (r_{i-1} - r_{i+2}) =$$
$$= r_i - r_{i-1} - (r_{i+2} - r_{i+1}) + (r_{i+1} + r_{i+2})$$

then,

$$\frac{V_{n,2} - V_{n+1,3}}{V_{n+1,2}} = r_1 - r_0 - (r_3 - r_2) + (r_0 + r_3)$$

then,

$$r_{i+2} + r_{i+1} = r_i - r_{i-1} = r_{i+2} - r_{i+1} = Cte, \quad \forall i \in \mathbb{N}.$$

Hence, the finite difference of the sequence $(r_j)_{j \in \mathbb{N}}$ is a constant $Cte$. This means that the sequence $(r_j)_{j \in \mathbb{N}}$ is an arithmetic sequence of the form

$$r_j = bj - a \quad j \in \mathbb{N}.$$

For the **W**-weights

$$\begin{aligned}
W_{\lambda_j} &= r_{\lambda_{j-1}} \times r_{\lambda_{j-2}} \times \ldots r_1 \\
&= \prod_{j=1}^{\lambda_j - 1} (b(\ell - 1) + b - a) \\
&= (b - a)_{\lambda_j \uparrow b}.
\end{aligned}$$

Let $\sigma = \frac{a}{b}$,

$$\begin{aligned}
&= b^{\lambda_j}(1 - \sigma)_{\lambda_j \uparrow 1} \\
&= b^{\lambda_j} \frac{\Gamma(\lambda_j - \sigma)}{\Gamma(1 - \sigma)}.
\end{aligned}$$

$\square$

After deriving the **W**-weights and if they are substituted in the EPPF of Equation (2.17), the additional constants $b^{\lambda_j}$ can be added to the $V_{n,k}$. For this reason, the functional form for the **W**-weights can be reduced to a single parameter $\sigma \in (-\infty, 1]$. The backward recursion (2.19) does not have a unique solution but there exists a set of solutions which is a convex set. The parameter $\sigma$ determines the type of elements in this set. Theorem 12 of Gnedin and Pitman (2006) states that each Gibbs partition of type $\sigma \in (-\infty, 1]$ is a mixture of an extreme element of such convex set. There are three qualitatively different cases of interest

i) When $\sigma \in (-\infty, 0)$, the set of extreme points is discrete and corresponds to the two parameter Chinese restaurant family, with parameters $(\sigma, |\sigma|m)$, $m \in \mathbb{N}$ and

$$V_{n,k} = \frac{|\sigma|^k \Gamma(m + 1) \Gamma(m|\sigma|)}{\Gamma(m - k + 1) \Gamma(m|\sigma| + n)}.$$

ii) When $\sigma = 0$, the set is continuous and corresponds to the Chinese restaurant

family, with parameter $\theta$, $\theta \in \mathbb{R}^+$ and

$$V_{n,k} = \frac{\theta^k \Gamma(\theta)}{\Gamma(\theta + n)}.$$

iii) When $\sigma \in (0,1)$, it corresponds to Poisson–Kingman partitions, which are not extreme points, with

$$V_{n,k} = \int_{\mathbb{R}^+} \int_0^t t^{-n}(t-s)^{n-1-k\sigma} h(t) f_\sigma(s) dt ds.$$

These three cases lead to different asymptotic behaviours for the expected number of blocks of a partition of $[n]$ when $n \to \infty$. Figure 2.2 shows some families and subfamilies of Bayesian nonparametric priors that are Gibbs-type, see Gnedin et al. (2007) for further details. All these priors could have additional parameters apart from the $\sigma$ parameter, but each additional parameter only affects the $V_{n,k}$ weight. For this reason, all Gibbs-type priors share the same functional form for the **W**-weights.

Let the cardinalities of the blocks be integrated out in Equation (2.17), another property of Gibbs-type priors is that the marginal distribution for the number of blocks $K_n$ in a partition of $[n]$ can be obtained as

$$\Pr(K_n = k) = V_{n,k} B_{n,k}(\mathbf{W}).$$

The distribution of an exchangeable partition, conditional on a given number of blocks, is the following

$$\Pr(\Pi_n = \{A_1, \ldots, A_k\} \mid K_n = k) = \frac{\prod_{j=1}^k W_{|A_j|}}{B_{n,k}(\mathbf{W})}. \tag{2.20}$$

Figure 2.1 depicts the general form for the backward recursion of Equation (2.19), each node corresponds to a $V_{n,k}$ parameter. The first index $n$ denotes the set of integers for a partition of $[n]$ with $k$ blocks. In the Gibbs-type case, this graph is called a generalised Stirling triangle with $\gamma_{n,k} = n - \sigma k$ and $\delta_{n,k} = 1$ which correspond to the coefficients of the backward recursion from Equation (2.19).

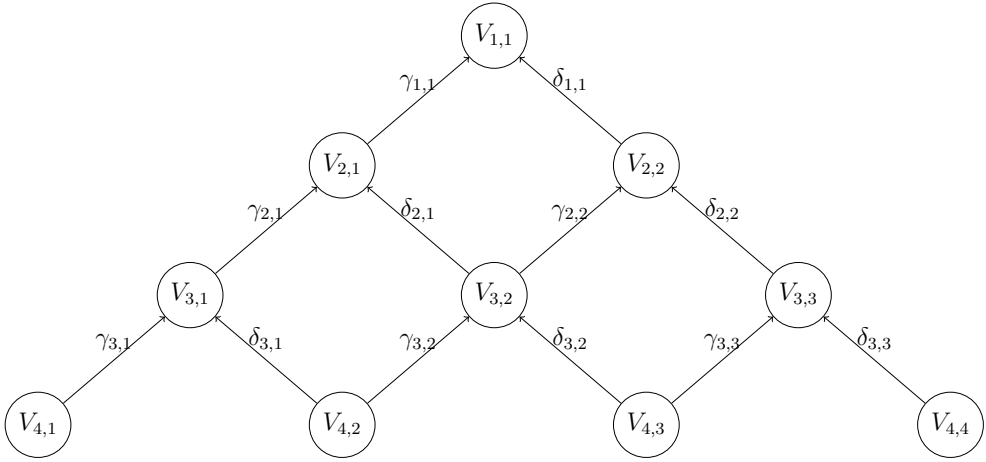

Figure 2.1: Infinite simplex $\mathcal{V}_\sigma$ and the general case for the backward recursion of Lemma 1. This graph is called Pascal or Tartaglia triangle.

## 2.4 Two types of random probability measures

In the previous sections, a compact metric on the space of mass partitions was stated such that, once we endow this space with any probability measure, this measure is tight since the compactness is inherited (Parthasarathy, 1967, Chapter 6, Theorem 6.4). The tightness property implies that convergence of the finite dimensional distributions is equivalent to weak convergence of probability measures (Billingsley, 1999). For this reason, we only need to review how to construct a probability measure on the space of mass partitions and these are guaranteed to be well-behaved. A mass partition can be sampled if we start with a discrete random measure on $(0, \infty)$. After we are given a particular realisation of a mass partition, we can then associate a discrete random variable that takes infinitely many values with probabilities given by the mass partition. Hence, a random mass partition is also referred to as a random discrete distribution that takes infinitely many values, although the latter will not necessarily be arranged in decreasing order. The tightness of probability measures is still preserved since the underlying space for the random discrete distributions is the space numerical sequences in $[0,1]$, denoted by $\mathcal{S}_{[0,1]}$ which is compact with respect to the metric $d_{\mathcal{S}_{[0,1]}}$ from Equation (2.11).

There are two possible random discrete distributions that can be obtained from a specific discrete random measure called Completely Random measure (CRM). Infor-mally, a CRM is a discrete random measure with an independence property. One of them is the well known random discrete distributions called Normalised Random Measures with Independent Increments (Regazzini et al., 2003; James et al., 2009). They are obtained from CRMs after a suitable normalisation operation. Another random discrete distributions, obtained by conditioning rather than normalising the corresponding CRMs, are called Poisson–Kingman Random Probability Measure.

**Definition 14.** *(Completely Random Measure) Let $\mathbb{X}$ be a complete and separable metric space endowed with the Borel $\sigma$-field $\mathcal{B}(\mathbb{X})$. A Completely Random Measure $\mu$ is a random element taking values on the space of boundedly finite measures on $\mathbb{X}$ such that, for any collection of measurable subsets, $A_1, \ldots, A_n$ in $\mathcal{B}(\mathbb{X})$, with $A_i \cap A_j = \varnothing$ for $i \neq j$, the random variables $\mu(A_1), \ldots, \mu(A_n)$ are mutually independent.*

CRMs were first proposed and studied by Kingman (1967), who showed that a CRM $\mu$ can always be decomposed into a sum of three independent parts

$$\mu = \mu_0 + \sum_{k \geqslant 1} u_k \delta_{\phi_k} + \sum_{\ell=1}^{N} v_\ell \delta_{\psi_\ell}$$

where $\mu_0$ is a (non-random) measure over $\mathbb{X}$, $\{\psi_\ell\}_{\ell \in [N]} \subset \mathbb{X}$ is a collection of $N$, $1 \leqslant N \leqslant \infty$, atoms at fixed locations and independent random masses $(v_\ell)_{\ell \in [N]}$, and $(u_k, \phi_k)_{k \geqslant 1}$ is a collection of atoms with random masses and random locations.

In applications of Bayesian nonparametrics it is usually assumed that $\mu_0 = 0$ and $N = 0$, so that $\mu$ consists only of the atoms with random masses and locations. However, the posterior distribution of $\mu$ given data would typically contain atoms at fixed locations, hence, the usefulness of the larger class of CRMs.

The distribution of the random atoms and their masses under a CRM is characterized by the Lévy–Khintchine representation of its Laplace functional transform. Specifically,

$$\mathbb{E}\left[e^{-\int g(y)\mu(dy)}\right] = \exp\left\{-\int_{\mathbb{R}^+ \times \mathbb{X}} \left(1 - e^{-sg(y)}\right) \rho(ds) H_0(dy)\right\}, \qquad (2.21)$$

for any measurable function $g : \mathbb{X} \to \mathbb{R}$ such that $\int_{\mathbb{X}} |g(x)|\mu(dx) < +\infty$ almost-surely. The underlying measure $\nu = \rho \times H_0$ is uniquely characterises the random atoms in $\mu$. The only intensity measures that are considered herein are those that factorise $\nu(ds, dy) = \rho(ds)H_0(dy)$ for some measure $\rho$ on $\mathbb{R}^+$ absolutely continuous with respect to the Lebesgue measure, and some non-atomic probability measure $H_0$ on $\mathbb{X}$. The corresponding CRM is said to be *homogeneous* and write $\text{CRM}(\rho, H_0)$ for the law of $\mu$. Succesively, the measure $\rho$ is referred to as the Lévy measure, while $H_0$ is the base distribution. Homogeneity implies independence between $(u_k)_{k \geqslant 1}$ and $(\phi_k)_{k \geqslant 1}$, where $(\phi_k)_{k \geqslant 1}$ is a sequence of random variables independent and identically distributed according to $H_0$ while the law of $(u_k)_{k \geqslant 1}$ is governed by $\rho$. Intuitively, the point process $(u_k, \phi_k)_{k \geqslant 1}$ is described by a Poisson process over $\mathbb{R}^+ \times \mathbb{X}$ with intensity measure $\nu = \rho \times H_0$, and (2.21) is the characteristic functional of the Poisson process evaluated at the function $(s, y) \mapsto sg(y)$.

It is required that $\mu$ has almost-surely finite total mass, Equation (2.21) with $g(y) = 1$ shows that the Lévy measure is required to satisfy the property

$$\int_{\mathbb{R}^+ \times \mathbb{X}} \left(1 - e^{-s}\right) \rho(ds) H_0(dy) = \int_{\mathbb{R}^+} (1 - e^{-s}) \rho(ds) < \infty.$$

Furthermore, the expected number of random atoms in $\mu$ is obtained by Campbell's Theorem to be the total mass of the Lévy measure $\rho(\mathbb{R}^+)$. In typical applications of Bayesian nonparametrics this is infinite, so we can work with mixture models with infinite number of components. This also guarantees that the total mass is positive almost-surely.

**Definition 15.** *(Normalised Random Measure)* *Let $\mu$ be a homogeneous CRM, with Lévy measure $\rho$ and base distribution $H_0$, with almost-surely positive and finite total mass. A* Normalised Random Measure *(NRM) is an almost-surely discrete random probability measure $P$ on $\mathbb{X}$ obtained by normalising $\mu$*

$$P = \frac{\mu}{T} = \sum_{k \geqslant 1} p_k \delta_{\phi_k}$$

*with $T = \sum_{k \geqslant 1} u_k$ and $p_k = u_k / T$. Since $\mu$ is homogeneous, the law of $(p_k)_{k \geqslant 1}$ is governed by the Lévy measure $\rho$ and the atoms $(\phi_k)_{k \geqslant 1}$ are a sequence of random variables independent of $(p_k)_{k \geqslant 1}$, and independent and identically distributed according to $H_0$. We denote it by $P \sim \text{NRM}(\rho, H_0)$.*

Let $\mu \sim \text{CRM}(\rho, H_0)$ with an almost-surely finite and positive total mass $T$. Suppose that $T$ is positive and finite almost-surely, and absolutely continuous with respect to Lebesgue measure with density $f_\rho(t)$, see Sato (2013); Applebaum (2009) and Ap-pendix A.3 for details on how to compute the density for the total mass $T$, denoted by $f_\rho$, using the Lévy measure $\rho$.

Let $(X_i)_{i \geqslant 1}$ be a sequence of random variables that, given $P$, are independent and identically distributed according to $P$. Since $\mu$ is almost-surely discrete, there is a positive probability that $X_i = X_j$ for each pair $i \neq j$, i.e. when both are assigned to the same atom in $P$. This induces a partition $\Pi$ on $\mathbb{N}$, where $i$ and $j$ are in the same block in $\Pi$ if and only if $X_i = X_j$. The random partition $\Pi$ is exchangeable, and its exchangeable partition probability function (EPPF) can be deduced from the law of the NRM. Kingman (1975) demonstrates that there is a correspondence between the law of $\Pi$ and that of the random probability measure, this fact is reviewed in the previous sections.

**Definition 16.** *(Poisson–Kingman random probability measure)*

*Poisson–Kingman RPMs were introduced in Pitman (2003) as a generalisation of homogeneous NRMs. Let $\mu \sim \text{CRM}(\rho, H_0)$ and let $T = \mu(\mathbb{X})$ be finite, positive almost-surely, and absolutely continuous with respect to Lebesgue measure. For any $t \in \mathbb{R}^+$, let us consider the conditional distribution of $\mu/t$ given that the total mass $T \in dt$. This distribution is denoted by $\text{PK}(\rho, \delta_t, H_0)$, were $\delta_t$ denotes the Dirac delta function. Poisson–Kingman RPMs form a class of RPMs whose distributions are obtained by mixing $\text{PK}(\rho, \delta_t, H_0)$, over $t$, with respect to some distribution $\gamma$ on the positive real line. Specifically, a* Poisson–Kingman RPM *has the hierarchical representation*

$$T \sim \gamma$$
$$P|T = t \sim \text{PK}(\rho, \delta_t, H_0). \qquad (2.22)$$

*The RPM $P$ is referred to as the Poisson–Kingman RPM with Lévy measure $\rho$, base distribution $H_0$ and mixing distribution $\gamma$. The distribution of $P$ is denoted by $\text{PK}(\rho, \gamma, H_0)$. If the density for the total mass equals the density obtained from its Lévy measure $f_\rho$, i.e. $\gamma(dt) = f_\rho(t)dt$, then the distribution $\text{PK}(\rho, f_\rho, H_0)$ coincides with $\text{NRM}(\rho, H_0)$. See Appendix A.3 for details on how to compute the density $f_\rho(t)$ from the Lévy measure $\rho$. Since $\mu$ is homogeneous, the atoms $(\phi_k)_{k \geqslant 1}$ of $P$ are independent of their masses $(p_k)_{k \geqslant 1}$. They form a sequence of independent random variables identically distributed according to $H_0$. Finally, the masses of $P$ have distribution governed by the Lévy measure $\rho$ and the distribution $\gamma$.*

## 2.5 Poisson–Kingman characterisations

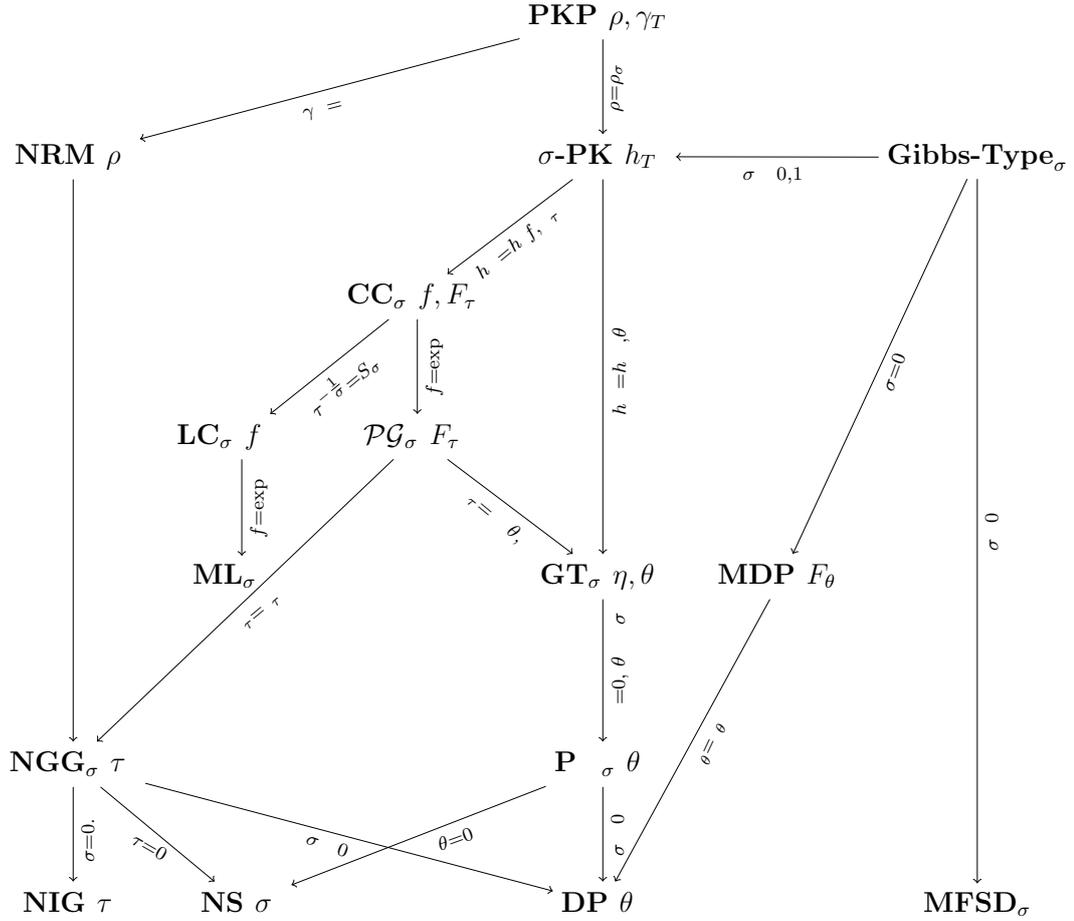

Figure 2.2: Poisson–Kingman processes

Figure 2.2 shows various Poisson–Kingman random probability measures, this is a very wide class of Bayesian nonparametric priors. Indeed, this family of priors encompasses mostly all of the Bayesian nonparametric priors used in the literature. Some interesting relationships between different classes and subclasses are shown together with other not so well-known priors. These include the Gamma tilted (GT) class, the Poisson-Gamma ($\mathcal{PG}$) class (James, 2013), the composition of classes (CC) and Mittag–Leffler classes (ML) (Ho et al., 2008), among others. See the examples section in Chapter 3 to see the specific form for the mixing distribution $\gamma$, which governs the behaviour of the total mass $T$ random variable.

However, the conditioning operation from the generative process from Equation (2.22) is not well defined *a priori* but the following proposition from Perman et al. (1992) helps to bypass this difficulty. It also states an interesting Markovian property.

**Proposition 1.** *(Perman et al., 1992) The sequence of surplus masses $(V_k)_{k \geqslant 0}$ forms a Markov chain, where $V_k := T - \sum_{\ell=1}^{k} \tilde{J}_\ell$, with initial distribution and transition kernels given as follows*

$$Pr_{\rho,H_0}(V_0 \in dv_0) = f_\rho(v_0)dv_0$$

$$Pr_{\rho,H_0}(V_k \in dv_k | V_1 \in dv_1, \ldots, V_{k-1} \in dv_{k-1}) = Pr_{\rho,H_0}(V_k \in dv_k | V_{k-1} \in dv_{k-1})$$
$$= \frac{(v_k - v_{k-1})\rho(d(v_k - v_{k-1}))}{v_{k-1}} \frac{f_\rho(v_k)}{f_\rho(v_{k-1})}.$$

This proposition can be proven by induction, with each step being an application of the Palm formula (Kingman, 1993) for Poisson processes. The proof is similar to the proof of Proposition 2.4 in Bertoin (2006, Chapter 2).

*Proof.* **Induction over $k$, where $k$ is the number of size-biased picks.** Let $\mu = \sum_{k=1}^{\infty} u_k \delta_{X_k}$ be a homogeneous completely random measure. A size-biased sample is obtained in the following way: first pick the $k$-th atom with probability $\frac{u_k}{T}$ where $T = \sum_{k=1}^{\infty} u_k$, the $k$-th surplus is $V_k = T - \sum_{\ell=1}^{k} \tilde{J}_\ell$ and set $\tilde{J}_k = s_k$.

i) **Case k = 1**. We are interested in the following conditional expectation

$$\mathbb{E}\left(\mathbb{I}_{\{\tilde{J}_1 \in ds_1\}} \mid T \in [t, t+\epsilon]\right) = \frac{\Pr\left(\tilde{J}_1 \in ds_1, T \in [t, t+\epsilon]\right)}{\Pr(T \in [t, t+\epsilon])}. \quad (2.23)$$

Indeed, we can formally define the conditional expectation of interest as a weak limit of the conditional expectations indexed by $\epsilon$, when $\epsilon \to 0+$. In this proof, it is only required for a simple function of the form $f = \delta_{\tilde{J}_1}(ds_1)$. See Laha and Rohatgi (1979) for the formal definition of conditional expectation for an arbitrary measurable function $f$. First, we draw from the random measure

$$\Pr\left(\tilde{J}_1 \in ds_1, T \in dt \mid \mu\right) = \sum_{k=1}^{\infty} \frac{u_k}{T} \delta_{u_k}(ds_1) \delta_T(dt).$$

Successively, if we average over it, we obtain the numerator in (2.23), by the tower rule for expectations (Laha and Rohatgi, 1979),

$$\Pr\left(\tilde{J}_1 \in ds_1, T \in [t, t+\epsilon]\right) = \Pr\left(\sum_{k=1}^{\infty} \frac{u_k}{T} \delta_{u_k}(ds_1), T \in [t, t+\epsilon]\right)$$

$$= \int \rho(y) y \delta_y(ds_1) \mathbb{E}\left((y+T)^{-1}\mathbb{I}_{\{y+T\in[t,t+\epsilon]\}}\right)$$

$$\to \frac{\epsilon}{t}\rho(ds_1) s_1 f_\rho(t-s_1) \quad \text{as} \quad \epsilon \to 0+.$$

For the second equality, we use Palm's Formula (Kingman, 1993) where $G(M,f) = \frac{1}{T}\delta_y(ds_1)\mathbb{I}_{[t,t+\epsilon]}(T)$. For the third,

$$\mathbb{E}\left((y+T)^{-1}\mathbb{I}_{\{y+T\in[t,t+\epsilon]\}}\right) = \int_{[t-y,t-y+\epsilon]} (y+z)^{-1} f_\rho(z)dz$$

$$\to \epsilon \frac{f_\rho(t-y)}{t} \quad \text{as} \quad \epsilon \to 0+.$$

Again, we use the fact that

$$\lim_{\epsilon\to 0+} \frac{1}{\epsilon}\int_{[x,x+\epsilon]} p(z)dz = \int \delta_z(x)p(z)dz = p(x)$$

to obtain the denominator, hence,

$$\Pr\left(\tilde{J}_1 \in ds_1 \mid T \in [t,t+\epsilon]\right) \to \frac{s_1 \rho(ds_1) f_\rho(t-s_1)}{t f_\rho(t)} \quad \text{as} \quad \epsilon \to 0+.$$

ii) **Suppose the statement holds true for $k > 1$, i.e.**

$$\Pr\left(\tilde{J}_k \in ds_k \mid T \in [t,t+\epsilon], \tilde{J}_\ell \in ds_\ell, \forall \ell \in [k-1]\right) =$$

$$\Pr\left(\tilde{J}_k \in ds_k \mid T - \sum_{\ell=1}^{k-1}\tilde{J}_\ell \in \left[t - \sum_{\ell=1}^{k-1} ds_\ell, t - \sum_{\ell=1}^{k-1} ds_\ell + \epsilon\right]\right)$$

$$\frac{s_k f_\rho(t - \sum_{\ell=1}^k s_\ell)}{(t - \sum_{\ell=1}^{k-1} s_\ell) f_\rho(t - \sum_{\ell=1}^{k-1} s_\ell)} \rho(ds_k) \quad \text{as} \quad \epsilon \to 0+.$$

iii) **Case $k+1$.**

$$\Pr\left(\tilde{J}_{k+1} \in ds_{k+1} \mid T \in [t,t+\epsilon], \tilde{J}_\ell \in ds_\ell, \forall \ell \in [k]\right) =$$

$$\frac{\Pr\left(\tilde{J}_{k+1} \in ds_{k+1}, \tilde{J}_\ell \in ds_\ell, \forall \ell \in [k], T \in [t,t+\epsilon]\right)}{\Pr\left(\tilde{J}_\ell \in ds_\ell, \forall \ell \in [k], T \in [t,t+\epsilon]\right)}.$$

The denominator is given by the induction hypothesis (IH) and the numerator is

$$\Pr\left(\tilde{J}_{k+1} \in ds_{k+1}, \tilde{J}_\ell \in ds_\ell, \forall \ell \in [k], T \in [t, t+\epsilon]\right) =$$

$$\Pr\left(\sum_{k=1}^{\infty} \frac{u_k}{T} \delta_{u_k}(ds_\ell), \forall \ell \in [k+1], T \in [t, t+\epsilon])\right) \stackrel{IH}{=}$$

$$\int \rho(y) y \delta_y(ds_{k+1}) \mathbb{E}\left((y + T - \sum_{\ell=1}^{k} \tilde{J}_\ell)^{-1} \mathbb{I}_{\{y+T-\sum_{\ell=1}^{k} \tilde{J}_\ell \in [t-\sum_{\ell=1}^{k} s_\ell, t-\sum_{\ell=1}^{k} s_\ell+\epsilon]\}}\right) \to$$

$$\frac{\epsilon}{(t - \sum_{\ell=1}^{k} s_\ell)} \rho(ds_{k+1}) s_{k+1} f_\rho(t - \sum_{\ell=1}^{k+1} s_\ell) \quad \text{as} \quad \epsilon \to 0+.$$

Again, the IH is used in the second equality. Since the sequence of the first $k$ surplus masses satisfies the Markov property so it is enough to condition on the last surplus mass. Finally,

$$\Pr\left(\tilde{J}_{k+1} \in ds_{k+1} \mid T \in [t, t+\epsilon], \tilde{J}_\ell \in ds_\ell \forall \ell \in [k]\right) =$$

$$\frac{s_{k+1} f_\rho(t - \sum_{\ell=1}^{k+1} s_\ell)}{(t - \sum_{\ell=1}^{k} s_\ell) f_\rho(t - \sum_{\ell=1}^{k} s_\ell)} \rho(ds_{k+1}) \quad \text{as} \quad \epsilon \to 0+.$$

$\square$

A second object induced by a Poisson–Kingman RPM is a size-biased permutation of its atoms. Specifically, order the blocks in $\Pi$ by increasing order of the least element in each block, and for each $\ell \in \mathbb{N}$ let $Z_\ell$ be the least element of the $\ell$-th block. $Z_\ell$ is the index among $(X_i)_{i \geqslant 1}$ of the first appearance of the $\ell$-th unique value in the sequence. Let $\tilde{J}_\ell = \mu(\{X_{Z_\ell}\})$ be the mass of the corresponding atom in $\mu$. Then $(\tilde{J}_\ell)_{\ell \geqslant 1}$ is a size-biased permutation of the masses of atoms in $\mu$, with larger masses tending to appear earlier in the sequence. It is easy to see that $\sum_{\ell \geqslant 1} \tilde{J}_\ell = T$, and that the sequence can be understood as a stick-breaking construction: start with a stick of length $T_0 = T$; break off the first piece of length $\tilde{J}_1$; the surplus length of stick is $V_1 = T - \tilde{J}_1$; then the second piece with length $\tilde{J}_2$ is broken off, etc.

Proposition 1 (Perman et al., 1992) states that the sequence of surplus masses $(V_\ell)_{\ell \geqslant 0}$ forms a Markov chain and gives the corresponding initial distribution and transition kernels. The density of $T$ is denoted by $\gamma(t) \propto h(t) f_\rho(t)$. The PKP generative process for the sequence $(X_i)_{i \geqslant 1}$ goes as follows, where parts of the PK random measure $\mu$ are simulated as required.

i) Draw from the total mass from its distribution $\Pr_{\rho,H_0}(T \in dt) \propto h(t) f_\rho(t) dt$.

ii) The first draw $X_1$ from $\mu/T$ is a size-biased pick from the masses of $\mu$. The actual value of $X_1$ is simply $X_1^* \sim H_0$, while the mass of the corresponding atom in $\mu$ is $\tilde{J}_1$, with conditional distribution given by

$$\Pr_{\rho,H_0}(\tilde{J}_1 \in ds_1 | T \in dt) = \frac{s_1}{t} \rho(ds_1) \frac{f_\rho(t - s_1)}{f_\rho(t)}.$$

The leftover mass is $V_1 = T - \tilde{J}_1$.

iii) For subsequent draws $i \geq 2$:

- Let $k$ be the current number of distinct values among $X_1, \ldots, X_{i-1}$, and let $X_1^*, \ldots, X_k^*$ be the unique values, i.e., atoms in $\mu$. The masses of these first $k$ atoms are denoted by $\tilde{J}_1, \ldots, \tilde{J}_k$ and the leftover mass is $V_k = T - \sum_{\ell=1}^k \tilde{J}_\ell$.
- For each $\ell \leq k$, with probability $\tilde{J}_\ell/T$, we set $X_i = X_\ell^*$.
- With probability $V_k/T$, $X_i$ takes on the value of an atom in $\mu$ besides the first $k$ atoms. The actual value $X_{k+1}^*$ is drawn from $H_0$, while its mass is drawn from

$$\Pr_{\rho,H_0}(\tilde{J}_{k+1} \in ds_{k+1} | V_k \in dv_k) = \frac{s_{k+1}}{v_k} \rho(ds_{k+1}) \frac{f_\rho(v_k - s_{k+1})}{f_\rho(v_k)}.$$

The leftover mass is again $V_{k+1} = V_k - \tilde{J}_{k+1}$.

By multiplying the above infinitesimal probabilities one obtains the joint distribution of the random elements $T$, $\Pi$, $(\tilde{J}_i)_{i \geq 1}$ and $(X_i^*)_{i \geq 1}$. Such a joint distribution was first obtained in Pitman (2003), it is recalled in the next proposition, see also Pitman (2006) for details.

**Proposition 2.** *Let $\Pi_n$ be the exchangeable random partition of $[n] := \{1, \ldots, n\}$ induced by a sample $(X_i)_{i \in [n]}$ from $P \sim \mathrm{PK}(\rho, h, H_0)$. Let $(X_\ell^*)_{\ell \in [k]}$ be the $k$ distinct values in $(X_i)_{i \in [n]}$ with masses $(\tilde{J}_\ell)_{\ell \in [k]}$. Then*

$$Pr_{\rho,H_0}(T \in dt, \Pi_n = (c_\ell)_{\ell \in [k]}, X_\ell^* \in dx_\ell^*, \tilde{J}_\ell \in ds_\ell \text{ for } \ell \in [k]) \qquad (2.24)$$

$$= h(t) t^{-n} f_\rho(t - \sum_{\ell=1}^k s_\ell) dt \prod_{\ell=1}^k s_\ell^{|c_\ell|} \rho(ds_\ell) H_0(dx_\ell^*),$$

where $(c_\ell)_{\ell \in [k]}$ denotes a particular partition of $[n]$ with $k$ blocks, $c_1, \ldots, c_k$, ordered by increasing least element and $|c_\ell|$ is the cardinality of block $c_\ell$.

The distribution (2.24) is invariant to the size-biased order. The usefulness of this proposition can be illustrated with the following examples of particular EPPFs.

**Example 1.** (James et al., 2009) The EPPF of the exchangeable partition $\Pi$ induced by $\mathrm{NRM}(\rho, H_0)$ is given by

$$\mathrm{Pr}_{\rho, H_0}(\Pi_n = (c_\ell)_{\ell \in [k]}) = \int_{\mathbb{R}^+} \frac{u^{n-1}}{\Gamma(n)} e^{-\psi_\rho(u)} du \prod_{\ell=1}^k \kappa_\rho(|c_\ell|, u)$$

where

$$\psi_\rho(u) = -\log \int_{\mathbb{R}^+} e^{-ut} f_\rho(t) dt = \int_{\mathbb{R}^+} (1 - e^{-us}) \rho(ds)$$

$$\kappa_\rho(m, u) = \int_{\mathbb{R}^+} v^m e^{-uv} \rho(dv).$$

Example 1 can be obtained from Proposition 2 by introducing the following disintegration

$$T^{-n} = \int_{\mathbb{R}^+} \frac{u^{n-1} e^{-uT}}{\Gamma(n)} du,$$

performing the change of variables $V = T - \sum_{\ell=1}^k \tilde{J}_\ell$, and marginalising out $V$ and $(\tilde{J}_\ell)_{\ell \in [k]}$.

**Example 2.** The EPPF of the exchangeable partition $\Pi$ induced by $\mathrm{NGG}(\sigma, \tau)$ is given by

$$\mathrm{Pr}_{\rho, H_0}(\Pi_n = (c_\ell)_{\ell \in [k]}) = \int_{\mathbb{R}^+} \frac{u^{n-1}}{\Gamma(n)} e^{-\psi_\rho(u)} du \prod_{\ell=1}^k \kappa_\rho(|c_\ell|, u)$$

where

$$\psi_\rho(u) = -\log \int_{\mathbb{R}^+} e^{-ut} f_\rho(t) dt = \int_{\mathbb{R}^+} (1 - e^{-us}) \rho(ds)$$

$$= \int_{\mathbb{R}^+} (1 - e^{-us}) \frac{a}{\Gamma(1-\sigma)} s^{-\sigma-1} e^{-\tau s} ds$$

$$\kappa_\rho(m, u) = \int_{\mathbb{R}^+} v^m e^{-uv} \rho(dv)$$
$$= \frac{a}{\Gamma(1-\sigma)} \frac{\Gamma(m-\sigma)}{(u+\tau)^{m-\sigma}}.$$

Example 2 can be obtained substituting the NGG's Lévy measure in Example 1.

**Example 3.** The EPPF of the exchangeable partition $\Pi$ induced by $PY(\theta, \sigma)$ is given by

$$\Pr\left(\Pi_n = (c_\ell)_{\ell \in [k]}\right) = \sigma^k \frac{\Gamma(\theta)}{\Gamma(\frac{\theta}{\sigma})} \int_{\mathbb{R}^+} \frac{r^{\theta+k\sigma-1} \exp(-r^\sigma)}{\Gamma(\theta+n)} dr \prod_{\ell=1}^k \frac{\Gamma(|c_\ell|-\sigma)}{\Gamma(1-\sigma)}$$
$$= \sigma^k \frac{\Gamma(\theta)}{\Gamma(\frac{\theta}{\sigma})} \frac{\Gamma(\frac{\theta}{\sigma}+k)}{\Gamma(\theta+n)} \prod_{\ell=1}^k \frac{\Gamma(|c_\ell|-\sigma)}{\Gamma(1-\sigma)}.$$

The EPPF can be obtained from Proposition 2 by introducing the disintegration

$$T^{-(n+\theta)} = \int_{\mathbb{R}^+} \frac{r^{n+\theta-1} e^{-rT}}{\Gamma(n+\theta)} dr,$$

substituting the Lévy measure for the $\sigma$-Stable case, performing the change of variables $V = T - \sum_{\ell=1}^k \tilde{J}_\ell$, $W = R^\sigma$ and marginalising out $V$ and $(\tilde{J}_\ell)_{\ell \in [k]}$. See the Appendix A.4 for the detailed derivations.

**Example 4.** The EPPF of the exchangeable partition $\Pi$ induced by the $DP(\theta)$ with concentration parameter $\theta$ is given by

$$\Pr\left(\Pi_n = (c_\ell)_{\ell \in [k]}\right) = \int_0^\infty \int_{\mathcal{V}_k} t^{-n} f_\rho(t - \sum_{\ell=1}^k s_\ell) dt \prod_{\ell=1}^k s_\ell^{|c_\ell|} \rho(ds_\ell)$$
$$= \frac{\theta^k \Gamma(\theta)}{\Gamma(n+\theta)} \prod_{\ell=1}^k \Gamma(|c_\ell|)$$

where

$$\rho(x) = \theta x^{-1} \exp(-x)$$
$$f_\rho(x) = \frac{1}{\Gamma(\theta)} x^{\theta-1} \exp(-x)$$

are the Lévy measure of the Gamma Process and the corresponding density function, respectively.

Example 4 can be obtained from Proposition 1 by performing the change of variables $\pi_\ell = \tilde{J}_\ell/T$ for $\ell = 1, \ldots, k-1$, and marginalising out $T$ and $(\pi_\ell)_{\ell \in [k]}$.

## 2.6 MCMC for Bayesian nonparametric models

Constructing MCMC schemes for models with one or more Bayesian nonparametric components is an active research area since dealing with the infinite dimensional component $P$ forbids the direct use of standard simulation-based methods. These methods usually require a finite-dimensional representation. Figure 2.3 shows the graphical model corresponding to an infinite mixture model where the top level is any prior from the Poisson–Kingman class. The latent variables required are countably infinitely many, hence, this component makes the model computationally intractable. The general idea for designing inference schemes is to find finite dimensional representations to be able to store the model in a computer with finite capacity. Once this representation is found, standard simulation-based methods can be used for inference.

There are two main sampling approaches to facilitate simulation in the case of Bayesian nonparametric models: random truncation and marginalisation. These two schemes are known in the literature as conditional and marginal samplers. Over the years, many MCMC methods, which rely on a tailored representation of the underlying process, have been proposed (Escobar and West, 1995; Escobar, 1994; Neal, 2000; Papaspiliopoulos and Roberts, 2008; Walker, 2007; Barrios et al., 2013; Favaro and Teh, 2013).

In the next section, different sampling schemes are described for two basic nonparametric models: the Dirichlet and the Pitman–Yor mixture models. In Chapters 3 and 4, two MCMC samplers are proposed for the $\sigma$-Stable Poisson–Kingman class of priors. This is a large class of infinite mixture models which encompasses almost all nonparametric mixture models previously explored in the literature. The first is a marginalised sampler, while the second, is a novel hybrid MCMC scheme which benefits from the advantages of both the marginal and the conditional MCMC samplers. In Chapter 5, we propose an SMC sampler in the same generic flavour.

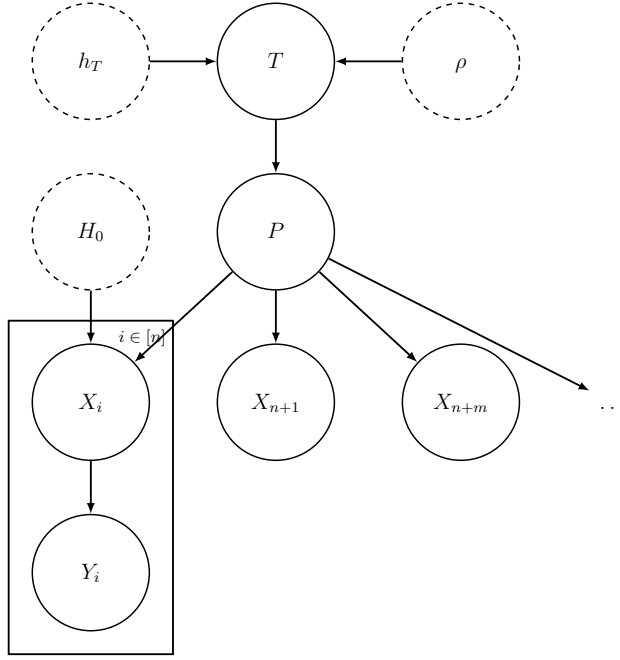

Figure 2.3: PKP intractable graphical model: the latent variables are countably infinitely many which makes the model computationally intractable.

### 2.6.1 Marginal or marginalised MCMC samplers

Marginal samplers bypass the need to represent the infinite-dimensional component by marginalising it out. These schemes have lower storage requirements than conditional samplers because they only store the induced partition, but could potentially have worse mixing properties.

A marginal sampler for the Pitman–Yor process can be derived starting with the EPPF of Example 3. The corresponding joint distribution of the block assignments $(c_\ell)_{\ell \in [k]}$, where $c_\ell$ denotes the $\ell$-th set in the partition, the parameters $\{X_\ell^*\}_{\ell \in [k]}$ and observations $\{Y_i\}_{i \in [k]}$ is

$$\Pr(\Pi_n = (c_\ell)_{\ell \in [k]}, X_\ell^* \in dx_\ell^* \text{ for } \ell \in [k], Y_i \in dy_i \text{ for } i \in [n])$$
$$= \frac{\Gamma(\theta)}{\Gamma(\frac{\theta}{\sigma})} \frac{\Gamma(\frac{\theta}{\sigma} + k)}{\Gamma(\theta + n)} \times \frac{\sigma^k}{\Gamma(n - \sigma k)} \prod_{\ell=1}^k (1-\sigma)_{(n_\ell - 1) \uparrow 1} H_0(dx_\ell^*) \prod_{i \in c_\ell} F(Y_i \in dy_i | x_\ell^*) \quad (2.25)$$

where $(1-\sigma)_{(n_\ell - 1) \uparrow 1}$ denotes the ascending factorial which is defined as a ratio of incomplete Gamma functions. See Appendix A.2 for details about this quantity. We say that the base distribution $H_0$ is *conjugate* to the likelihood of the observations if the parameter $X_c^*$ of cluster $c$ can be integrated out. Specifically, the following integrals can be solved analytically

$$\int H_0(dx) F(Y \in dy|x) \prod_{j \in c} F(Y \in dy_j|x) \text{ and } \int H_0(dx) \prod_{j \in c} F(Y \in dy_j|x).$$

Then, the corresponding likelihood term of observation $y$ is the conditional density of $y$ under cluster $c$ given the observations $\mathbf{Y}_c = (Y_j)_{j \in c}$ currently assigned to that cluster is denoted by $F(Y \in dy|\mathbf{y}_c)$ and given by

$$F(Y \in dy|\mathbf{y}_c) = \frac{\int H_0(dx) F(Y \in dy|x) \prod_{j \in c} F(Y \in dy_j|x)}{\int H_0(dx) \prod_{j \in c} F(Y \in dy_j|x)}.$$

The conditional probability of the cluster assignment for the $i$-th observation given the remaining variables can be computed from the joint distribution of Equation (2.25)

$$\Pr(i \text{ joins cluster } c' \mid \text{rest}) \propto (|c'| - \sigma) F(Y_i \in dy_i|\mathbf{y}_{c'})$$
$$\Pr(i \text{ joins new cluster c'} \mid \text{rest}) \propto \frac{\theta + \sigma k}{M} F(Y_i \in dy_i). \qquad (2.26)$$

The above conditional probability states that the $i$-th customer joins an existing cluster $c$ with probability proportional to the size of cluster $c$ minus $\sigma$, denoted by $|c|$, times a likelihood term which depends on all observations currently assigned to cluster $c$. Alternatively, the $i$-th customer joins a new cluster $c'$ with probability proportional to $\theta$ minus $\sigma$ times the number of clusters in the partition, denoted by $k$, multiplied by a likelihood term. In the non-conjugate case, the conditional probability of the cluster assignment for the $i$-th observation can be directly obtained from the joint distribution of Equation (2.25), it is given by

$$\Pr(i \text{ joins cluster c'} \mid \text{rest}) \propto (|c_k^{-i}| - \sigma) F(Y_i \in dy_i|x_{c'}^*)$$
$$\Pr(i \text{ joins new cluster c'} \mid \text{rest}) \propto \frac{\theta + \sigma k}{M} F(Y_i \in dy_i|x_{c'}^e). \qquad (2.27)$$

The paremeter $x_{c'}^e$, $e \in \{1, \ldots, M\}$, denotes the instantiated parameters for the $M$ empty clusters. Algorithm 1 contains the pseudocode for the conjugate case and Algorithms 2 and 3, for the non-conjugate case. Algorithm 2 corresponds to Algorithm 8 of Neal (2000), also described in Escobar (1994) for the Dirichlet process mixture model. Algorithm 3 corresponds to the Reuse algorithm from Favaro and Teh (2013) which was derived for all NRMI mixture models.

In Algorithm 2 (Neal, 2000), the $M$ empty cluster's parameters are drawn from the prior $H_0$ every time an observation is reassigned to a cluster and after being reassigned, the $M$ parameters are thrown away. The main difference with Algorithm 3 is that Algorithm 3 "reuses" the cluster parameters while reassigning observations, effectively reducing the computational cost. Indeed, only when one of the empty cluster's parameters is moved to an occupied cluster, a new empty cluster parameter is drawn from the prior $H_0$ to always have $M$ parameters for the empty clusters. A more detailed algorithmic description of the Reuse algorithm is given in Chapter 3.

---

**Algorithm 1** AlgorithmConjugate($\Pi_n, \{Y_i\}_{i\in[n]}, H_0$)

    **for** $i = 1 \to n$ **do**
        Let $c \in \Pi_n$ be such that $i \in c$
        $c \leftarrow c\setminus\{i\}$
        **if** $c = \varnothing$ **then**
            $\Pi_n \leftarrow \Pi_n \setminus \{c\}$
        **end if**
        Set $c'$ according to $\Pr[i \text{ joins cluster } c' \mid \{Y_i\}_{i\in c}]$     ▷ from (2.26)
        **if** $|c'| = 1$ **then**
            $\Pi_n \leftarrow \Pi_n \cup \{\{i\}\}$
        **else**
            $c' \leftarrow c' \cup \{i\}$
        **end if**
    **end for**

---

### 2.6.2 Conditional MCMC samplers

Conditional samplers replace the infinite-dimensional prior by a finite-dimensional representation chosen according to a truncation level. Since these samplers do not integrate out the infinite-dimensional component, their output provides a more comprehensive representation of the random probability measure. This property is useful when computing expectations of interest with respect to the posterior. Some existing conditional samplers for Dirichlet process are those by Ishwaran and James (2001), Papaspiliopoulos and Roberts (2008), Walker (2007) and the efficient slice version of Kalli et al. (2011). Favaro and Walker (2012) propose a conditional slice sampler for the $\sigma$-Stable Poisson–Kingman class of priors. The latter one is the first generic algorithm proposed in the literature; this scheme is reviewed in detail in Chapter 3. In this section, the algorithm of Kalli et al. (2011) for a Pitman–Yor mixture model is explained in some

**Algorithm 2** AlgorithmNon-conjugate($\Pi_n, M, \{Y_i\}_{i\in[n]}, \{X_c^*\}_{c\in\Pi_n}, H_0 \mid \text{rest}$)
- **for** $i = 1 \to n$ **do**
  - Draw $\{X_j^e\}_{j=1}^M \overset{i.i.d.}{\sim} H_0$
  - Let $c \in \Pi_n$ be such that $i \in c$
  - $c \leftarrow c\setminus\{i\}$
  - **if** $c = \varnothing$ **then**
    - $\ell \sim \text{UniformDiscrete}(\frac{1}{M})$
    - $X_\ell^e \leftarrow X_c^*$
    - $X_{(1:M)\neg \ell}^e \sim H_0$
    - $\Pi_n \leftarrow \Pi_n \setminus \{c\}$
  - **else**
    - $X_{1:M}^e \sim H_0$
  - **end if**
  - Set $c'$ according to $\Pr[i \text{ joins cluster } c' \mid \{Y_i\}_{i\in c}, X_c^*, \text{rest}]$ ▷ from (2.27)
  - **if** $|c'| = 1$ **then**
    - $\Pi_n \leftarrow \Pi_n \cup \{\{i\}\}$
    - $X_{\{i\}}^* \leftarrow X_{c'}^e$
  - **else**
    - $c' \leftarrow c' \cup \{i\}$
  - **end if**
  - $X_{1:M}^e = []$
- **end for**
- **for** $c \in \Pi_n$ **do**
  - Sample from the posterior $X_c^* \sim F(\cdot \mid \{Y_i\}_{i\in c})$
- **end for**

---

**Algorithm 3** ReUse($\Pi_n, M, \{Y_i\}_{i\in[n]}, \{X_c^*\}_{c\in\Pi_n}, H_0 \mid \text{rest}$)
- Draw $\{X_j^e\}_{j=1}^M \overset{i.i.d.}{\sim} H_0$
- **for** $i = 1 \to n$ **do**
  - Let $c \in \Pi_n$ be such that $i \in c$
  - $c \leftarrow c\setminus\{i\}$
  - **if** $c = \varnothing$ **then**
    - $\ell \sim \text{UniformDiscrete}(\frac{1}{M})$
    - $X_\ell^e \leftarrow X_c^*$
    - $\Pi_n \leftarrow \Pi_n \setminus \{c\}$
  - **end if**
  - Set $c'$ according to $\Pr[i \text{ joins cluster } c' \mid \{Y_i\}_{i\in c'}, X_{c'}^*, \text{rest}]$ ▷ from (2.27)
  - **if** $|c'| = 1$ **then**
    - $\Pi_n \leftarrow \Pi_n \cup \{\{i\}\}$
    - $X_{\{i\}}^* \leftarrow X_{c'}^e$
    - $X_{c'}^e \sim H_0$
  - **else**
    - $c' \leftarrow c' \cup \{i\}$
  - **end if**
- **end for**
- **for** $c \in \Pi_n$ **do**
  - Sample from the posterior $X_c^* \sim F(\cdot \mid \{Y_i\}_{i\in c})$
- **end for**

detail. The corresponding pseudocode is provided in Algorithm 4.

---

**Algorithm 4** SliceConditionalPY($N, \pi, \mathbf{X}, \phi, \mathbf{Y}; \text{Rest}$)
---
   **for** $i = 1 \to n$ **do**
      Update slice variables $S_i \sim \mathcal{U}(0, \exp(-x_i))$
      $N_i = \lfloor -\log(s_i) \rfloor$
      $M_{\max} = \max\{M_{\max}, N_i\}$
   **end for**
   $N = \max\{N_i\}$
   **if** $M_{\max} > N$ **then**
      **for** new $= N + 1 \to M_{\max}$ **do**
         $Z_{\text{new}} = \text{Beta}(1 - \sigma, \theta + \sigma(N + \text{new}))$
         $\mathbf{Z} = [\mathbf{Z}, Z_{\text{new}}]$
      **end for**
      $(\pi_1, \ldots, \pi_N, \pi_{N+1}, \ldots, \pi_{M_{\max}}) = \text{StickBreaking}(\mathbf{Z})$
   **end if**
   $N = M_{\max}$
   Update $V_1 \ldots, V_N \stackrel{\text{ind}}{\sim} \text{Beta}(n_\ell - \sigma, \theta + N\sigma), \ell = 1, \ldots, N$
   **for** $j = 1 \to N$ **do**
      Update parameters $\phi_j \sim f(\phi \mid \{Y_i\}_{\{i:x_i=j\}}, \mathbf{x})$       ▷ from (3.11)
   **end for**
   **for** $i = 1 \to n$ **do**
      Update $x_i \sim \Pr(X_i = x \mid \mathbf{x}_{-i}, \text{Rest})$       ▷ from (2.28)
   **end for**

---

All that is required is a finite but random number of stick-breaking weights. These weights correspond to the ones that fall above the slice variables, sampled at every iteration. Then, the cluster assignment variable for each observation is chosen from a categorical distribution, where the number of categories given by the number of represented sticks. For this reason, both the number of occupied and empty clusters is stored at every iteration. These numbers could be potentially large.

First, the $j$-th stick breaking weight of the Pitman–Yor process $\pi_j$, $j \in \mathbb{N}$ can be easily constructed a priori by sampling $j$ independent Beta random variables $Z_r = \text{Beta}(1 - \sigma, \theta + \sigma r)$, $r = 1, \ldots, j$, and constructing the corresponding weight is $\pi_j = Z_j \prod_{r<j}(1 - Z_r)$. A posteriori, it is also easy to construct the stick breaking weights for the Pitman–Yor, they only differ in the parameters for the Beta random variables, namely, $Z_r = \text{Beta}(n_r - \sigma, \theta + \sigma N)$, $r = 1 \ldots, N$ where $n_r$ is the number of observations currently assigned to cluster $r$ and $N$ is the current number of represented sticks. Then, the cluster parameters are sampled from

$$f(\phi_j \mid \{Y_i\}_{\{i:x_i=j\}}, \mathbf{x}) \propto \begin{cases} H_0(d\phi_j) \prod_{\{i:x_i=j\}} f(Y_i \mid \phi_j) & \text{if } A_{s_i}\setminus\{j\} \neq \varnothing \\ H_0(d\phi_j) & \text{o.w.} \end{cases} \quad (2.28)$$

where $A_{s_i} = \{\mathbf{x} : s_i < \exp(-x)\} \subseteq \{1,\ldots,N\}$, $N_i = \lfloor -\log(s_i)\rfloor$, $N = \text{Max}\{N_i\}$. Finally, to sample the allocation variables

$$\Pr(X_i = x \mid \mathbf{x}_{-i}, \text{Rest}) \propto \pi_x \exp(x) \text{Normal}(y_i \mid \phi_x) \mathbb{I}\{s_i \leqslant \exp\{-x\}\} \quad (2.29)$$

However, once we decide to pick a different nonparametric prior, the nice tractable construction of the Pitman–Yor stick breaking weights is no longer preserved. We introduce a way to deal with this issue for the $\sigma$-Stable Poisson–Kingman priors in Chapter 3.

## 2.7 Introduction to Sequential Monte Carlo

Sequential Monte Carlo (SMC) falls under the umbrella of non-iterative Monte Carlo approximation methods (Murphy, 2012). As its name indicates, it is useful in sequential scenarios: it generates samples from a sequence of posterior distributions in order to recursively compute expectations of interest with respect to each posterior, as more data is obtained. For instance, it allows to compute the predictive posterior of a new data point given previous ones since it can be expressed as an expectation with respect to the posterior.

In this section, importance sampling is reviewed first. This method can be viewed as a batch method for obtaining samples from a posterior given that the whole dataset is observed at once. Then, sequential importance sampling (SIS) is reviewed as a sequential version of importance sampling, as its name indicates. Finally, sequential Monte Carlo (SMC) is reviewed, it can be understood as SIS with an additional resampling step. See Doucet et al. (2001) for an introduction to SMC methods.

### 2.7.1 Importance sampling

Importance sampling is an example of a Monte Carlo sampling scheme that provides approximately independent and identically distributed samples from a distribution of interest or target distribution, such as a posterior distribution, denoted by $f$, by generating a candidate sample from a proposal or importance distribution $g$. The fact that the weights $w^p = \frac{f(x^p)}{g(x^p)}$, with $p = 1, \ldots, L$, can be computed is exploited, and samples from the target are obtained by sampling from a weighted empirical distribution. The pseudocode for the case when the target distribution $f$ is the posterior distribution when $n$ data points have been obtained is given in Algorithm 5.

---
**Algorithm 5** ImportanceSampling($f, g$)

    **for** $p \in [L]$ **do**
        Draw $X^p \sim g(x)$
        Compute weights $w^p = \frac{f(x^p | y_1, \ldots, y_n)}{g(x^p)}$
    **end for**
    Normalise weights $\tilde{w}^p = \frac{w^p}{\sum_{p=1}^{L} w^p}$

    Sample $L$ times from $\sum_{p=1}^{L} \tilde{w}^p \delta_{x^p}$

---

The output of Algorithm 5 is the following consistent estimate of the posterior distribution

$$\hat{\pi}(x \mid y_1, \ldots, y_n) = \sum_{p=1}^{L} \tilde{w}^p \delta_{x^p}.$$

One problem with this method is that it is not easy to choose the proposal distribution $g$. A good proposal should share most of the support of the target distribution and have the same number of modes, i.e. it should be "close" to the target. However, since the expectation of a specific function needs to be computed, one possibility to choose the proposal distribution $g$ is to minimise the variance of its estimate, as suggested by Murphy (2012). A second problem is that it is a batch estimation method. To tackle this latter issue, in the next section, an extension to a sequential scenario is described.

### 2.7.2 Sequential importance sampling

Let us assume that the importance distribution $g$ at time $i$ depends on all data points until time $i$ and not on the future data points, the joint posterior can be written in the following factorised form

$$g(x_0, \ldots, x_n \mid y_1, \ldots, y_n) = g(x_0) \prod_{i=1}^{n} g(x_i \mid y_1, \ldots, y_i, x_1, \ldots, x_{i-1})$$

The corresponding importance weight is

$$w_n = \frac{f(x_0,\ldots,x_n \mid y_1,\ldots,y_n)}{g(x_0,\ldots,x_n \mid y_1,\ldots,y_n)}$$
$$= \frac{f(x_0)\prod_{i=1}^{n} f(x_i \mid y_1,\ldots,y_i,x_1,\ldots,y_{i-1})}{g(x_0)\prod_{i=1}^{n} g(x_i \mid y_1,\ldots,y_i,x_1,\ldots,x_{i-1})} \quad (2.30)$$

The target distribution is the posterior distribution up to time $i$, which changes sequentially as we observe more data.

For the sake of illustration, consider a state-space model with the usual structure in which the latent variable $X$ follows a Markov process, in a non-Gaussian non-linear state space model

$$f(x_0)$$
$$f(x_i \mid x_{i-1}) \quad i \geq 1$$
$$f(y_i \mid x_i) \quad i \geq 1$$

Then the posterior distribution is

$$f(x_0,\ldots,x_i \mid y_1,\ldots,y_i) \propto f(y_1,\ldots,y_i \mid x_0,\ldots,x_i) f(x_0,\ldots,x_i)$$

and can be estimated recursively due to

$$f(x_0,\ldots,x_{i+1} \mid y_1,\ldots,y_{i+1}) = f(x_0,\ldots,x_i \mid y_1,\ldots,y_i)\frac{f(y_{i+1} \mid x_{i+1})f(x_{i+1} \mid x_i)}{f(y_{i+1} \mid y_1,\ldots,y_i)}. \quad (2.31)$$

Substituting the numerator of Equation (2.31) in (2.30) we obtain a recursive equation for the importance weight at time $i+1$

$$w_{i+1} = w_i \times \frac{f(y_{i+1} \mid x_{i+1})f(x_{i+1} \mid x_i)}{g(x_{i+1} \mid y_1,\ldots,y_i,x_1,\ldots,x_i)}.$$

Even if we are not dealing with a state space model, SIS can be used as a general-purpose algorithm. The data are assumed to be observed sequentially so the observation's index is the time index.

### 2.7.3 Sequential Monte Carlo

The main problem with SIS is that the weights become more skewed as the number of data points increases (Doucet et al., 2001), after a few steps only one particle will have a significant weight. To remedy this, a resampling step can be introduced which allows to eliminate particles with small weights and replicate particles with high weights. This selection step can be introduced only occasionally or at every step of the algorithm. Thus, we resample from the empirical posterior

$$\hat{f}(x \mid y_1, \ldots, y_i) = \sum_{p=1}^{L} \tilde{w}_i^p \delta_{x_i^p}$$

to obtain

$$\hat{f}(x \mid y_1, \ldots, y_i) = \frac{1}{L} \sum_{p=1}^{L} n^p \delta_{x_i^p}.$$

where $n^p$ is the number of copies of particle $p$, $p = 1, \ldots, L$ and $L$ is the number of particles. If the selection step is to be performed occasionally, a possible criterion is when the effective sample size (ESS) is below a given threshold, which is a function of the number of particles, a popular choice is $0.5L$. The ESS for the unnormalised weights is given by

$$ESS_i = \frac{\left(\sum_{p=1}^{L} w_i^p\right)^2}{\sum_{p=1}^{L} w_i^{p2}}.$$

The SMC sampler's performance varies depending on the choice of resampling scheme. Some improved resampling schemes have been proposed after the original multinomial scheme of Gordon et al. (1993). For instance, the minimum variance or systematic resampling (Kitagawa, 1996), stratified resampling (Carpenter et al., 1999; Kitagawa and Gersh, 1996), residual resampling (Liu and Chen, 1998; Higuchi, 1997), among others. The systematic resampling scheme is the most widely used scheme in the literature. Algorithms 6 and 7 provide the pseudocode for the systematic and residual resampling schemes, respectively.

**Algorithm 6** SystematicResample$\left(\{w_i^\ell\}_{\ell=1}^L, L\right)$

---
Sample $U^1 \sim \text{Uniform}(0, \frac{1}{L})$

$U^p = U^1 + \frac{p-1}{L}$ for $p = 2, \ldots, L$.

$N_i^p = \left\{ U^j : \sum_{p=1}^{r-1} w_{i-1}^p \leqslant U^j \leqslant \sum_{p=1}^r w_i^p \right\}$

---

**Algorithm 7** ResResample$(\{w_i^p\}_{p=1}^L, L)$

---
Set $\tilde{N}_i^p = \lfloor L \times w_i^p \rfloor$

Sample $\bar{N}_i^p \sim \text{Multinomial}(\{\bar{w}_p^{1:L}\}_{p=1}^L, L)$

where $\bar{W}_p^{1:L} \propto W_i^p - \frac{\tilde{N}_i^p}{L}$

$N_i^p = \tilde{N}_i^p + \bar{N}_i^p$.

---

### 2.7.4 SMC for BNP

Recently, various SMC schemes for Bayesian nonparametric models have been proposed. Fearnhead (2004) and Wood and Black (2008) propose an SMC scheme for Dirichlet process mixture modes while an SMC for time varying DPMs is introduced in Caron et al. (2007). Griffin (2011) introduces an SMC for NRM mixture models, and Chopin (2002) an SMC for finite mixture models.

**Algorithm 8** SMC

---
$\Pi_1^p = \{\{1\}\}, \forall p \in \{1, \ldots, L\}$
**for** $i = 2 : n$ **do**
    **for** $p = 1 : L$ **do**
        Set $c'$ according to $\Pr\left(\text{i joins cluster } c' \mid \Pi_{i-1}^p, \mathbf{y}_{1:i-1}\right)$         ▷ from (2.32)
        **if** $|c'| = 1$ **then**
            $\Pi_i^p = \Pi_{i-1}^\ell \cup \{\{i\}\}$
        **else**
            $c' = c' \cup \{i\}$, $c' \in \Pi_{i-1}^p$
            $\Pi_i^p = \Pi_{i-1}^p$
        **end if**
        Compute weight $w_i^p \propto p(y_i \mid \Pi_i^p, \mathbf{y}_{1:i-1})$         ▷ from (2.33)
    **end for**
    Normalise the weights $\tilde{w}_i^p = \frac{w_i^p}{\sum_{j=1}^L w_i^j}$
    Resample $p' \sim \text{Multinomial}\left(\tilde{w}_i^1, \ldots, \tilde{w}_i^L\right), \forall p \in \{1, \ldots, L\}$
    $\Pi_i^p = \Pi_i^{p'}$ ▷ The $p$-th particle inherits the entire path of the $p'$-th particle, we are resampling at every step.
**end for**

---

In this section an SMC sampler for a Pitman–Yor mixture model is explained in detail. The pseudocode is provided in Algorithm 8, it describes a modified version of the SMC from Wood and Black (2008). Let $\Pi_n^p = (c_\ell)_{\ell=1}^k$ denote a partition for the $p$-th particle with $k$ blocks, when $n$ data points have been observed. The resampling step is performed at every iteration of the algorithm and the corresponding weights are set to one. In the Pitman–Yor case, the proposal distribution is the posterior up to the $i-1$-th data point is

$$\Pr\left(\text{i joins cluster } c' \mid \Pi_{i-1}^p, \mathbf{y}_{1:i-1}\right) \propto \frac{|c'| - \sigma}{\theta + i - 1} F(Y_i \in dy_i \mid \{y_j\}_{j \in c'})$$
$$\Pr\left(\text{i joins a new cluster} \mid \Pi_{i-1}^p, \mathbf{y}_{1:i-1}\right) \propto \frac{\theta + \sigma|\Pi_{i-1}^p|}{\theta + i - 1} F(Y_i \in dy_i) \qquad (2.32)$$

where

$$F(Y_i \in dy_i) = \int F(Y_i \in dy_i \mid \phi) H_0(d\phi)$$
$$F(Y_i \in dy_i) \mid \{y_j\}_{j \in c}) = \int F(Y_i \in dy_i \mid \phi) f(\phi \mid \{y_j\}_{j \in c}) d\phi.$$

and, the incremental weight is given by

$$\Pr\left(Y_i \in dy_i \mid \Pi_i^p, \mathbf{y}_{1:i-1}\right) = \frac{\theta + \sigma|\Pi_i^p|}{\theta + i} F(Y_i \in dy_i) + \frac{1}{\theta + i} \sum_{c' \in \Pi_i^p} (|c'| - \sigma) F(Y_i \in dy_i \mid \{y_j\}_{j \in c'}).$$
$$(2.33)$$

In Wood and Black (2008), rather than using the posterior as a proposal they sample from the prior with the corresponding weights

$$\Pr\left(\text{i joins cluster } c' \mid \Pi_{i-1}^p, \mathbf{y}_{1:i-1}\right) \propto \frac{|c'| - \sigma}{\theta + i - 1}$$
$$\Pr\left(\text{i joins a new cluster} \mid \Pi_{i-1}^p, \mathbf{y}_{1:i-1}\right) \propto \frac{\theta + \sigma|\Pi_{i-1}^p|}{\theta + i - 1}$$

with incremental weight given by

$$\Pr\left(Y_i \in dy_i \mid \Pi_i^p, \mathbf{y}_{1:i-1}\right) \propto F(Y_i \in dy_i) \mid \{y_j\}_{j \in c}), \quad i \in c. \qquad (2.34)$$

In the case of the Pitman–Yor mixture model, the marginal likelihood can be com-puted exactly. This can be used as a benchmark to asses the performance of the SMC algorithm. The prior for the random partition is given by

$$\Pr\left(\Pi_n = \pi\right) = \frac{\sigma^k \Gamma(\theta) \Gamma(\frac{\theta}{\sigma} + k)}{\Gamma(\frac{\theta}{\sigma}) \Gamma(\theta + n)} \prod_{\ell=1}^{k} \frac{\Gamma(|c_\ell| - \sigma)}{\Gamma(1 - \sigma)}.$$

The target posterior distribution is then

$$\Pr\left(\Pi_n = \pi, \{Y_i \in dy_i\}_{i=1}^n\right) = \frac{\sigma^k \Gamma(\theta) \Gamma(\frac{\theta}{\sigma} + k)}{\Gamma(\frac{\theta}{\sigma}) \Gamma(\theta + n)} \prod_{\ell=1}^{k} \frac{\Gamma(|c_\ell| - \sigma)}{\Gamma(1 - \sigma)} \Pr\left(\{y_j\}_{j: j \in c_\ell} \mid \Pi_n = \pi\right)$$

where

$$\Pr\left(\{Y_j \in dy_j\}_{\{j \in c_\ell\}} \mid \Pi_n = \pi\right) = \tau_1^{\frac{n_\ell}{2}} (2\pi)^{-\frac{n_\ell}{2}} \tau_0^{1/2} (\tau_0 + n_\ell)^{1/2}$$

$$\times \exp\left[-\frac{\tau_1}{2}\left(\sum_{i=1}^n y_i\right) + \frac{1}{2} \frac{(\tau_0 \mu_0 + \tau_1 \sum_{i=1}^{n_\ell} y_i)^2}{\tau 0 + n_\ell \tau_1}\right]$$

$$\Pr\left(\{Y_i \in dy_i\}_{i=1}^n\right) = \sum_{\{\Pi_n = \pi\}} \Pr\left(\Pi_n = \pi, \{Y_i \in dy_i\}_{i=1}^n\right).$$

where $k = |\Pi_n|$ and $n_\ell$ is the number of observations assigned to cluster $c_\ell$. The above computations can be carried out by combinatorial enumeration for a small sample size to test the correctness of the SMC sampler, as suggested by Bouchard-Côté et al. (2015). We can then verify that the Monte Carlo estimates converge to these quantities.

Even though we can analytically evaluate this quantity, we should be careful when enumerating the number of possible partitions that the SMC sampler visits. Indeed, the number of possible partitions of $n$ is given by the Bell number. Sometimes, different orderings of the sets of a partition are considered to be different partitions. This is the case *a posteriori* in a Bayesian nonparametric mixture model because the posterior partition is conditioned on the observations that belong to each block. The number of ordered partitions is given by the ordered Bell numbers. However, an SMC sampler without a rejuvenation step explores a space which is smaller: each observation $j$ will be assigned to a block whose index is $j \leq i$, we call this an order constraint. See Martinez and Mena (2014) for an alternative way to tackle this issue based on a distribution defined on the set compositions that uses the time induced order.

In Appendix A.5, the explicit formulae for an SMC algorithm are derived in generic terms. These expressions are useful to see the sequence of posterior distributions that are targeted every time a new datapoint is observed.

## 2.8 Summary of this Chapter

Firstly, some fundamental definitions are presented in order to understand the interplay between random probability measures and random partitions. This justifies why it is valid to use the induced partition obtained from a random probability measure whose restrictions to $[n]$ have the advantage of being of finite size. This idea is useful for deriving the finite dimensional representations used in the algorithms presented in the next chapters. Secondly, the Gibbs type random partitions are reviewed and the generative process of Poisson–Kingman partitions. Both are key elements that are repeatedly used as building blocks in the next chapters. Finally, both MCMC and SMC existing inference schemes for infinite mixtures models with a Pitman–Yor prior are surveyed as a starting point for the more general inference schemes presented later in this thesis.

# Chapter 3

# MCMC scheme I: A marginal sampler

The aim of this chapter aim is to present a marginal MCMC scheme for $\sigma$-Stable Poisson–Kingman mixture models, a subclass of the Poisson–Kingman mixture models, discussed in Chapter 2. This interesting subclass, also referred to as Gibbs-type mixture models with parameter $\sigma \in (0,1)$, encompasses most of the Bayesian nonparametric priors previously explored in the literature. See Gnedin and Pitman (2006) for further details about the equivalence between $\sigma$-Stable Poisson–Kingman priors and Gibbs-type priors with parameter $\sigma \in (0,1)$. Firstly, the family of priors is described in terms of some of its most useful properties. Specifically, certain characterisations are reviewed which lead us to derive the sampling algorithm. Secondly, the updates to some of the other variables are discussed together with some reparametrisations which improve the numerical stability. Thirdly, the competing conditional sampler scheme from Favaro and Walker (2012) is described and compared against in the experiments section. Finally, a multidimensional experiment is performed with an interesting dataset.

## 3.1 $\sigma$-Stable Poisson–Kingman RPMs

If we focus on the class of $\sigma$-Stable Poisson–Kingman RPMs, we obtain a noteworthy subclass which encompasses most of the popular discrete RPMs used in Bayesian nonparametrics, for instance, the Pitman–Yor process and the normalised generalised Gamma process. For any $\sigma \in (0,1)$, the density function of a positive $\sigma$-Stable random variable is $f_\sigma(t) = \frac{1}{\pi} \sum_{j=0}^{\infty} \frac{(-1)^{j+1}}{j!} \sin(\pi\sigma j) \frac{\Gamma(\sigma j+1)}{t^{\sigma j+1}}$, it is denoted by $f_\sigma$. A $\sigma$-Stable

Poisson–Kingman RPMs is a Poisson–Kingman RPM with Lévy measure given by

$$\rho(dx) = \rho_\sigma(dx) := \frac{\sigma}{\Gamma(1-\sigma)} x^{-\sigma-1} dx, \tag{3.1}$$

base distribution $H_0$. The mixing distribution for the total mass $T$ takes the following factored form $\gamma(dt) = h(t) f_\sigma(dt) / \int_0^{+\infty} h(t) f_\sigma(t) dt$, for any non-negative measurable function $h$ such that $\int_0^{+\infty} h(t) f_\sigma(t) dt < \infty$. Accordingly, $\sigma$-Stable Poisson–Kingman RPMs form a class of discrete RPMs indexed by the parameters $(\sigma, h)$. The Dirichlet process can be recovered as a limiting case, if $\sigma \to 0$, for some choices of $h$. We denote by $\text{PK}(\rho_\sigma, h, H_0)$ the distribution of a $\sigma$-Stable Poisson–Kingman RPM with parameters $(\sigma, h)$ and base distribution $H_0$.

The following examples of $\sigma$-Stable Poisson–Kingman RPM are obtained by specifying the tilting function $h$.

**Example 5. Normalised $\sigma$-Stable process** (NS) (Kingman, 1975). It corresponds to $h(t) = 1$.

**Example 6. Normalised Generalised Gamma process** (NGG) (James, 2002; Pitman, 2003). It corresponds to $h(t) = \exp\{\tau - \tau^{1/\sigma} t\}$, for any $\tau > 0$. See also Lijoi et al. (2005), Lijoi et al. (2007b), Lijoi et al. (2008), James et al. (2009), Barrios et al. (2013) and James (2013).

**Example 7. Pitman–Yor process** (PY), (Perman et al., 1992). It corresponds to $h(t) = \frac{\Gamma(\theta+1)}{\Gamma(\theta/\sigma+1)} t^{-\theta}$ with $\theta \geqslant -\sigma$, see Pitman and Yor (1997).

**Example 8. Gamma-tilted process** (GT). It corresponds to $h(t) = t^{-\theta} \exp\{-\eta t\}$, for any $\eta > 0$ or $\eta = 0$ and $\theta > -\sigma$.

**Example 9. Poisson-Gamma class** ($\mathcal{PG}$), (James, 2013). It corresponds to $h(t) = \int_{\mathbb{R}^+} \exp\{\tau - \tau^{1/\sigma} t\} F(d\tau)$, for any distribution function $F$ over the positive real line, see also Pitman and Yor (1997) and James (2002).

**Example 10. Composition of Classes class** (CC), (Ho et al., 2008). Let $\mathcal{T}$ a positive random variable, this class corresponds to $h(t) = \frac{\mathbb{E}[f(t\mathcal{T}^{1/\sigma})]}{\int_0^{+\infty} \mathbb{E}[f(t\mathcal{T}^{1/\sigma})] f_\sigma(t) dt}$, where $f$ is any positive function, see James (2002) for details.

**Example 11. Lamperti class** (LT), (Ho et al., 2008). Let $S_\sigma$ a positive $\sigma$-Stable

random variable, the class corresponds to the choice

$$h(t) = L_\sigma \mathbb{E}\left(g(S_\sigma t^{-1})\right), \tag{3.2}$$

where

$$\frac{1}{L_\sigma} = \frac{\sin(\pi\sigma)}{\pi} \int_0^{+\infty} \frac{f(y) y^{\sigma-1}}{y^{2\sigma} + 2y^\sigma \cos(\pi\sigma) + 1} dy$$

and $g$ is any positive function such that Equation (3.2) is well-defined, see James (2002).

**Example 12. Mittag–Leffler class** (ML), (Ho et al., 2008). It corresponds to the choice of $g(x) = \exp\{-x^\sigma\}$ for the tilting function in Equation (3.2).

The relationship between these examples of $\sigma$-Stable Poisson–Kingman RPMs is shown in Figure 2.2 of Chapter 2.

The distribution of the exchangeable random partition induced by a sample from a $\sigma$-Stable Poisson–Kingman RPMs is obtained by a direct application of Proposition 2. The next proposition specializes Proposition 2 to $\sigma$-Stable Poisson–Kingman RPMs.

**Proposition 3.** *Let $\Pi_n$ be the exchangeable random partition of $[n]$ induced by a sample from $P \sim \mathrm{PK}(\rho_\sigma, h, H_0)$. Then,*

$$Pr_{\rho_\sigma, h, H_0}(\Pi_n = (c_\ell)_{\ell \in [k]}) = V_{n,k} \prod_{\ell=1}^{k} W_{|c_\ell|} \tag{3.3}$$

*where*

$$V_{n,k} = \frac{\sigma^k}{\Gamma(n - k\sigma)} \int_{\mathbb{R}^+} \int_0^t t^{-n}(t-v)^{n-1-k\sigma} h(t) f_\sigma(v) dt dv$$

*and*

$$W_m = \Gamma(m - \sigma)/\Gamma(1 - \sigma) = (1 - \sigma)_{m-1\uparrow 1} := \prod_{i=0}^{m-2} (1 - \sigma + i).$$

*Proof.* Let us start with the conditional distribution of a random partition of $[n]$ of size $|\Pi_n| = k$ given the total mass, which is distributed according to a Poisson–Kingman partition. This conditional distribution can be obtained from the joint distribution given by Proposition 2 if it is divided by the density of the total mass, which we assume is proportional to $h(t)f(t)$. It is given by

$$\Pr\left(\Pi_n = \pi, \tilde{J}_1 \in ds_1, \ldots, \tilde{J}_k \in ds_k \mid T = t\right) = \frac{t^{-n}f(t - \sum_{\ell=1}^{k} s_\ell)dt \prod \rho(s_\ell)s_\ell^{n_\ell} ds_\ell}{f(t)}.$$

Then, the following change of variables is performed: $P = \frac{\sum_{i=1}^{k} \tilde{J}_\ell}{T}$ and $U_\ell = \frac{\tilde{J}_\ell}{PT}$ for $\ell = 1, \ldots, k-1$ with corresponding Jacobian $|J| = P^{k-1}T^k$, and we obtain

$$\Pr\left(\Pi_n = \pi, U_1 \in du_1, \ldots, U_k \in du_k, P \in dp \mid T = t\right) = \frac{t^k p^{n+k-1} f(t-tp)dt \prod_{\ell=1}^{k} \rho(u_\ell pt)u_\ell^{n_\ell} du_\ell}{f(t)}.$$

Let $\mathcal{S}_k = \{(u_1, \ldots, u_k) : u_i \geqslant 0, u_1 + \ldots + u_k = 1\}$, if we substitute the corresponding Lévy measure of a $\sigma$-Stable random variable,

$$\Pr\left(\Pi_n = \pi, P \in dp \mid T = t\right) = \frac{t^k p^{n+k-1} f_\sigma(t-tp)dt}{f_\sigma(t)} \int_{\mathcal{S}_k} \prod_{\ell=1}^{k} \frac{\sigma}{\Gamma(1-\sigma)} u_\ell^{n_\ell - \sigma - 1} p^{-\sigma-1} t^{-\sigma-1}$$

$$= t^{k-k(\sigma+1)} p^{n+k-1-k(\sigma+1)} \frac{f_\sigma(t(1-p))dt}{f_\sigma(t)} \left(\frac{\sigma}{\Gamma(1-\sigma)}\right)^k$$

$$\times \int_{\mathcal{S}_k} \left[\prod_{\ell=1}^{k-1} u_\ell^{n_\ell - \sigma - 1}\right] \times \left(1 - \sum_{\ell=1}^{k-1} u_\ell\right)^{n - \sum_{\ell<k} n_\ell - \sigma - 1}$$

$$= t^{-k\sigma} p^{n-k\sigma-1} \frac{f_\sigma(t(1-p))dt}{f_\sigma(t)} \frac{\sigma^k}{\Gamma(n-k\sigma)} \prod_{\ell=1}^{k} \frac{\Gamma(n_\ell - \sigma)}{\Gamma(1-\sigma)}.$$

The $(U_\ell)_{\ell=1}^{k}$ are Dirichlet distributed with parameters $(n_1 - \sigma, \ldots, n_{k-1} - \sigma, n - \sum_{\ell<k} n_\ell - \sigma)$. Then, we obtain

$$\Pr\left(\Pi_n = \pi\right) = \int_{\mathbb{R}^+} \int_0^1 t^{-k\sigma} p^{n-k\sigma-1} f_\sigma(t(1-p))h(t)dtdp$$

$$\times \frac{\sigma^k}{\Gamma(n-k\sigma)} \prod_{\ell=1}^{k} \frac{\Gamma(n_\ell - \sigma)}{\Gamma(1-\sigma)}.$$

This expression corresponds to Equation (66) of Pitman (2003). Finally, if an additional change of variables is performed, with $V = (1-P)T$ and Jacobian $|J| = T^{-1}$, then

$$\Pr\left(\Pi_n = \pi\right) = \frac{\sigma^k}{\Gamma(n-k\sigma)} \int_{\mathbb{R}^+} \int_0^t t^{-n}(t-v)^{n-k\sigma-1} f_\sigma(v)h(t)dt$$

$$\times dv \prod_{\ell=1}^{k} \frac{\Gamma(n_\ell - \sigma)}{\Gamma(1-\sigma)}.$$

$\square$

Proposition 3 provides one of the main tools to derive the marginal MCMC sampler. We refer to the Chapter 2 for a quick review of MCMC for BNP, where the two existing types of marginal algorithms are described in detail for a Pitman–Yor mixture model. A brief overview of Gibbs-type random partitions with parameter $\sigma \in (0,1)$ is given in Chapter 2. Gnedin and Pitman (2006) and Pitman (2006) give a more comprehensive study of exchangeable random partitions with distribution of the form given in Equation (3.3).

## 3.2 Marginal samplers for $\sigma$-Stable Poisson–Kingman mixture models

In this section a marginal sampler is derived such that it can be effectively applied to all members of the $\sigma$-Stable Poisson–Kingman family. The sampler does not require any numerical integrations, nor evaluations of special functions, for instance, the density $f_\sigma$ of the positive $\sigma$-Stable distribution (Wuertz et al., 2013). It applies to non-conjugate hierarchical mixture models based on $\sigma$-Stable Poisson–Kingman RPMs by extending the *Reuse data augmentation scheme* of Favaro and Teh (2013).

Figure 2.3 from Chapter 2 shows the graphical model for a Poisson–Kingman mixture model and its corresponding intractable component. If the $\sigma$-Stable Poisson–Kingman subclass is picked, then it is possible to obtain a tractable representation, given by Figure 3.1, in terms of some random variables and the partition induced by the RPM. This representation is based on an augmented version of its corresponding exchangeable partition probability function. In the next section, the corresponding augmentations are discussed in detail.

## 3.3 Representation with data augmentation

If we start with a member of the $\sigma$-Stable Poisson–Kingman family, the distribution for the total mass is proportional to $\gamma(dt) \propto h(t) f_\sigma(t)$. Then, the joint distribution over the induced partition $\Pi_n$, total mass $T$ and surplus mass $V$ is given by

$$\Pr\nolimits_{\rho_\sigma, \gamma, H_0}(T \in dt, V \in dv, \Pi_n = (c_\ell)_{\ell \in [k]}) \tag{3.4}$$
$$= t^{-n}(t-v)^{n-1-k\sigma} f_\sigma(v) h(t) dt \frac{\sigma^k}{\Gamma(n-k\sigma)} \prod_{\ell=1}^{k} \frac{\Gamma(|c_\ell| - \sigma)}{\Gamma(1-\sigma)} \mathbb{I}_{(0,t)}(v) \mathbb{I}_{(0,\infty)}(t).$$

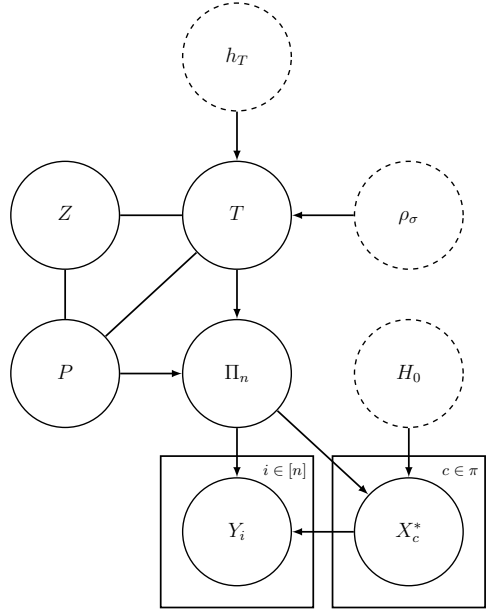

Figure 3.1: Marginal sampler's graphical model. Each node represents a variable used in the augmented representation for the joint distribution from Equation 3.4. The boxes represent that there is a number of repeated nodes of the same type, $n$ for the observations and $|\pi|$ for the cluster parameters. The nodes with dashed lines represent the algorithmic inputs.

This joint distribution is available to construct marginal samplers with the exception of two difficulties. The first difficulty is that it is necessary to evaluate $f_\sigma$ pointwise if we work with an MCMC scheme using the above representation. Current software packages compute this density using numerical integration techniques, which can be unnecessarily expensive. However, the following integral representation (Kanter, 1975; Zolotarev, 1966) can be used to introduce an auxiliary variable into the system. Thus, the need to repeatedly evaluate the integral numerically is removed. Let $\sigma \in (0, 1)$. Then,

$$f_\sigma(t) = \frac{1}{\pi} \frac{\sigma}{1-\sigma} \left(\frac{1}{t}\right)^{\frac{1}{1-\sigma}} \int_0^\pi A(z) \exp\left(-\left(\frac{1}{t}\right)^{\frac{\sigma}{1-\sigma}} A(z)\right) dz \qquad (3.5)$$

where the function $A$ is called Zolotarev function and is given by

$$A(z) = \left[\frac{\sin(\sigma z)}{\sin(z)}\right]^{\frac{1}{1-\sigma}} \left[\frac{\sin((1-\sigma)z)}{\sin(\sigma z)}\right], \qquad z \in (0, \pi).$$

Zolotarev's representation has been used by Devroye (2009) to construct a random number generator for polynomially and exponentially tilted $\sigma$-Stable random variables and in a rejection sampling scheme within the conditional slice MCMC sampler by Favaro and Walker (2012). Our proposal here is to introduce an auxiliary variable $Z$, an example of a data augmentation scheme (Tanner and Wong, 1987), with conditional distribution given $T \in dt$ described by the integrand in (3.5).

The second difficulty is that the variables $T$ and $V$ are dependent and that computations with small values of $T$ and $V$ might not be numerically stable. To address these problems, the following reparameterisation is proposed: $W = \frac{\sigma}{1-\sigma} \log T$, and $R = V/T$ where $W \in \mathbb{R}$ and $R \in (0,1)$. This gives our final representation

$$\Pr\nolimits_{\rho_\sigma, \gamma, H_0}(W \in dw, R \in dr, Z \in dz, \Pi_n = (c_\ell)_{\ell \in [k]})$$
$$= \frac{1}{\pi} e^{-w(1+(1-\sigma)k)} (1-r)^{n-1-k\sigma} r^{-\frac{1}{1-\sigma}} h(e^{\frac{1-\sigma}{\sigma}w}) A(z) \exp\left\{-e^{-w} r^{-\frac{\sigma}{1-\sigma}} A(z)\right\}$$
$$\times \frac{\sigma^k}{\Gamma(n-\sigma k)} \prod_{\ell=1}^{k} (1-\sigma)_{(|c_\ell|-1)\uparrow 1}.$$

## 3.4 $\sigma$-Stable Poisson–Kingman mixture models

In this section, a $\sigma$-Stable Poisson–Kingman RPM is considered as the mixing distribution in a Bayesian nonparametric mixture model

$$T \sim \gamma$$
$$P | T \sim \text{PK}(\rho_\sigma, \delta_T, H_0)$$
$$X_i | P \overset{\text{iid}}{\sim} P$$
$$Y_i | X_i \overset{\text{ind}}{\sim} F(\cdot | X_i).$$

$F(\cdot | X)$ is the observation's distribution and our dataset consists of $n$ observations $(y_i)_{i \in [n]}$ of the corresponding variables $(Y_i)_{i \in [n]}$. We will assume that $F(\cdot | X)$ is smooth.

Two marginal samplers for infinite mixture models are derived. The random variables $W$, $R$, $Z$ are also included in the final representation, where the unique values for the component parameters are denoted by $(X_\ell^*)_{\ell \in [k]}$. The joint distribution over all

variables is given by

$$\Pr(W \in dw, R \in dr, Z \in dz, \Pi_n = (c_\ell)_{\ell\in[K]}, X_\ell^* \in dx_\ell^* \text{ for } \ell \in [k], Y_i \in dy_i \text{ for } i \in [n])$$
$$= \frac{1}{\pi} e^{-w(1+(1-\sigma)k)}(1-r)^{n-1-k\sigma} r^{-\frac{1}{1-\sigma}} h(e^{\frac{1-\sigma}{\sigma}w}) A(z) e^{-r^{-\frac{\sigma}{1-\sigma}} e^{-w} A(z)} dw\, dr\, dz$$
$$\times \frac{\sigma^k}{\Gamma(n-\sigma k)} \prod_{\ell=1}^{k} (1-\sigma)_{(n_\ell-1)\uparrow 1} H_0(dx_\ell^*) \prod_{i\in c_\ell} F(dy_i|x_\ell^*). \tag{3.6}$$

The system of predictive distributions governing the distribution over partitions, given all the other variables, can be read from the joint distribution (3.6). Specifically, the conditional distribution of a new observation $Y_{n+1}$ is

$$\Pr(Y_{n+1} \in dy_{n+1} \mid \Pi_n = (c_\ell)_{\ell\in[k]}, X_\ell^* \in dx_\ell^* \text{ for } \ell \in [k], W \in dw, R \in dr, Z \in dz)$$
$$\propto \sigma e^{(\sigma-1)w}(1-r)^{-\sigma} \frac{\Gamma(n+1-\sigma k)}{\Gamma(n+1-\sigma(k+1))} \int F(Y_{n+1} \in dy_{n+1} \mid x) H_0(dx)$$
$$+ \sum_{c\in\Pi_n} (|c|-\sigma) F(Y_{n+1} \in dy_{n+1} \mid x_c^*).$$

The conditional probability of the next observation joining an existing cluster $c$ is proportional to $|c|-\sigma$, which is the same for all exchangeable random probability measures based on the $\sigma$-Stable CRM. The conditional probability of joining a new cluster is more complex and depends on the other random variables. Such system of predictive distributions was first studied by Blackwell and McQueen (1973) for the generalised Polýa urn, a sequential construction equivalent to the Chinese restaurant process. See Aldous (1985) for details and Ewens (1972) for an early account in population genetics.

## 3.5 Sampler updates

Firstly, the Gibbs updates to the partition $\Pi_n$ are described, conditioned on the other variables $W, R, Z$. Then, the corresponding updates to the rest of the random variables are explained. A non-conjugate case is presented first where the component parameters $(X_\ell^*)_{\ell\in[k]}$ cannot be marginalised out, based on an extension of Favaro and Teh (2013). In the case where the base distribution $H_0$ is conjugate to the observation's distribution $F$, the component parameters can be marginalised out as well, which leads to an extension to Algorithm 3 of Neal (2000).

In the non-conjugate case, the state space of the Markov chain consists of $(c_\ell)_{\ell\in[k]}$,

$W, R, Z$, as well as the cluster parameters $(X_\ell^*)_{\ell\in[k]}$. The Gibbs updates to the partition involve updating the cluster assignment of one observation, say the $i$-th one, at a time.

The ReUse algorithm (Favaro and Teh, 2013) goes as follows: a fixed number $M > 0$ of potential new clusters is maintained along with those in the current partition $(c_\ell)_{\ell\in[k]}$, the parameters for each of these potential new clusters are denoted by $(X_\ell^{\text{new}})_{\ell\in[M]}$. When updating the cluster assignment of the $i$-th observation, potential new clusters are considered as well as those clusters in the current partition after the $i$-th observation is removed, denoted by $(c_\ell^{-i})_{\ell\in[k^{-i}]}$. If one of the potential new clusters is chosen, it is moved into the partition, and in its place a new cluster is generated by drawing a new parameter from $H_0$. Conversely, when a cluster in the partition is emptied, it is moved into the list of potential new clusters, displacing a randomly chosen one. The parameters of the potential new clusters are refreshed by i.i.d. draws from $H_0$ after a full iteration through the dataset, see Algorithm 10 for details and Algorithm 3, in Chapter (2), for the Pitman–Yor mixture model case. The conditional probability of the cluster assignment of the $i$-th observation is

$$\Pr(i \text{ joins cluster } c_\ell^{-i} \mid \text{rest}) \propto (|c_\ell^{-i}| - \sigma) F(Y_i \in dy_i | x_\ell^*) \tag{3.7}$$

$$\Pr(i \text{ joins new cluster } c \mid \text{rest}) \propto \frac{1}{M} \sigma e^{(\sigma-1)w}(1-r)^{-\sigma} \frac{\Gamma(n-\sigma k^{-i})}{\Gamma(n-\sigma(k^{-i}+1))} F(Y_i \in dy_i | x_c^{\text{new}}).$$

If $H_0$ is conjugate, we can replace the likelihood term in the cluster assignment rule by the conditional density of $y$ under cluster $c$, denoted by $F(Y \in dy|\mathbf{y}_c)$, given the observations $\mathbf{Y}_c = (Y_j)_{j\in c}$ currently assigned to that cluster

$$F(Y \in dy|\mathbf{y}_c) = \frac{\int H_0(dx) F(Y \in dy|x) \prod_{j\in c} F(Y \in dy_j|x)}{\int H_0(dx) \prod_{j\in c} F(Y \in dy_j|x)}.$$

The updates to the other variables $W, R$ and $Z$ are straightforward. Their complete conditional densities can be read off the joint density (3.6)

$$\Pr(W \in dw \mid \text{rest}) \propto e^{-w(1+(1-\sigma)k)} h(e^{\frac{1-\sigma}{\sigma}w}) e^{-r^{-\frac{\sigma}{1-\sigma}}e^{-w}A(z)} dw, \quad w \in \mathbb{R}$$

$$\Pr(R \in dr \mid \text{rest}) \propto (1-r)^{n-1-k\sigma} r^{-\frac{1}{1-\sigma}} e^{-r^{-\frac{\sigma}{1-\sigma}}e^{-w}A(z)} dr, \quad r \in (0,1)$$

$$\Pr(Z \in dz \mid \text{rest}) \propto A(z) e^{-r^{-\frac{\sigma}{1-\sigma}}e^{-w}A(z)} dz, \quad z \in (0,\pi).$$

These are not in standard form: each complete conditional distribution does not correspond to a known probability density function. Hence, there are no available random number generators to obtain an independent and identically distributed sample from these. However, their states can be updated easily using generic MCMC methods. In our implementation, the slice sampling algorithm (Neal, 2003) is used within an overall Gibbs sampling scheme (Geman and Geman, 1984). If we have a prior on the index parameter $\sigma$, it can be updated as well. Due to the heavy-tailed nature of the conditional distribution, we recommend transforming $\sigma$ to $\log \frac{\sigma}{1-\sigma}$. However, some of the complete conditional distributions for this parameter include an intractable normalisation constant, such as the Mittag–Leffler priors. Even if one is able to obtain an unbiased estimator the inverse is no longer unbiased. Some debiasing technique needs to be applied then but its non-negativity has to be enforced simultaneously, which is problematic, see Jacob and Thiery (2015) for further details.

---

**Algorithm 9** MarginalSamplerNonConj($h_T, \sigma, M, H_0, \{Y_i\}_{i\in[n]}$)

    Initialize all random variables.
    **for** $t = 2 \to iter$ **do**
        Update $z^{(t)}$: Slice sample $\tilde{\Pr}(Z \in dz \mid \text{rest})$
        Update $p^{(t)}$ : Slice sample $\tilde{\Pr}(P \in dp \mid \text{rest})$
        Update $w^{(t)}$: Slice sample $\tilde{\Pr}(W \in dw \mid \text{rest})$
        Update $\pi^{(t)}, \{x_c^*\}_{c\in\pi}^{(t)}$: **ReUse**($\Pi_n, M, \{Y_i\}_{i\in[n]}, \{X_c^*\}_{c\in\Pi_n}, H_0 \mid \text{rest}$)
    **end for**

---

**Algorithm 10** ReUse($\Pi_n, M, \{Y_i\}_{i\in[n]}, \{X_c^*\}_{c\in\Pi_n}, H_0 \mid \text{rest}$)

    Draw $\{X_j^e\}_{j=1}^M \overset{\text{i.i.d.}}{\sim} H_0$
    **for** $i = 1 \to n$ **do**
        Let $c \in \Pi_n$ be such that $i \in c$
        $c \leftarrow c\backslash\{i\}$
        **if** $c = \varnothing$ **then**
            $k \sim \text{UniformDiscrete}(\frac{1}{M})$
            $X_k^e \leftarrow X_c^*$
            $\Pi_n \leftarrow \Pi_n \backslash \{c\}$
        **end if**
        Set $c'$ according to $\Pr[i \text{ joins cluster } c' \mid \{Y_i\}_{i\in c}, X_c^*, \text{rest}]$     ▷ from Equation (3.7)
        **if** $|c'| = 1$ **then**
            $\Pi_n \leftarrow \Pi_n \cup \{\{i\}\}$
            $X_{\{i\}}^* \leftarrow X_{c'}^e$
            $X_{c'}^e \sim H_0$
        **else**
            $c' \leftarrow c' \cup \{i\}$
        **end if**
        **for** $c \in \Pi_n$ **do**
            Sample from the posterior $X_c^* \sim F(\cdot \mid \{Y_i\}_{i\in c})$
        **end for**
    **end for**

## 3.6 Conditional sampler for $\sigma$-Stable Poisson–Kingman mixture models

Figure 2.3 from Chapter 2 shows the graphical model for a Poisson–Kingman mixture model and its corresponding intractable component. An alternative to marginal samplers, which also leads to a tractable representation, is to use a conditional MCMC sampler. Figure 3.2 shows such representation for the $\sigma$-Stable Poisson–Kingman family for the conditional slice MCMC sampler (Favaro and Walker, 2012).

The generative model for the conditional slice MCMC sampler from Favaro and Walker (2012) can be derived starting with Proposition 2: after the following two change of variables $\pi_j = \frac{\tilde{J}_j}{T - \sum_{\ell<j} \tilde{J}_\ell}$ and $Z_j = \frac{\pi_j}{1 - \sum_{\ell<j} \pi_\ell}$ we obtain the corresponding joint distribution in terms of $N$ $(0,1)$-valued stick variables $\{Z_j\}_{j=1}^N$. We call them stick variables because they are the building blocks of the stick breaking weights $\{\pi_j\}_{j=1}^N$, where $\pi_j = Z_j \prod \ell < j (1 - Z_\ell)$. This joint distribution corresponds to Equation (19) of Favaro and Walker (2012). The truncation level $N$ needs to be randomised to have an exact MCMC scheme. The authors propose to use the efficient slice sampler (Kalli et al., 2011). In our implementation of the conditional sampler an overall Gibbs sampling scheme is used. The complete conditional distributions are not in standard form so there are no ready-made random number generators to sample from them. We used a slice sampling step (Neal, 2003) within the Gibbs, they are given as follows

$$\Pr\left(T \in dt, (Z_j \in dz_j)_{j=1}^N \mid \text{Rest}\right) \propto h(t) \left[t \prod_{j=1}^N (1-z_j)\right]^{-\frac{3}{2}} \exp\left[-\frac{1}{4} t^{-1} \left(\prod_{j=1}^N (1-z_j)\right)^{-1}\right]$$

$$\times t^{-\frac{N}{2}} \prod_{j=1}^N z_j^{n_j - \frac{1}{2}} (1-z_j)^{m_j - \frac{N-j}{2}} \qquad (3.8)$$

where

$$n_j = \sum_{i=1}^n \mathbb{I}\{x_i = j\}, \quad m_j = \sum_{i=1}^n \mathbb{I}\{x_i > j\}$$

$$\Pr(T \in dt \mid \text{Rest}) \propto h(t) \exp\left[-\frac{1}{4} t^{-1} \left(\prod_{j=1}^N (1-z_j)\right)^{-1}\right] t^{-\frac{(N+3)}{2}}$$

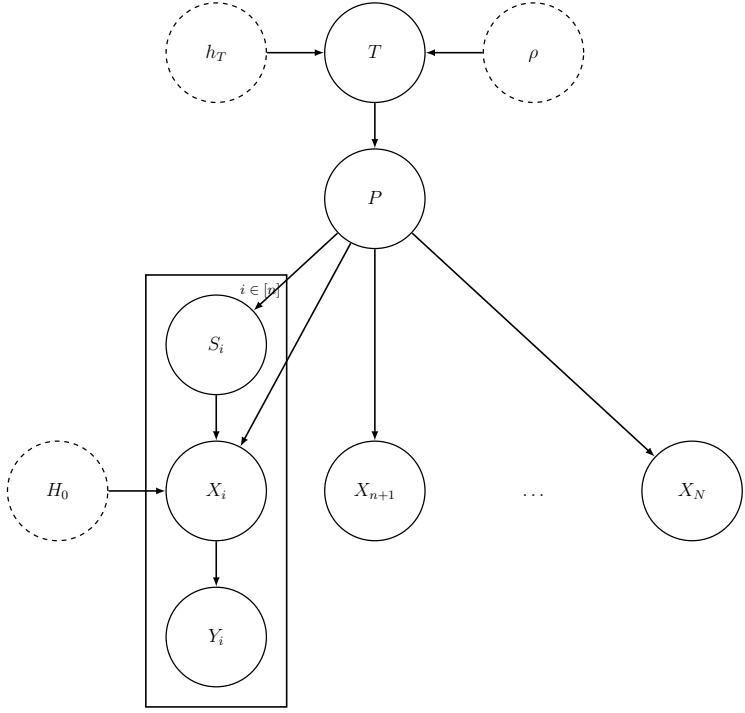

Figure 3.2: Conditional sampler's graphical model. Each node represents a variable used in the augmented representation for the joint distribution. The latent variables represent the number of occupied and unoccupied components.

and

$$\Pr\left(Z_j \in dz_j \mid \mathbf{z}_{-j}, t, \text{Rest}\right) \propto \exp\left[-\frac{1}{4}t^{-1}\left(\prod_{\ell \neq j}^{N}(1-z_\ell)\right)^{-1}\frac{1}{1-z_j}\right]$$
$$\times z_j^{n_j - \frac{1}{2}}(1-z_j)^{m_j - \frac{(N-j+3)}{2}} \quad (3.9)$$

If $\sigma = 0.5$, the complete conditional for the $k+1$ stick variable simplifies to

$$\Pr\left(Z_{N+1} \in dz_{N+1} \mid \text{Rest}\right) \propto z_{N+1}^{-\frac{1}{2}}(1-z_{N+1})^{-\frac{1}{2}}(1-z_{N+1})^{-1}$$
$$\times \exp\left[\frac{-1}{4t}\left(\prod_{j=1}^{N}(1-z_j)\right)^{-1}\left(\frac{1}{1-z_{N+1}}-1\right)\right].$$

The above conditional density corresponds to a random variable such which satisfies the following distributional identity

$$Z_{N+1} \mid Z_1, \ldots, Z_N, T \stackrel{d}{=} \frac{X_{N+1}}{X_{N+1} + Y_{N+1}}.$$

The random variables $X_{N+1}$ and $Y_{N+1}$ are independent and

$$X_{N+1}^2 \sim \text{Gamma}\left(\frac{3}{4}, 1\right)$$

$$Y_{N+1}^2 \sim \text{Inverse Gamma}\left(\frac{1}{4}, \frac{1}{4^3 t^2}\left[\prod_{j=1}^{N}(1-z_j)\right]^{-2}\right).$$

This identity in distribution was introduced in Favaro et al. (2014a). If $\sigma = \frac{u}{v} < \frac{1}{2}$ where $u, v$ are coprime integers, then, the complete conditional for the $N+1$ stick variable has a more complicated form

$$\Pr(Z_{N+1}, \in dz_{N+1} \mid \text{Rest}) \propto z_{N+1}^{1-\sigma} f_\sigma\left(t \prod_{j<N+1}(1-z_j)(1-z_{N+1})\right).$$

The above conditional density corresponds to a random variable such that

$$Z_{N+1} \mid Z_1, \ldots, Z_N, T \stackrel{d}{=} \frac{X_{N+1}}{X_{N+1} + Y_{N+1}}.$$

The random variables $X_{N+1}$ and $Y_{N+1}$ are independent and

$$\frac{1}{X_{N+1}} \sim \mathcal{E}_T\left(\frac{\frac{u^2}{tv^{\frac{v}{u}}}}{\prod_{j=1}^{N}(1-z_j)}, \frac{1}{L_{\frac{u}{v}}^{1/u}}\right)$$

with

$$L_{\frac{u}{v}} = \prod_{i=0}^{u-2} \text{Beta}\left(\frac{2i+2}{v}, \frac{i+1}{u} - \frac{2i+2}{v}\right)$$

$$\times \prod_{i=0}^{u-2} \text{Beta}\left(\frac{2i+1}{v}, \frac{i+1}{u} - \frac{2i+2}{v}\right)$$

$$\times \prod_{i=1}^{v-2u} \text{Gamma}\left(\frac{2(u-1)+i}{v}, 1\right)$$

$$Y_{N+1} \sim \text{Inverse Gamma}\left(1 - \frac{u}{v}, \frac{\frac{u^2}{tv^{\frac{v}{u}}}}{\prod_{j=1}^{N}(1-z_j)}\right).$$

Let $\mathcal{E}_T(b, x)$ denote an exponentially tilted distribution of $X$ with tilting parameter $b$ with corresponding density given by $\frac{\exp(-bx)f_X(x)}{\int \exp(-bx')f_X(x')dx'}$. See Favaro et al. (2014a) for details about an improved rejection sampling method to sample this random variables.

The preceding identities in distribution are useful for our variation of the slice conditional MCMC sampler (Favaro and Walker, 2012). If the resulting slice is lower that the previous one, additional stick breaking weights are then instantiated. The existing stick variables are sampled from each complete conditional given the other variables from Equation (3.9) with a slice sampling step. In contrast, Favaro and Walker (2012) introduce an augmentation to decorrelate these random variables. Finally, the cluster assignment variables and cluster parameters are sampled according to Equations (3.10) and (3.11). See Algorithm 11 for the corresponding pseudocode.

$$f(\phi_j \mid \{Y_i\}_{\{i:z_i=j\}}, \mathbf{z}) \propto \begin{cases} H_0(d\phi_j) \prod_{\{i:z_i=j\}} f(Y_i \mid \phi_j) & \text{if } A_{s_i} \setminus \{j\} \neq \emptyset \\ H_0(d\phi_j) & \text{o.w.} \end{cases} \quad (3.10)$$

where $A_{s_i} = \{\mathbf{x} : s_i < \exp(-x)\} \subseteq \{1, \ldots, N\}$, $N_i = \lfloor -\log(s_i) \rfloor$, $N = \text{Max}\{N_i\}$, and

$$\Pr(X_i = x \mid \mathbf{x}_{-i}, \text{Rest}) \propto P_x \exp(x) \text{Normal}(y_i \mid \phi_x) \mathbb{I}\{s_i \leqslant \exp\{-x\}\} \quad (3.11)$$

---

**Algorithm 11** SliceConditional$\frac{1}{2}$PK($N, \mathbf{P}, \mathbf{X}, \phi, \mathbf{Y}; \text{Rest}$)

    **for** $i = 1 \to n$ **do**
        Update slice variables $S_i \sim \mathcal{U}(0, \exp(-x_i))$
        $N_i = \lfloor -\log(s_i) \rfloor$
        $M_{\max} = \text{Max}\{M_{\max}, N_i\}$
    **end for**
    **if** $M_{\max} > N$ **then**
        **for** new $= N + 1 \to M_{\max}$ **do**
            $G_{new} = \textbf{Gamma}\left(\frac{3}{4}, 1\right)$
            $IG_{new} \mid Z_1, \ldots, Z_{new}, T = \textbf{InverseGamma}\left(\frac{1}{4}, \frac{1}{4^3 t^2}\left[\prod_{j=1}^{new}(1-z_j)\right]^{-2}\right)$
            $Z_{\text{new}} = \frac{G_{new}^{1/2}}{G_{new}^{1/2} + IG_{new}^{1/2}}$
            $\mathbf{Z} = [\mathbf{Z}, Z_{\text{new}}]$
        **end for**
        $(\pi_1, \ldots, \pi_{M_{max}}) = \text{StickBreaking}(\mathbf{Z})$
    **end if**
    $N = M_{\max}$
    Update $Z_1 \ldots, Z_N \mid T, \text{Rest} \sim g_{\mathbf{Z}}(\mathbf{z} \mid \text{Rest})$
    Update total mass $T \mid \mathbf{Z}, \text{Rest} \sim g_T(t \mid \text{Rest})$
    **for** $j = 1 \to N$ **do**
        Update parameters $\phi_j \sim f(\phi \mid \{Y_i\}_{\{x_i=j\}})$         ▷ from Equation (3.10)
    **end for**
    **for** $i = 1 \to n$ **do**
        Update $x_i \sim \Pr(X_i = x \mid \mathbf{x}_{-i}, \text{Rest})$         ▷ from Equation (3.11)
    **end for**

The number of represented stick variables is a random quantity. For this reason, some additional stick variables need to be instantiated and sometimes these variables will not be associated with any of the observations. This could potentially lead to slower running times and larger memory requirements to store these quantities when the number of data points is large. In our implementation, we found that some of the stick variables are highly correlated which leads to a slow mixing of the chain. A quantitative comparison is presented in Table 3.2 in terms of running times in seconds and effective sample size (ESS).

## 3.7 Numerical illustrations

In this section, we illustrate the algorithm on unidimensional and multidimensional data sets. We applied our MCMC sampler for density estimation of a $\sigma$-Stable Poisson–Kingman mixture model, denoted by PK $(\sigma, H_0, h_T(t))$, for various choices of $h(t)$. In all of our experiments whether a conjugate or non-conjugate prior is used for the mixture component parameters, we sample the parameters rather than integrating them out. We kept the hyperparameters of each $h$-tilting function fixed. The correctness of our posterior simulators was tested using the "Getting it right" test (Geweke, 2004), see Appendix A.10 for further details about this procedure.

### 3.7.1 Unidimensional experiment

The dataset from Roeder (1990) consists of measurements of velocities in km/sec of 82 galaxies from a survey of the Corona Borealis region. We chose the following base distribution $H_0$ and the corresponding likelihood $F$ for cluster $c$

$$H_0(d\mu_c, d\tau_c) = F_\mu(d\mu_c \mid \mu_0, \tau_0\tau_c) F_\tau(d\tau_c \mid \alpha_0, \beta_0)$$
$$F(dY_1 \in y_1, \ldots, Y_{|c|} \in dy_{|c|} \mid \mu_c, \tau_c) = \prod_{i \in c} \mathcal{N}\left(y_i \mid \mu_c, \tau_c^{-1}\right)$$

where $Y_1, \ldots, Y_{|c|}$ are the observations currently assigned to cluster $c$. $\mathcal{N}$ denotes a Normal distribution with given mean $\mu_c$ and variance $\tau_c^{-1}$. In the first sampler (*Marg-Conj I*), we used $H_0(d\mu_c, d\tau_c) = \mathcal{N}\left(d\mu_c \mid \mu_0, \tau_0^{-1}\right) \delta\{\tau_c = \tau\}$ with a common precision parameter $\tau$ among all clusters, set to $\frac{1}{4}$ of the range of the data. In the second sampler (*Marg-Conj II*), we used $H_0(d\mu_c, d\tau_c) = \mathcal{N}\left(d\mu_c \mid \mu_0, \tau_0^{-1}\tau_c^{-1}\right) \text{Gamma}(\tau_c \mid \alpha_0, \beta_0)$. In the third sampler (*Marg-NonConj*), we used a non conjugate distribution for the mean per cluster, $\mu = \log \varphi$ where $\varphi \sim \text{Gamma}(\varphi \mid a_0, b_0)$, and for the precision parameter, we used $\tau_c \sim \text{Gamma}(\tau_c \mid \alpha_0, \beta_0)$.

| Algorithm | $\sigma$ | M | Running time | ESS($\pm$std) |
|---|---|---|---|---|
| Pitma-Yor process ($\theta = 10$) | | | | |
| Marginal-Conj II | 0.5 | 4 | 23770.3(2098.22) | 4857.644(447.583) |
| Marginal-NonConj | 0.5 | 4 | 46352.4(252.27) | 5663.696(89.264) |
| Normalised Generalised Gamma process ($\tau = 1$) | | | | |
| Marginal-Conj II | 0.5 | 4 | 22434.1(78.191) | 3400.855(110.420) |
| Marginal-NonConj | 0.5 | 4 | 28933.5(133.97) | 5361.945(88.521) |

Table 3.1: Comparison between a conjugate base distribution for the cluster parameters versus a non-conjugate case. Running times in seconds and ESS for the number of clusters, averaged over 5 chains. Unidimensional dataset, 30,000 iterations, 10,000 burn in.

| Algorithm | $\sigma$ | M | Running time | ESS($\pm$std) |
|---|---|---|---|---|
| Pitman–Yor process ($\theta = 50$) | | | | |
| Marginal-Conj I | 0.5 | 2 | 1.1124e+04 | 4121.94(821.562) |
| Marginal-Conj I | 0.5 | 6 | 1.1216e+04 | 11215.55(596.249) |
| Marginal-Conj I | 0.5 | 10 | 1.1355e+04 | 12469.87(548.981) |
| Marginal-Conj I | 0.5 | 15 | 1.1385e+04 | 13087.92(504.595) |
| Marginal-Conj I | 0.5 | 20 | 1.1415e+04 | 12792.78(391.123) |
| Conditional-Conj I | 0.5 | - | 1.5659e+04 | 707.82 (95.754) |
| Normalised Generalised Gamma process ($\tau = 50$) | | | | |
| Marginal-Conj I | 0.5 | 2 | 1.1617e+04 | 4601.63(574.339) |
| Marginal-Conj I | 0.5 | 6 | 1.1650e+04 | 10296.85(425.333) |
| Marginal-Conj I | 0.5 | 10 | 1.1692e+04 | 11415.41(377.418) |
| Marginal-Conj I | 0.5 | 15 | 1.1795e+04 | 11473.44(374.031) |
| Marginal-Conj I | 0.5 | 20 | 1.1875e+04 | 11461.08(506.744) |
| Conditional-Conj I | 0.5 | - | 1.5014e+04 | 848.73 (135.138) |

Table 3.2: Effect of increasing the algorithmic parameter $M$, which denotes the number of empty clusters. As $M$ is increased, the ESS increases but also the running time which implies a higher computational cost. Running times in seconds and ESS for the number of clusters, averaged over 10 chains. Unidimensional dataset, 50,000 iterations per chain, 20,000 burn in.

In Table 3.1, we reported an increase in the running times if we use a non-conjugate prior (*Marg-NonConj*) versus a conjugate prior (*Marg-Conj II*). In Table 3.2, the sensitivity to the number of new components $M$ is tested and compared against a conditional MCMC sampler (Favaro and Walker, 2012). As we increase the marginal sampler's number of new components per iteration, the ESS increases. Intuitively, the computation time increases but it also leads to a better mixing of the algorithm. In contrast, we found that the conditional sampler was not performing very well due to high correlations between the stick variables. Finally, in Table 3.3, it is shown how different values for $\sigma$ can be effectively chosen without modifying the algorithm as opposed to Favaro and Walker (2012), which is only available for $\sigma = 0.5$.

| Algorithm | $\sigma$ | $M$ | Running time | ESS($\pm$std) |
|---|---|---|---|---|
| Pitma-Yor process ($\theta = 10$) | | | | |
| Marginal-Conj I | 0.3 | 4 | 4685.7(84.104) | 2382.799(169.359) |
| Marginal-Conj I | 0.5 | 4 | 4757.2(37.077) | 2944.065(195.011) |
| Marginal-Conj I | 0.7 | 4 | 4655.2(52.514) | 2726.232(132.828) |
| Conditional-Conj I | 0.5 | - | 10141.6(237.735) | 905.444(41.475) |
| Normalised Stable process | | | | |
| Marginal-Conj I | 0.3 | 4 | 7658.3(193.773) | 2630.264(429.877) |
| Marginal-Conj I | 0.5 | 4 | 8203.1(106.798) | 3139.412(351.788) |
| Marginal-Conj I | 0.7 | 4 | 8095.7(85.2640) | 2394.756(295.923) |
| Conditional-Conj I | 0.5 | - | 10033.1(22.647) | 912.382(167.089) |
| Normalised Generalised Gamma process ($\tau = 1$) | | | | |
| Marginal-Conj I | 0.3 | 4 | 7685.8(208.98) | 3587.733(569.984) |
| Marginal-Conj I | 0.5 | 4 | 8055.6(93.164) | 4443.905(367.297) |
| Marginal-Conj I | 0.7 | 4 | 8117.9(113.188) | 4936.649(411.568) |
| Conditional-Conj I | 0.5 | - | 10046.9(206.538) | 1000.214(70.148) |

Table 3.3: Comparison of the marginal sampler for which different values of $\sigma$ are available versus the conditional sampler for which $\sigma = 0.5$ is the only available value. Running times in seconds and ESS for the number of clusters, averaged over 5 chains. Unidimensional dataset, 30,000 iterations per chain, 10,000 burn in.

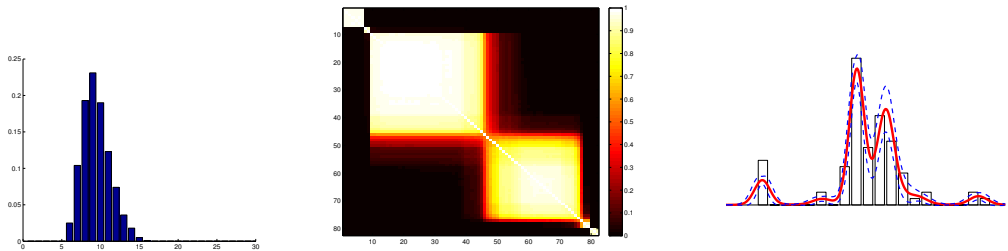

Figure 3.3: From left to right: posterior distribution for the number of clusters, coclustering probability matrix and density estimate. All of this quantities were computed with 210,000 iterations, 10,000 burn in and 20 thinning factor.

After assessing the algorithmic performance, we used our MCMC algorithm for posterior inference with a nonparametric mixture model where the top level is a prior from the $\sigma$-Stable Poisson–Kingman class. Since any prior in this class can be chosen, one possible criterion for model selection is predictive performance. In Table 3.4, we reported the average leave-one-out predictive probabilities, see Appendix A.9 for details on how to compute these quantities. We can see that all priors in this class have similar average predictive probabilities, the NGG slightly outperforms the rest on average.

| Examples | Average predictive probability |
|---|---|
| PY | 0.1033(0.052) |
| **NGG** | **0.1228(0.065)** |
| Gamma Tilted | 0.1186(0.065) |
| NS | 0.1123(0.057) |

Table 3.4: Unidimensional experiment's average leave-one-out predictive probabilities, see Appendix A.7 for details on how to compute .

In Figure 3.3, the mode of the posterior distribution is reported, the mean number of components is around ten clusters, and there are six clusters in the coclustering probability matrix. Indeed, a good estimate of the density might include superfluous components having vanishingly small weights as explained in Miller and Harrison (2013). The third plot shows the corresponding density estimate which is consistent under certain conditions as shown in De Blasi et al. (2015).

### 3.7.2 Multidimensional experiment

The dataset from de la Mata-Espinosa et al. (2011) consists of $n$ $D$-dimensional triacylglyceride profiles of different types of oils where $n = 120$ and $D = 4000$. The observations consist of profiles of olive, monovarietal vegetable and blends of oils. Within each type, there could be several subtypes, so we cannot know the number of varieties a priori. For this reason, a Bayesian nonparametric mixture model is a good choice to bypass the need to pick the number of different varieties. We preprocessed the data by applying Principal Component Analysis (PCA) (Jolliffe, 2002) to capture the relevant dimensions in it; a useful technique when the signal to noise ratio is small. We used the first $d = 8$ principal components which explained 97% of the variance and encoded sufficient information for the mixture model to recover distinct clusters.

Then, a $\sigma$-Stable Poisson–Kingman mixture of multivariate Normal distributions with an unknown covariance matrix and mean vector was chosen, for different $h$-tilting functions. A multivariate Normal-Inverse Wishart was chosen as a base measure and the corresponding likelihood $F$ for cluster $c$ is

$$H_0(d\mu_c, d\Sigma_c) = \mathcal{N}_d\left(d\mu_c \mid \mu_0, r_0 \Sigma_c^{-1}\right) \mathcal{IW}_d\left(d\Sigma_c \mid \nu_0, S_0\right)$$
$$F(Y_1 \in dy_1, \ldots, Y_{|c|} \in dy_{|c|} \mid \mu_c, \Sigma_c) = \prod_{i \in c} \mathcal{N}_d\left(dy_i \mid \mu_c, \Sigma_c\right)$$

where $Y_1, \ldots, Y_{|c|}$ are the observations currently assigned to cluster $c$. $\mathcal{N}_d$ denotes a $d$-variate Normal distribution with given mean vector $\mu_c$ and covariance matrix $\Sigma_c^{-1}$, $\mathcal{IW}_d$ denotes an inverse Wishart over $d \times d$ positive definite matrices with given degrees of freedom $\nu_0$ and scale matrix $S_0$. The Inverse Wishart is parameterised as in Gelman et al. (1995). $S_0$ was chosen to be a diagonal matrix with each diagonal element given by the maximum range of the data across all dimensions and degrees of freedom $\nu_0 = d+3$, a weakly informative case.

| Examples | Average predictive probability |
|---|---|
| DP | 5.5484e-12 (7.6848e-13) |
| PY | 4.1285e-12 (7.5549e-13) |
| **NGG** | **9.6266e-12 (3.4035e-12)** |
| Gamma tilted | 6.7099e-12 (1.5767e-12) |
| NS | 8.3328e-12 (9.7106e-13) |
| Lamperti tilted | 5.4251e-12 (1.0538e-12) |

Table 3.5: Multidimensional experiment's average 5-fold predictive probabilities, see Appendix A.7 for details on how to compute it. We used the standard ReUse algorithm based on the Polýa urn for the DP case.

In Table 3.5, the average five-fold predictive probabilities are reported, see the Appendix A.9 for details about how to compute these quantities. Again, we observe that all priors in this class have similar average predictive probabilities. In Figure 3.4, the mean curve per cluster and the coclustering probability matrix are reported. This mean curve reflects the average triacylglyceride profile per oil type. The coclustering probability matrix was used as an input to an agglomerative clustering algorithm, in order to obtain a hierarchical clustering representation (Medvedovic and Sivaganesan, 2002). In certain contexts, it is useful to think of a hierarchical clustering rather than a flat one since it might be natural to think of superclasses. See Appendix A.8 for details and the corresponding pseudocode.

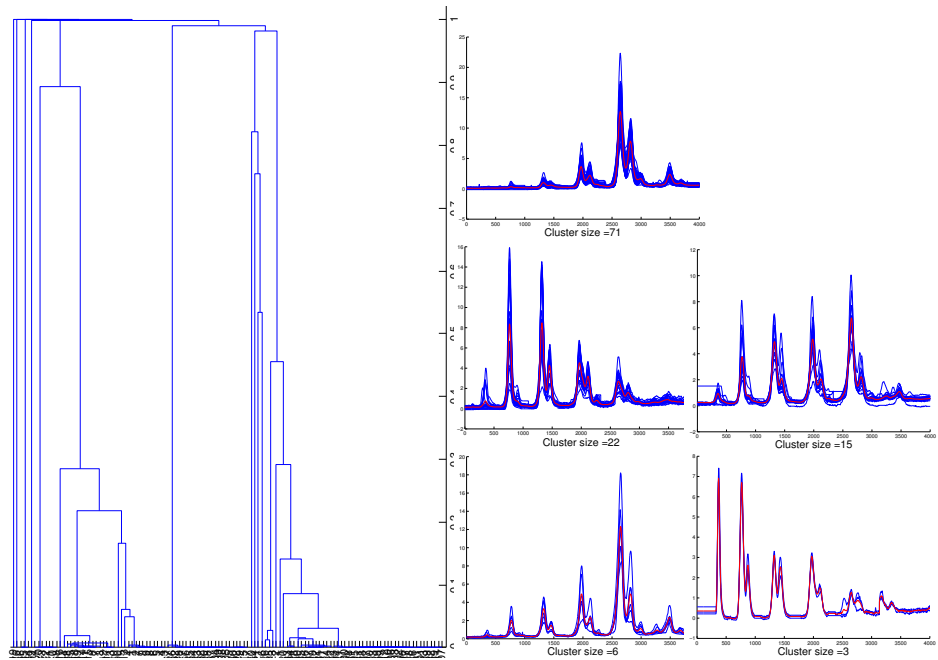

Figure 3.4: Dendrogram and mean profile per cluster (in red), profiles in each cluster (blue) using a NGG prior.

In Figure 3.4, the mean curves per cluster are shown. These were found by thresh-olding the hierarchy and ignoring clusters of size one. The first plot corresponds to the olive oil cluster, it is well represented by the mean curve. The last two plots correspond to data that belongs to non-olive blends of oil. The second and third plots correspond to non-olive monovarietal oil clusters. We could interpret these last clusters as different varieties of vegetable oil, monovarietal or not, since their corresponding mean curves are indeed different. In the dendrogram it is clear that most of the data belongs to three large clusters and that 60% of the triacylglycerides are olive oil.

## 3.8 Summary of this Chapter

In this chapter, the marginal MCMC sampler is introduced for inference in mixture models with priors from the $\sigma$-Stable Poisson–Kingman class. We show how we can use many different priors without changing the main algorithm and its corresponding algorithmic performance in both unidimensional and multidimensional datasets. Indeed, it is a competitive alternative against a conditional MCMC sampler in terms of running times and effective sample size. This is the first of three other computational contributions which allow us to explore inference with larger classes of mixture models. In the next Chapter, a hybrid MCMC scheme, which exploits the marginal and conditional samplers advantages is presented.

# Chapter 4

# MCMC scheme II: a hybrid sampler

The aim of this chapter is to introduce a novel MCMC sampler for Poisson–Kingman mixture models, one of the largest classes of Bayesian nonparametric mixture models that encompasses all previously explored priors in the literature and some unknown ones. In Chapter 2, a brief overview about this class of priors is given, and Figure 2.2 shows some subfamilies related to this class. The approach presented in this chapter is a hybrid MCMC scheme which combines the main strengths of both conditional and marginal MCMC samplers. If we start from the generative process of Poisson–Kingman processes from Proposition 2 in Chapter 2, a useful characterisation can be derived, based on a compact representation of the process, to construct the sampling scheme. The necessary derivations are described herein in detail. In particular, various versions of the algorithm are presented for all $\sigma$-Stable Poisson–Kingman mixture models. All versions differ only in one of the MCMC steps which updates a random variable of interest. However, this choice has an effect on the running times and effective sample size of the different algorithmic versions. Successively, another family of priors, referred to as the -logBeta Poisson–Kingman priors, which has not been explored before in the context of mixture models, is described. This class has similar asymptotic properties as the Dirichlet process prior. Finally, some comparisons are presented against the marginal MCMC sampler from Chapter 3 (Lomeli et al., 2016), and the conditional sampler of Favaro and Walker (2012).

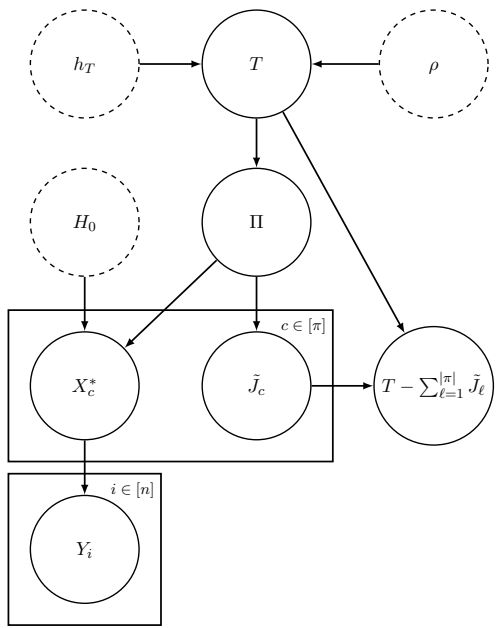

Figure 4.1: PK hybrid sampler's graphical model. Each node represents a variable used in the augmented representation for the joint distribution from Equation 4.1. The boxes represent that there is a number of repeated nodes of the same type, $n$ for the observations and $|\pi|$ for the cluster parameters. The nodes with dashed lines represent the algorithmic inputs.

## 4.1 Hybrid Sampler

Figure 4.1 shows the tractable and compact representation in which the hybrid MCMC sampler for a Poisson–Kingman mixture model is based. This is a good alternative to the tractable representations presented in Chapter 3 for the marginal and for conditional MCMC samplers, Figures 3.1 and 3.2 respectively. The hybrid sampler's tractable representation is obtained in the following way: let us start from Proposition 2, which is written in terms of the first $k$ size-biased weights. In order to obtain a complete representation of the RPM, we would need to size-bias sample from it a countably infinite number of times. However, since the induced partition is also included in the MCMC, it allows us to get a complete but finite representation of the infinite dimensional object. The joint distribution of interest, after being reparameterised in terms of the suplus mass random variable and after the likelihood terms have been

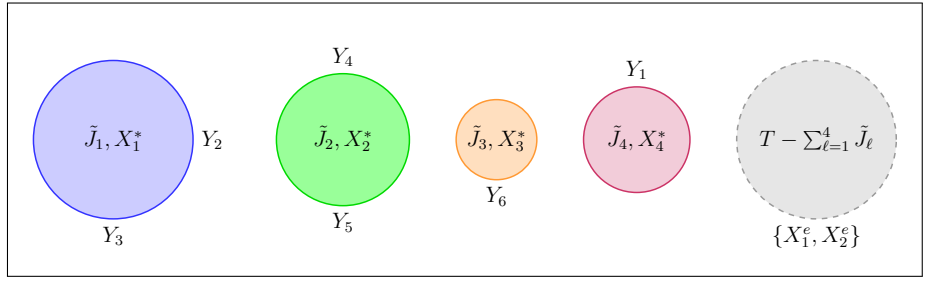

Figure 4.2: Hybrid sampler's restaurant representation for observations $\{Y_i\}_{i=1}^{9}$ clustered in $k = 4$ occupied tables (coloured) with corresponding size-biased sizes $\{\tilde{J}_\ell\}_{\ell=1}^{4}$ and surplus mass in grey.

added, is

$$\Pr\nolimits_{\rho,h,H_0}(\Pi_n = (c_\ell)_{\ell \in [|\Pi_n|]}, X_c^* \in dx_c^*, \tilde{J}_c \in ds_c \text{ for } c \in \Pi_n, T - \sum_{c \in \Pi_n} \tilde{J}_c \in dv, Y_i \in dy_i \text{ for } i \in [n])$$

(4.1)

$$= (v + \sum_{c \in \Pi_n} s_c)^{-n} h\left(v + \sum_{c \in \Pi_n} s_c\right) f_\rho(v) \prod_{c \in \Pi_n} s_c^{|c|} \rho(ds_c) H_0(dx_c^*) \prod_{i \in c} F(dy_i | x_c^*).$$

Indeed, this distribution encodes a complete, but finite, representation of the infinite dimensional part of the model. The size-biased weights associated to occupied clusters are the only ones needed to be explicitly represented plus a surplus mass term, associated to the rest of the empty clusters, as shown in Figure 4.2. The cluster reassignment step can be seen as a retrospective sampling scheme: we explicitly represent and update the weights associated to occupied clusters and create a new size-biased weight only when a new cluster appears. To make this possible, we use the induced partition. In the next section, the complete conditional distributions of each random variable are computed to implement an overall Gibbs sampling MCMC scheme (Geman and Geman, 1984).

## 4.2 Complete conditional distributions

If we star from Equation (4.1), we obtain the following complete conditional distributions for the Gibbs sampler

$$\Pr\left(T - \sum_{c \in \Pi_n} \tilde{J}_c \in dv \mid \text{Rest}\right) \propto \left(v + \sum_{c \in \Pi_n} s_c\right)^{-n} f_\rho(v) h\left(v + \sum_{c \in \Pi_n} s_c\right) dv \qquad (4.2)$$

$$\Pr\left(\tilde{J}_c \in ds_c \mid \text{Rest}\right) \propto \left(v + s_c + \sum_{c' \in \Pi_n \setminus \{c\}} s_{c'}\right)^{-n}$$

$$\times h\left(v + s_c + \sum_{c' \in \Pi_n \setminus \{c\}} s_{c'}\right) s_c^{|c|} \rho(ds_c) \mathbb{I}_{(0, \text{Surpmass}_c)}(s_c) ds_c$$

where $\text{Surpmass}_c = T - \sum_{\{c' \in \Pi_n : c' < c\}} \tilde{J}_{c'}$ and we denote by $c' < c$ all the indices of the blocks of the partition which where updated before $c$. The cluster assignment rule is

$$\Pr(i \text{ joins cluster } c' \mid \{\tilde{J}_c\}_{c \in \Pi_n}, V, \{X_c^*\}_{c \in \Pi_n}, \{Y_i\}_{i \in [n]}) \begin{cases} \tilde{J}_{c'} F(Y_i \in dy_i \mid \{Y_i\}_{i \in c'}, X_{c'}^*) & \text{if } c' \in \Pi_n \\ \frac{v}{M} F(Y_i \in dy_i \mid X_{c'}^*) & \text{o.w.} \end{cases}$$

$$(4.3)$$

According to this rule, the $i$-th observation will be either reassigned to an existing cluster or to one of the $M$ new clusters. If it is assigned to a new cluster $c_{\text{new}}$, then we need to sample a new size-biased weight from the following complete conditional distribution

$$\Pr\left(\tilde{J}_{c_{\text{new}}} \in ds_{c_{\text{new}}} \mid \text{Rest}\right) \propto f_\rho(v - s_{c_{\text{new}}}) \rho(s_{c_{\text{new}}}) s_{c_{\text{new}}} \mathbb{I}_{(0,v)}(s_{c_{\text{new}}}) ds_{c_{\text{new}}}. \qquad (4.4)$$

Every time a new cluster is created, its corresponding size-biased weight needs to be sampled. This could happen $1 \leqslant R \leqslant n$ times per MCMC iteration. Hence, it has a significant contribution to the overall computational cost. For this reason, an independent and identically distributed (i.i.d.) sample drawn from its complete conditional distribution from Equation (4.4) is highly desirable rather than having to run an MCMC chain to obtain a correlated sample. In the next section, how to obtain i.i.d. samples is presented for two different subclasses of Poisson–Kingman priors. Finally, for updating the cluster parameters $\{X_c^*\}_{c \in \Pi_n}$, in the case where $H_0$ is non-conjugate to the likelihood, an extension of the ReUse algorithm (Favaro and Teh, 2013) is used, see

Algorithm 14 for details.

The rest of the complete conditional distributions in Equation (4.2) do not have a standard form so there are no available random number generators to obtain an independent and identically distributed sample from these. However, a generic MCMC method can be applied to sample from each within the Gibbs sampler. The slice sampling MCMC (Neal, 2003) is used to update the size-biased weights and the surplus mass. However, in the $\sigma$-Stable Poisson–Kingman subclass of priors, the random variable for the surplus mass has an intractable component due to the $\sigma$-Stable density. Hence, an extra step is introduced to sample the surplus mass. In the next section, we present two alternative ways to overcome this difficulty.

## 4.3 Example of classes of Poisson–Kingman priors

**a) $\sigma$-Stable Poisson–Kingman processes (Pitman, 2003).** Let $\sigma \in (0,1)$ and

$$f_\sigma(t) = \frac{1}{\pi} \sum_{j=0}^{\infty} \frac{(-1)^{j+1}}{j!} \sin(\pi\sigma j) \frac{\Gamma(\sigma j + 1)}{t^{\sigma j + 1}} \mathbb{I}_{(0,\infty)}(t)$$

the density function of a positive $\sigma$-Stable random variable and $\rho(dx) = \rho_\sigma(dx) := \frac{\sigma}{\Gamma(1-\sigma)} x^{-\sigma-1} dx$. This class of RPMs is denoted by $\text{PK}(\rho_\sigma, h_T, H_0)$ where $h$ is a function that indexes each member of the class. See Figure 2.2 in Chapter 2 for an overall picture of all families and subfamilies included in it. This class includes all Gibbs type priors with parameter $\sigma \in (0,1)$, other choices of $h$ are possible, see Gnedin and Pitman (2006) and De Blasi et al. (2015) for a noteworthy account of this class of Bayesian nonparametric priors and the examples given in Chapter 3. Since the density for the surplus mass is intractable, two ways of dealing with this are proposed herein. Firstly, an integral representation for the $\sigma$-Stable density (Kanter, 1975) is exploited in order to introduce an auxiliary variable $Z$ and Gibbs sample each variable. In Chapter 3 (Lomeli et al., 2016), this was used for the construction of the marginal MCMC sampler. The relevant complete conditional distributions, after the augmentation, are as follows

$$\Pr\left(T - \sum_{c\in\Pi_n} \tilde{J}_c \in dv \mid \text{Rest}\right) \propto \left(v + \sum_{c\in\Pi_n} s_c\right)^{-n} v^{-\frac{\sigma}{1-\sigma}} \exp\left[-v^{\frac{-\sigma}{1-\sigma}} A(z)\right]$$

$$\times h\left(v + \sum_{c\in\Pi_n} s_c\right) dv \mathbb{I}_{(0, t-\sum_{c\in\Pi_n} s_c)}(v)$$

$$\Pr(Z \in dz \mid \text{Rest}) \propto A(z) \exp\left[-v^{(-\frac{\sigma}{1-\sigma})} A(z)\right] dz \mathbb{I}_{(0,\pi)}(z),$$

see Algorithm 12 for details.

---

**Algorithm 12** HybridSampler$\sigma$-PK$\left(T - \sum_{c\in\Pi_n} \tilde{J}_c, \Pi_n, \{Y_i\}_{i\in[n]}, \{X_c^*\}_{c\in\Pi_n}, \{\tilde{J}_c\}_{c\in\Pi_n}, H_0, M\right)$

    **for** $t = 2 \to iter$ **do**
        Slice sample $\tilde{\Pr}\left(T - \sum_{c\in\Pi_n} \tilde{J}_c \in dv \mid \text{rest}\right)$
        **for** $c \in \Pi_n$ **do**
            Slice sample $\tilde{\Pr}\left(\tilde{J}_c \in ds_c \mid \text{rest}\right)$
            $(\Pi_n, \{X_c^*\}_{c\in\Pi_n}) = \textbf{AddTable\&ReUse}\left(V, \Pi_n, M, \{Y_i\}_{i\in[n]}, \{X_c^*\}_{c\in\Pi_n}, \{\tilde{J}_c\}_{c\in\Pi_n}, H_0 \mid \text{rest}\right)$
        **end for**
    **end for**

---

Alternatively, we can bypass the evaluation of the density of the total mass by updating the surplus mass with a Metropolis-Hastings step with an independent proposal from a Stable or from an Exponentially Tilted Stable, with tilting parameter $\lambda$. It is straightforward to obtain i.i.d. draws from these proposals, see Devroye (2009) and Hofert (2011) for an improved rejection sampling method for the Exponentially tilted case. This leads to the following acceptance ratio

$$\frac{\Pr\left((T - \sum_{c\in\Pi_n} \tilde{J}_c)' \in dv' \mid \text{Rest}\right) f_\sigma(v) \exp(-\lambda v)}{\Pr\left(T - \sum_{c\in\Pi_n} \tilde{J}_c \in dv \mid \text{Rest}\right) f_\sigma(v') \exp(-\lambda v')} = \frac{\left(v' + \sum_{c\in\Pi_n} s_c\right)^{-n} h\left(v' + \sum_{c\in\Pi_n} s_c\right) dv' \exp(-v)}{\left(v + \sum_{c\in\Pi_n} s_c\right)^{-n} h\left(v + \sum_{c\in\Pi_n} s_c\right) dv \exp(-v')},$$

which allows us to bypass the point wise evaluation of the $\sigma$-Stable density altogether. See Algorithm 13 and (14) for details .

Finally, to sample a new size-biased weight, the corresponding complete conditional dsitribution is

$$\Pr\left(\tilde{J}_{c_{\text{new}}} \in ds_{c_{\text{new}}} \mid \text{Rest}\right) \propto f_\sigma(v - s_{c_{\text{new}}}) s_{c_{\text{new}}}^{-\sigma} \mathbb{I}_{(0,v)}(s_{c_{\text{new}}}) ds_{c_{\text{new}}}. \tag{4.5}$$

**Algorithm 13** HybridSampler-MH-$\sigma$PK$\left(T - \sum_{c \in \Pi_n} \tilde{J}_c, \Pi_n, \{Y_i\}_{i \in [n]}, \{X_c^*\}_{c \in \Pi_n}, \{\tilde{J}_c\}_{c \in \Pi_n}, H_0, M\right)$
___
for $t = 2 \to iter$ do
    for $c \in \Pi_n$ do
        Slice sample $\tilde{\Pr}\left(\tilde{J}_c \in ds_c \mid \text{rest}\right)$
        M-H step for $\tilde{\Pr}\left(T - \sum_{c \in \Pi_n} \tilde{J}_c \in dv \mid \text{rest}\right)$ with independent proposal `Stablernd`$(\sigma)$ or `ExpTiltStablernd`$(\lambda, \sigma)$ .
        $(\Pi_n, \{X_c^*\}_{c \in \Pi_n}) = $ **AddTable&ReUse**$\left(T - \sum_{c \in \Pi_n} \tilde{J}_c, \Pi_n, M, \{Y_i\}_{i \in [n]}, \{X_c^*\}_{c \in \Pi_n}, \{\tilde{J}_c\}_{c \in \Pi_n}, H_0\right)$
    end for
end for
___

**Algorithm 14** AddTable&ReUse$\left(V := T - \sum_{c \in \Pi_n} \tilde{J}_c, \Pi_n, M, \{Y_i\}_{i \in [n]}, \{X_c^*\}_{c \in \Pi_n}, \{\tilde{J}_c\}_{c \in \Pi_n}, H_0 \mid \text{rest}\right)$
___
Draw $\{X_j^e\}_{j=1}^M \overset{\text{i.i.d.}}{\sim} H_0$
for $i = 1 \to n$ do
    Let $c \in \Pi_n$ be such that $i \in c$
    $c \leftarrow c \backslash \{i\}$
    if $c = \emptyset$ then
        $k \sim \text{UniformDiscrete}(\frac{1}{M})$
        $X_k^e \leftarrow X_c^*$
        $\Pi_n \leftarrow \Pi_n \backslash \{c\}$
        $V \leftarrow V + \tilde{J}_c$
    end if
    Set $c'$ according to Pr(i joins cluster $c'$ |, Rest)         ▷ from Equation 4.3
    $\tilde{J}_{\text{new}} \leftarrow$ **ExactSampleNewTableSize**$(V, \sigma \mid \text{rest})$
    $V \leftarrow V - \tilde{J}_{\text{new}}$
    if $|c'| = 1$ then
        $\Pi_n \leftarrow \Pi_n \cup \{\{i\}\}$
        $X_{\{i\}}^* \leftarrow X_{c'}^e$
        $X_{c'}^e \sim H_0$
    else
        $c' \leftarrow c' \cup \{i\}$
    end if
end for
for $c \in \Pi_n$ do
    Sample from the posterior $X_c^* \sim F(\cdot \mid \{Y_i\}_{i \in c})$
end for
___

Fortunately, we can obtain an i.i.d. draw from Equation (4.5) due to our identity in distribution for the usual stick breaking weights for any prior in this class such that $\sigma = \frac{u}{v}$ where $u < v$ are coprime integers. These weights are then reparameterised to obtain the new size-biased weight, see Algorithm 15 and Chapter 3, Section 3.6 for details about these distributional identities (Favaro et al., 2014a).

---

**Algorithm 15** ExactSampleNewTableSize($V, \sigma \mid$ rest)

---

if $\sigma = 0.5$ then
    $G \sim \text{Gamma}\left(\frac{3}{4}, 1\right)$
    $IG \sim \text{Inverse Gamma}\left(\frac{1}{4}, \frac{1}{4^3}V^{-2}\right)$
    $V_{stick} = \frac{\sqrt{G}}{\sqrt{G}+\sqrt{IG}}$
    $\tilde{J}_{new} = V_{stick}V$
else
    if $\sigma < 0.5$ && $\sigma = \frac{u_\sigma}{v_\sigma}, u_\sigma, v_\sigma \in \mathbb{N}$ then
        $\lambda = u_\sigma^2 / v_\sigma^{\frac{v_\sigma}{u_\sigma}}$
        $IG \sim \text{Inverse Gamma}\left(1 - \frac{u_\sigma}{v_\sigma}, \lambda\right)$
        $\frac{1}{G} \sim \mathcal{E}_{\mathcal{T}}\left(\lambda, L_{\frac{u_\sigma}{v_\sigma}}^{-1/u}\right)$ ▷ Samples an exponentially tilted random variable which consists of a product of Gamma and Beta random variables.
        $V_{stick} = \frac{G}{G+IG}$
        $\tilde{J}_{new} = V_{stick}V$
    end if
end if

---

**b) $-\log$Beta-Poisson–Kingman processes (Regazzini et al., 2003; von Renesse et al., 2008).** Let

$$f_\rho(t) = \frac{\Gamma(a+b)}{\Gamma(a)\Gamma(b)} \exp(-at) \left(1 - \exp(-t)\right)^{b-1}$$

be the density of a positive random variable $T \stackrel{d}{=} -\log Y$, where $Y \sim \text{Beta}(a,b)$ and $\rho(x) = \frac{\exp(-at)(1-\exp(-bt))}{t(1-\exp(-t))}$. This class of RPMs generalises the Gamma process but has the same asymptotic properties, namely $|\Pi_n| \sim a \log(n)$. Indeed, if we take $b = 1$ and the density function for $T$ is $\gamma(t) = f_\rho(t)$, we recover the Lévy measure of a Gamma process. In this case, to sample a new size-biased weight, the corresponding complete conditional distribution is

$$\Pr\left(\tilde{J}_{c_{\text{new}}} \in ds_{c_{\text{new}}} \mid \text{Rest}\right) \propto \frac{(1 - \exp(s_{c_{\text{new}}} - v))^{b-1}(1 - \exp(-bs_{c_{\text{new}}}))}{1 - \exp(-s_{c_{\text{new}}})} ds_{c_{\text{new}}} \mathbb{I}_{(0,v)}(s_{c_{\text{new}}}).$$

If $b > 1$, this complete conditional distribution is a monotone decreasing unnormalised density with maximum at $b$. We can easily get an i.i.d. draw with a simple rejection sampler (Devroye, 1986) with proposal is $U(0, v)$, where the rejection rate is $bv$. There is no other known sampler for this process.

## 4.4 Relationship to marginal and conditional MCMC samplers

If we start from Proposition 2, another strategy is to reparameterise the model in terms of the stick breaking weights. A random truncation level is chosen in order to represent finitely many sticks (Favaro and Walker, 2012) and there will be both sticks associated to empty and occupied clusters. In this representation of the model the partition is not sampled but some allocation variables are represented that encode the cluster assignment to one of the represented sticks after truncation per data point. Alternatively, we can integrate out the random probability measure and sample only the induced partition to construct a marginal MCMC sampler, as in Chapter 3 and in Lomeli et al. (2016). Conditional samplers have large memory requirements as often, the number of sticks needed can be very large. Furthermore, the conditional distribution of the stick length is quite involved and dependent on the other sticks for general $\sigma$-Stable Poisson–Kingman priors other than the Pitman–Yor or Dirichlet process cases. For this reason, the conditional MCMC samplers could have slow running times. Marginal samplers have less storage requirements than conditional samplers but could potentially have worse mixing properties. ??he hybrid sampler presented herein exploits the advantages of both marginal and conditional MCMC samplers. It has less memory requirements since it only represents the size-biased weights associated to occupied clusters as opposed to conditional samplers which represent both empty and occupied clusters. Also, since it does not integrate out all of the size-biased weights, we obtain a more comprehensive representation of the RPM.

## 4.5 Relationship between size-biased and stick breaking parameterisations

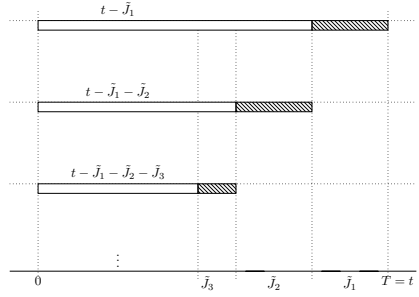

Figure 4.3: Generative process of PKP

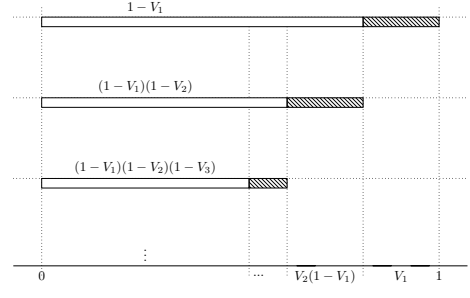

Figure 4.4: Pitman–Yor's stick breaking construction

$$T \sim \gamma$$
$$\tilde{J}_1 \mid T \sim \text{SBS}(T)$$
$$\tilde{J}_2 \mid T, \tilde{J}_1 \sim \text{SBS}\left(T - \tilde{J}_1\right)$$
$$\vdots$$
$$\tilde{J}_\ell \mid T, \tilde{J}_1, \ldots, \tilde{J}_{\ell-1} \sim \text{SBS}\left(T - \sum_{i<\ell} \tilde{J}_i\right)$$
$$\vdots$$

$$Z_1 \sim \text{Beta}(z_1 \mid 1 - \sigma, \theta + \sigma)$$
$$Z_2 \sim \text{Beta}(z_2 \mid 1 - \sigma, \theta + 2\sigma)$$
$$\vdots$$
$$Z_\ell \sim \text{Beta}(z_\ell \mid 1 - \sigma, \theta + \ell\sigma)$$
$$\vdots$$

the corresponding weights are:

$$\pi_\ell \stackrel{d}{=} \frac{\tilde{J}_\ell}{T - \sum_{j<\ell} \tilde{J}_j} \qquad \pi_\ell \stackrel{d}{=} Z_\ell \prod_{j<\ell}(1 - Z_j).$$

In Chapter 2 Section 2.5, we reviewed the PKP generative process from Figure 4.3, it is reminiscent of the well-known stick breaking construction from Ishwaran and James (2001), where a stick of length one is broken, as in Figure 4.4, but it is not the same. As mentioned previously, we can reparameterise the model, starting with Equation (2.24), and obtain the corresponding joint distribution in terms of $N$ $(0, 1)$-valued stick-breaking weights $\{\pi_j\}_{j=1}^N$, where $N$ is the number of represented sticks after trunctation (Favaro and Walker, 2012). This joint distribution is for a general Lévy measure $\rho$, density $f_\rho$ and it is conditioned on the valued of the random variable $T$. We can then recover the well-known Stick breaking representations for the Dirichlet and Pitman–Yor processes, for a specific choice of $\rho$, if we integrate out $T$. For instance, the Pitman–Yor stick breaking representation (Ishwaran and James, 2001) can be recovered from the size-biased sampling (SBS) generative process of Section 2.5 after integrating out the total mass $T$, a change of variables and the following distribution for the total mass is substituted

$$\gamma_{\text{PY}}(t) = \frac{\Gamma(\theta + 1)}{\Gamma(\frac{\theta}{\sigma} + 1)} t^{-\theta} f_\sigma(t) \mathbb{I}_{(0,\infty)}(t), \quad \theta > -\sigma$$

where $f_\sigma$ is the density of a $\sigma$-Stable random variable. However, in general, these stick random variables, denoted by $Z_j \stackrel{d}{=} \frac{\pi_j}{1-\sum_{\ell<j} \pi_\ell}$, form a sequence of dependent random variables with a complicated distribution, except for the two previously mentioned processes, see Pitman (1996) for details.

## 4.6 Performance assesssment

We illustrate the performance of our hybrid MCMC sampler on a range of Bayesian nonparametric mixture models, obtained by different specifications of $\rho$ and $\gamma$. These Bayesian nonparametric priors were chosen from both the $\sigma$-Stable and the -logBeta Poisson–Kingman classes, presented in Section 4.3. We chose the base distribution $H_0$ and the likelihood term $F$ for cluster $c$ to be

$$H_0(d\mu_c) = \mathcal{N}\left(d\mu_c \mid \mu_0, \sigma_0^2\right) \quad \text{and} \quad F(\{Y_i \in dy_i\}_{i \in c} \mid \mu_c, \tau_1) = \prod_{i \in c} \mathcal{N}\left(x_i \mid \mu_c, \sigma_1^2\right).$$

$\{Y_j\}_{j \in c}$ are the observations assigned to cluster $c$ at some iteration. $\mathcal{N}$ denotes a Normal distribution with mean $\mu_c$ and variance $\sigma_1^2$, a common parameter among all clusters. The mean's prior distribution is Normal, centered at $\mu_0$ and with variance $\sigma_0^2$. Although the base distribution is conjugate to the likelihood, we treated it as non-conjugate case and sampled the parameters at each iteration rather than integrating them out.

We used the dataset from Roeder (1990), introduced previously in Chapter 3, to test the algorithmic performance in terms of running time and effective sample size (ESS), as Table 4.1 shows. For the $\sigma$-Stable Poisson–Kingman class, we compared it against our implementation of the conditional MCMC sampler by Favaro and Walker (2012) and against the marginal sampler of Chapter 3 (Lomeli et al., 2016). We chose only to compare the hybrid sampler against these existing approaches since they also follow the same general purpose paradigm.

| Algorithm | $\sigma$ | Running time | ESS($\pm$std) |
|---|---|---|---|
| Pitman–Yor process ($\theta = 10$) | | | |
| Hybrid | 0.3 | 7135.1(28.316) | **2635.488(187.335)** |
| Hybrid-MH ($\lambda = 0$) | 0.3 | 5469.4(186.066) | 2015.625(152.030) |
| Conditional | 0.3 | NA | NA |
| Marginal | 0.3 | **4685.7(84.104)** | 2382.799(169.359) |
| Hybrid | 0.5 | **3246.9(24.894)** | **3595.508(174.075)** |
| Hybrid-MH ($\lambda = 50$) | 0.5 | 4902.3(6.936) | 3579.686(135.726) |
| Conditional | 0.5 | 10141.6(237.735) | 905.444(41.475) |
| Marginal | 0.5 | 4757.2(37.077) | 2944.065(195.011) |
| Normalized Stable process | | | |
| Hybrid | 0.3 | **5054.7(70.675)** | **5324.146(167.843)** |
| Hybrid-MH ($\lambda = 0$) | 0.3 | 7866.4(803.228) | 5074.909(100.300) |
| Conditional | 0.3 | NA | NA |
| Marginal | 0.3 | 7658.3(193.773) | 2630.264(429.877) |
| Hybrid | 0.5 | 5382.9(57.561) | **4877.378(469.794)** |
| Hybrid-MH ($\lambda = 50$) | 0.5 | **4537.2(37.292)** | 4454.999(348.356) |
| Conditional | 0.5 | 10033.1(22.647) | 912.382(167.089) |
| Marginal | 0.5 | 8203.1(106.798) | 3139.412(351.788) |
| Normalized Generalized Gamma process ($\tau = 1$) | | | |
| Hybrid | 0.3 | **4157.8(92.863)** | **5104.713(200.949)** |
| Hybrid-MH ($\lambda = 0$) | 0.3 | 4745.5(187.506) | 4848.560(312.820) |
| Conditional | 0.3 | NA | NA |
| Marginal | 0.3 | 7685.8(208.98) | 3587.733(569.984) |
| Hybrid | 0.5 | 6299.2(102.853) | **4646.987(370.955)** |
| Hybrid-MH ($\lambda = 50$) | 0.5 | **4686.4(35.661)** | 4343.555(173.113) |
| Conditional | 0.5 | 10046.9(206.538) | 1000.214(70.148) |
| Marginal | 0.5 | 8055.6(93.164) | 4443.905(367.297) |
| -logBeta ($a = 1, b = 2$) | | | |
| Hybrid | - | **2520.6(121.044)** | **3068.174(540.111)** |
| Conditional | - | NA | NA |
| Marginal | - | NA | NA |

Table 4.1: Running times in seconds and ESS averaged over 10 chains, 30,000 iterations, 10,000 burn in.

Table 4.1 shows that different choices of $\sigma$ result in differences in the running times and ESS. When $\sigma = 0.5$, there are readily available random number generators for the stick variables, and the computational cost does not increase. In contrast, when $\sigma = 0.3$, a rejection sampler method is needed every time a new size-biased weight is sampled. This increases the computational cost, see Favaro et al. (2014a) for details about the improved rejection sampler. In most cases, both marginal and conditional MCMC schemes are outperformed in terms of running times and in all cases, in terms of ESS. In the Hybrid-MH case, even thought the ESS and running times are competitive, we found that the acceptance rate is not optimal. Finally, for the -logBeta Poisson–Kingman processes of Example b), our approach is the only one available and it has good running times and ESS. This qualitative comparison confirms our previous statements about our novel approach.

## 4.7 Summary of this Chapter

In this Chapter, a novel MCMC is presented based on the generative process for Poisson–Kingman priors. The auxiliary sequence of size-biased weights can be sampled retrospectively since a new size-biased weight needs to be sampled only when we observe a new cluster. This property allows us to only represent those size-biased weights associated to observations, as opposed to a conditional MCMC sampler, were there is a need to represent weights for both empty and occupied clusters. In our experiments, our hybrid MCMC scheme outperforms both marginal and conditional MCMC samplers in running times in most cases and in ESS in all cases.

# Chapter 5

# SMC sampler for a sequential scenario

The aim of this chapter is to introduce various SMC samplers for Bayesian sequential inference. Specifically, in this sequential scenario, the dataset is not available in batch mode but each datapoint is obtained one by one. The posterior distribution is then computed every time a new datapoint is observed and a marginal likelihood estimate is repeatedly obtained. A Bayesian hypothesis test is then performed to compare two competing hypotheses. The first two SMC schemes are for an infinite mixture model with priors that belong to the $\sigma$-Stable Poisson–Kingman class. In Chapter 2, an infinite mixture model was motivated as a limit of a finite mixture model, when the total number of components tends to infinity. In other words, the prior assigns all its mass to an infinite total number of components. In contrast, the third SMC scheme presented herein can be used for inference in a mixture of finite mixtures (MFM): a model where the prior for the total number of components assigns a positive probability to any total number of mixture components $m \in \mathbb{N}$. The infinite mixture model is repeatedly tested against the MFM, as we observe more data with a Bayes factor test (Kass and Raftery, 1995). Alternatively, a predictive likelihood test (Geweke and Amisano, 2010) could be used. Our main goal is to answer a rather philosophical question: will the Bayesian nonparametric model's unbounded complexity benefit our posterior inferences about the number of components through better marginal likelihood estimates soon enough? Indeed, in an infinite mixture model, the number of occupied components is always finite but it is unbounded and grows as a function of the number of observations. In contrast, the MFM could saturate, under a misspecified model scenario, with respect to the mean number of occupied components. In other words, the MFM model might not adapt or learn more components as we observe more data, in the scenario where the true number of components is large but we assume a priori a small number of components. One way that the adaptability of the Bayesian nonparametric model can be observed is sequentially, regardless of whether we obtained all the data at once or successively. This is because the number of occupied components of the infinite mixture model grows sub linearly with respect to the data size. Hence, the only reasonable scenario to construct such a Bayesian hypothesis test is sequentially.

This question has been formulated before (Miller and Harrison, 2015) but we are not aware of any Bayesian tests for a quantitative comparison. Furthermore, in some cases, having a Bayesian nonparametric component could not be justified. Hence, the proposed alternative of using a MFM model with different prior distributions for the total number of components is a reasonable choice. Our aim is to potentially show that the latter alternative imposes too much structure and is computationally difficult or inexact for certain choices of prior. The asymptotic results from Gnedin et al. (2007), which justify the Bayesian nonparametric model's unbounded complexity, are very elegant but what are their statistical implications in terms of predictive performance?, will we ever reach the asymptotic regime?

Firstly, we construct the three SMC samplers. The first scheme can only be applied to two members of the $\sigma$-Stable Poisson–Kingman class, the Pitman–Yor and the Normalised Generalised Gamma process, due to a sampling difficulty; it relies on a result from Arbel et al. (2015). The second scheme is for a larger subclass, the $\mathcal{PG}$ class with $\sigma \leqslant 0.5$ (James, 2013), first reviewed in Example 9 of Chapter 3. It uses the same generative process and corresponding characterisations exploited in the proposed MCMC scheme of Chapter 3 (Lomeli et al., 2015). The third scheme is an approximate SMC which includes all Gibbs type priors with parameter $\sigma \in (-\infty, 1)$. It relies on unbiased and positive Monte Carlo estimators, one of which can be obtained due to a distributional identity given in Ho et al. (2008). Finally, we review the Bayes factor, an existing model criticism tool, and use it in a numerical illustration for our present context.

## 5.1 Gibbs type priors

In Chapter 2, the finite dimensional distributions which correspond to exchangeable random partitions of $\mathbb{N}$ were reviewed. These are called exchangeable partition probability functions (EPPF) and specify the law of the restriction of an infinite partition to the first $n$ integers, $n \in \mathbb{N}$, denoted by $\Pi_n$, with $k$ blocks and $n_\ell$ is the size of the $\ell$-th block, $\ell = 1, \ldots, k$. Furthermore, an EPPF is Gibbs type if it has the following mathematically convenient factorised form

$$\Pr(\Pi_n = \pi) = V_{n,k} \prod_{\ell=1}^{k} (1-\sigma)_{n_\ell - 1 \uparrow 1}.$$

$(1-\sigma)_{n_\ell - 1 \uparrow 1}$ denotes the ascending factorial, see the Appendix A.2 for details of how to compute this quantity. The first term specifies the probability of having $k$ blocks in a partition of $[n]$; it only depends on $n$ and $k$. Each term in the product specifies the probability that the $\ell$-th block has cardinality $n_\ell$. See De Blasi et al. (2015) for a comprehensive review about Gibbs type priors. The predictive distribution is given by

$$\Pr(X_{n+1} \in \cdot \mid X_1 \ldots, X_n) = \frac{V_{n+1,k+1}}{V_{n,k}} H_0(\cdot) + \frac{V_{n+1,k}}{V_{n,k}} \sum_{\ell=1}^{k} (n_\ell - \sigma) \delta_{X_\ell^*}(\cdot). \quad (5.1)$$

All Gibbs type priors have the remarkable property that the $V_{n,k}$ parameters satisfy the following backward recursive equation

$$\begin{aligned} V_{n,k} &= (n - \sigma k) V_{n+1,k} + V_{n+1,k+1} \\ 1 &= (n - \sigma k) \frac{V_{n+1,k}}{V_{n,k}} + \frac{V_{n+1,k+1}}{V_{n,k}}. \end{aligned} \quad (5.2)$$

In the species sampling literature, Gibbs type priors have been used to compute estimators for the species discovery probability from a Bayesian point of view. Some estimators have been obtained which are analogous to frequentist estimators of these probabilities, see Bunge and Fitzpatrick (1993) and Bunge et al. (2014) for a review of the species sampling problem and Lijoi et al. (2007a) for a Bayesian nonparametric approach and references therein.

Given a sample of $n$ observations $(X_1, \ldots, X_n)$ from a Gibbs type random probability measure, the probability that the $n + 1$-th draw coincides with an existing type in the sample with frequency $r$, $r \in \{0, \ldots, n\}$, is called the discovery probability, it is denoted by $D_n(r)$. Let $A_0$ be the set that specifies a new type, and $A_r$, the set that specifies all the existing types with frequency $r$ in a sample of size $n$, respectively. Specifically,

$$A_0 = \mathbb{X} \setminus \{X_1^*, \ldots, X_k^*\},$$
$$A_r = \{X_i^* : n_\ell = r\}, \quad r = 1, \ldots, n, \quad \ell = 1, \ldots, k$$
$$m_r = |A_r|. \tag{5.3}$$

Let $0 \leqslant k \leqslant n$ be the number of existing types in a sample of size $n$, each type is denoted by $X_\ell^*, \ell = 1, \ldots, k$, with corresponding frequency $n_\ell$; and $\mathbb{X}$ is the space of all types, which has infinitely countably many. The $r$-th discovery probability can be obtained using the system of predictive distributions of Equation (5.1) for any Gibbs type prior by evaluating it in the sets $A_0$ and $A_r$ from Equation (5.3), where

$$\begin{aligned} D_n(0) &= \Pr\left(X_{n+1} \in A_0 \mid X_1, \ldots, X_n\right) \\ &= \mathbb{E}_{\mathcal{P}}\left(P(A_0) \mid X_1, \ldots, X_n\right) \\ &= \frac{V_{n+1,k+1}}{V_{n,k}} \end{aligned} \tag{5.4}$$

and

$$\begin{aligned} D_n(r) &= \Pr\left(X_{n+1} \in A_r \mid X_1, \ldots, X_n\right) \\ &= \mathbb{E}_{\mathcal{P}}\left(P(A_r) \mid X_1, \ldots, X_n\right) \\ &= (r - \sigma) \times m_r \frac{V_{n+1,k}}{V_{n,k}}. \end{aligned}$$

Both quantities are expectations with respect to the law of a random discrete distribution, denoted by $P$ and its law is denoted by $\mathcal{P}$, namely $P \sim \mathcal{P}$, this fact will be used in the next section. The distributional identity from Arbel et al. (2015) is used to obtain an unbiased and positive estimator of the ratio given by Equation (5.4) and Equation (5.2) is used to compute its complement. These two estimators are then used as an input for an SMC sampler. So far, estimators for these ratios can only be obtained for two noteworthy members of the $\sigma$-Stable Poisson–Kingman class the Pitman–Yor and Generalised Gamma processes. This first scheme uses unbiased and positive estimators for both the proposal and for the weights of an SMC algorithm. The nested

Sequential Monte Carlo algorithm (Naesseth et al., 2015) is an example of the former. They propose to use an unbiased and positive estimator of the proposal distribution in an Sequential Monte Carlo algorithm provided that it is properly weighted with respect to the proposal distribution. They show that the Nested SMC is an example of a pseudo marginal MCMC, first introduced in Andrieu and Roberts (2009). The random weight Sequential Monte Carlo algorithm (Fearnhead et al., 2010) is an example of the latter. They propose to use an unbiased and positive estimator of the weights in an Sequential Monte Carlo algorithm. We believe that the results from these two references could be tailored to show that our algorithm results in an exact approximation but it is only a conjecture. In the next sections, two more SMC schemes are presented that can be used for a larger subclass of the $\sigma$-Stable Poisson–Kingman family of priors.

## 5.2 A first SMC sampler

---
**Algorithm 16** SMC
$\Pi_1^p = \{\{1\}\}, \forall p \in \{1, \ldots, L\}$
**for** $i = 2 : n$ **do**
    **for** $p = 1 : L$ **do**
        $\hat{R}_{i,|\Pi_{i-1}^p|} =$ MonteCarloEstimate$(\theta, \sigma, i, |\Pi_{i-1}^p|)$
        $\hat{R}_{i,|\Pi_{i-1}^p|+1} = 1 - (i - \sigma|\Pi_{i-1}^p|)\hat{R}_{i,|\Pi_{i-1}^p|}$
        Set $c'$ according to
$\Pr\left(i \text{ joins cluster } c' \mid \Pi_{i-1}^p, \mathbf{y}_{1:i-1}, \hat{R}_{i,|\Pi_{i-1}^p|}, \hat{R}_{i,|\Pi_{i-1}^p|+1}\right)$     ▷ from Equation (5.6)
        **if** $|c'| = 1$ **then**
            $\Pi_i^p = \Pi_{i-1}^p \cup \{\{i\}\}$
        **else**
            $c' = c' \cup \{i\}, c' \in \Pi_{i-1}^p$
            $\Pi_i^p = \Pi_{i-1}^p$
        **end if**
        Compute weight
$w_i^p \propto w_{i-1}^p \times \Pr\left(Y_i \in dy_i \mid \Pi_i^p, \mathbf{y}_{1:i-1}, \hat{R}_{i,|\Pi_{i-1}^p|}, \hat{R}_{i,|\Pi_{i-1}^p|+1}\right)$     ▷ from Equation (5.7)
    **end for**
    Normalise the weights $\tilde{w}_i^p = \frac{w_i^p}{\sum_{j=1}^L w_i^j}$
    ESS$= \frac{1}{\sum_{p=1}^L (\tilde{w}_i^p)^2}$
    **if** ESS $<$ thresh $\times L$ **then**
        Resample $p' \sim$ Multinomial $\left(\tilde{w}_i^1, \ldots, \tilde{w}_i^L\right), \forall p \in \{1, \ldots, L\}$
        $\Pi_i^p = \Pi_i^{p'}$ ▷ The $p$-th particle inherits the entire path of the $p'$-th particle, we are resampling at every step.
    **end if**
**end for**
---

In Algorithm 16, the pseudocode for the first SMC sampler proposed herein is presented. At first sight, it is ready to use for any Gibbs type prior but it is not clear how to obtain an unbiased and positive estimator for the ratio $\hat{R}_{i,|\Pi_{i-1}^p|}$. Let us denote the estimators for the ratios of interest as follows

$$\hat{R}_{i,|\Pi_{i-1}^p|} = \frac{\widehat{V_{i,|\Pi_{i-1}^p|}}}{V_{i-1,|\Pi_{i-1}^p|}}$$

$$\hat{R}_{i,|\Pi_{i-1}^p|+1} = \frac{\widehat{V_{i,|\Pi_{i-1}^p|+1}}}{V_{i-1,|\Pi_{i-1}^p|}} \tag{5.5}$$

If the posterior is used as the proposal, the corresponding cluster assignment conditional probability is

$$\Pr\left(\text{i joins cluster } c' \mid \Pi_{i-1}^p, \mathbf{y}_{1:i-1}, \hat{R}_{i,|\Pi_{i-1}^p|}, \hat{R}_{i,|\Pi_{i-1}^p|+1}\right) \propto$$
$$\left\{ \begin{array}{ll} \hat{R}_{i,|\Pi_{i-1}^p|}(|c'| - \sigma)F\left(Y_i \in dy_i \mid \{y_j\}_{j \in c'}\right) & \text{if } c' \in \Pi_{i-1}^p \\ \hat{R}_{i,|\Pi_{i-1}^p|+1}F(Y_i \in dy_i) & \text{o.w.} \end{array} \right\} \tag{5.6}$$

where $F\left(Y_i \in dy_i \mid \{y_j\}_{j \in c}\right) = \int F\left(Y_i \in dy_i \mid x\right) f\left(x \mid \{y_j\}_{j \in c}\right) dx$, and

$$\Pr\left(Y_i \in dy_i \mid \Pi_i^p, \mathbf{y}_{1:i-1}, \hat{R}_{i,|\Pi_{i-1}^p|}, \hat{R}_{i,|\Pi_{i-1}^p|+1}\right) = \hat{R}_{i,|\Pi_i^p|}F(Y_i \in dy_i)$$
$$+ \hat{R}_{i,|\Pi_i^\ell|} \sum_{c' \in \Pi_i^p} (|c'| - \sigma)F(Y_i \in dy_i \mid \{y_j\}_{j \in c'}). \tag{5.7}$$

If the prior is used as the proposal, the corresponding cluster assignment probability is

$$\Pr\left(\text{i joins cluster } c' \mid \Pi_{i-1}^p, \mathbf{y}_{1:i-1}, \hat{R}_{i,|\Pi_{i-1}^p|}, \hat{R}_{i,|\Pi_{i-1}^p|+1}\right) \propto \left\{ \begin{array}{ll} \hat{R}_{i,|\Pi_{i-1}^p|}(|c'| - \sigma) & \text{if } c' \in \Pi_{i-1}^p \\ \hat{R}_{i,|\Pi_{i-1}^p|+1} & \text{o.w.} \end{array} \right\}$$
$$\tag{5.8}$$

and

$$\Pr\left(Y_i \in dy_i \mid \Pi_i^p, \mathbf{y}_{1:i-1}\right) = F\left(Y_i \in dy_i \mid \{y_j\}_{j \in c}\right), \quad i \in c \tag{5.9}$$

There are two noteworthy cases, namely the Pitman–Yor and Generalised Gamma processes, where it is possible to obtain unbiased and positive estimators of the ratio of interest from Equation (5.5) (Arbel et al., 2015). The ratios of interest are expectations with respect to the law of a random discrete distribution, $\mathcal{P}$, either the Pitman–Yor or the Generalised Gamma processes, as mentioned in the previous section. We can then easily obtain a Monte Carlo estimate of this expectation due to the following identities in distribution. See Arbel et al. (2015) for the complete proof for this result.

The Pitman–Yor process $\mathcal{P}_{\text{PY}}$ is described in Chapter 3, Example 7, in terms of a specific tilting function which parameterises the $\sigma$-Stable Poisson–Kingman class of priors. A Monte Carlo estimate of the 0-th discovery probability can be obtained by repeatedly sampling from the distributional identity of Equation (5.10) as follows.

Let $W_{a,b}$ be a random variable that is a quotient of random variables

$$W_{a,b} = \frac{bR_{\sigma,b}}{bR_{\sigma,b} + G_{a,1}}$$

such that $G_{a,1}$ is independent of $R_{\sigma,b}$ and

$$G_{a,1} \sim \text{Gamma}(a,1)$$
$$R_{\sigma,b} \sim \mathcal{E}_T(b, S_\sigma)$$
$$S_\sigma \sim f_\sigma.$$

$\mathcal{E}_T(b,x)$ is an exponentially tilted distribution of $X$ with tilting parameter $b$. Let $G_{\frac{\theta}{\sigma}+k} \sim \text{Gamma}\left(\frac{\theta}{\sigma}+k, 1\right)$ is a Gamma random variable with shape parameter $\frac{\theta}{\sigma}+k$ and scale parameter 1, and $Z_{PY} \stackrel{d}{=} G_{\frac{\theta}{\sigma}+k}^{1/\sigma}$ is a random variable with density given by

$$f_{Z_{PY}}(z) = \frac{\sigma}{\Gamma(\frac{\theta}{\sigma}+k)} z^{\theta+\sigma k-1} \exp(-z^\sigma) \mathbb{I}_{(0,+\infty)}(z).$$

Then,

$$P_{PY}(A_0) \mid (X_1, \ldots, X_n) \stackrel{d}{=} W_{n-\sigma k, Z_{PY}}$$
$$\stackrel{d}{=} \beta_{\theta+k\sigma, n-\sigma k}$$
$$\widehat{D}_n^{\text{PY}}(0) = \frac{1}{N_s} \sum_{j=1}^{N_s} \beta_{\theta+k\sigma, n-\sigma k}(j) \qquad (5.10)$$

where $\beta_{a,b}$ is a Beta random variable with parameters $a$ and $b$ and $N_s$ is the number of Monte Carlo samples. Analogously, a Monte Carlo estimate of the $r$-th discovery probability is obtained by repeatedly sampling from the random variable

$$P_{PY}(A_r) \mid (X_1, \ldots, X_n) \stackrel{d}{=} (1 - W_{n-\sigma k, Z_{PY}}) \times \beta_{(r-\sigma)\times m_r, n-\sigma k-(r-\sigma)\times m_r}$$

$$\stackrel{d}{=} \beta_{(r-\sigma)\times m_r, \theta+n-(r-\sigma)\times m_r}$$

$$\widehat{D}_n(r) = \frac{1}{N_s} \sum_{j=1}^{N_s} \beta_{(r-\sigma)\times m_r, \theta+n-(r-\sigma)\times m_r}(j). \tag{5.11}$$

The Normalised Generalised-Gamma process, denoted by $\mathcal{P}_{GG}$, is a random discrete distribution with a Lévy measure for the random weights given by

$$\rho(ds) = \frac{1}{\Gamma(1-\sigma)} s^{-\sigma-1} \exp(-\tau s), \quad \sigma \in (0,1), \tau > 0.$$

Alternatively, the NGG process can be described as in Example 6 of Chapter 3, in terms of a specific tilting function which parameterises the $\sigma$-Stable Poisson–Kingman class of priors. A Monte Carlo estimate of the 0-th discovery probability can be obtained by repeatedly sampling from the distributional identity of Equation (5.13) as follows. Let $Z_{GG}$ be a random variable with density

$$f_{Z_{GG}}(z) = \frac{\sigma z^{\sigma k - n}(z-\tau)^{n-1} \exp(-z^\sigma) \mathbb{I}_{(\tau,+\infty)}(z)}{\sum_{i=0}^{n-1} \binom{n-1}{i}(-\tau)^i \Gamma(k-i\sigma; \tau^\sigma)}$$

where the incomplete Gamma function is given by

$$\Gamma(a;b) = \int_b^{+\infty} x^{a-1} \exp(-x) dx. \tag{5.12}$$

Then,

$$P_{GG}(A_0) \mid (X_1, \ldots, X_n) \stackrel{d}{=} W_{n-\sigma k, Z_{GG}}$$

$$\widehat{D}_n^{GG}(0) = \frac{1}{N_s} \sum_{j=1}^{N_s} W_{n-\sigma k, Z_{GG}}(j) \tag{5.13}$$

where $N_s$ is the number of Monte Carlo samples. Analogously, a Monte Carlo estimate of the $r$-th discovery probability is obtained by repeatedly sampling from the random

variable

$$P_{GG}(A_r) \mid (X_1, \ldots, X_n) \stackrel{d}{=} \beta_{(r-\sigma)\times m_r, n-\sigma k-(r-\sigma)\times m_r}$$
$$\times (1 - W_{n-\sigma k, Z_{GG}})$$
$$\widehat{D}_n(r) = \frac{1}{N_s} \sum_{j=1}^{N_s} \beta_{(r-\sigma)\times m_r, n-\sigma k-(r-\sigma)\times m_r}(j)$$
$$\times (1 - W_{n-\sigma k, Z_{GG}}(j)). \qquad (5.14)$$

If the change of variables is performed with the density of $\xi = Z_{GG}^{\sigma}$ given by Equation (5.12), with Jacobian given by $|\frac{dZ_{GG}}{d\xi}| = \frac{1}{\sigma}\xi^{1/\sigma-1}$, then, the corresponding density is

$$f_\xi(\xi) = \frac{\xi^{k(1-n)/\sigma-1}(\xi^{1/\sigma} - \tau)^{n-1} \exp(-\xi)\mathbb{I}_{(\tau^\sigma, +\infty)}(z)}{\sum_{i=0}^{n-1} \binom{n-1}{i}(-\tau)^i \Gamma(k - i\sigma; \tau^\sigma)}. \qquad (5.15)$$

The density from Equation (5.15) is log-concave so an adaptive rejection sampling scheme (Gilks and Wild, 1992) can be used to sample from it. In Equations (5.11) and (5.14) how to obtain the $r$-th discovery probabilities estimates is reviewed for completeness, but the 0-th case is the only case of interest, given in Equations (5.10) and (5.13). Furthermore, the complement can be obtained due to the $V_{n,k}$ parameter's backward recursion from Equation (5.2). In the next section, the second SMC scheme is presented as well as the corresponding algorithmic details.

## 5.3 An auxiliary SMC sampler for the $\mathcal{PG}(\sigma \leqslant 0.5)$ class

In this section, a particle filter based on an augmentation is proposed. Ideally, we should build exact algorithms so that we do not need to deal with the approximation error introduced by targeting an approximate posterior but a rigorous proof ought to be included to verify the exactness of the proposed scheme. This scheme is for the $\mathcal{PG}$ class with $\sigma \leqslant 0.5$, given in Example 9 of Chapter 3 (James, 2013), first used in Favaro et al. (2014b) for inference in mixture models with priors from this class. We again exploit the generative process for the $\sigma$-Stable Poisson–Kingman family of priors from Proposition 2 in Chapter 2, and the characterisations of Pitman (2003) used in the hybrid MCMC sampler of Chapter 4 (Lomeli et al., 2015) to build an auxiliary SMC algorithm, in the spirit of Griffin (2011) for NRMI mixture models. Indeed,

Griffin (2011) uses the auxiliary variable characterisation given in James et al. (2009) of NRMIs. Analogously, if we start with the generative process described in Chapter 2 for Poisson–Kingman priors, denoted by $\text{PK}(\rho, h_t, H_0)$, where $\rho$ is the Lévy measure, $h_t$ is the density for the total mass and $H_0$ is the base distribution, then the following joint distribution is obtained

$$\text{Pr}_{\rho,h}(\Pi_n = (c_\ell)_{\ell \in [k]}, \tilde{J}_\ell \in ds_\ell \text{ for } \ell \in [k], T - \sum_{\ell=1}^{k} \tilde{J}_\ell \in dv)$$
$$= (v + \sum_{\ell=1}^{k} s_\ell)^{-n} h\left(v + \sum_{\ell=1}^{k} s_\ell\right) f_\rho(v) \prod_{\ell=1}^{k} s_\ell^{|c_\ell|} \rho(ds_\ell).$$

Then,

$$\text{Pr}_{\rho,h}(\Pi_n = (c_\ell)_{\ell \in [k]}) \qquad (5.16)$$
$$= \mathbb{E}_{V, \tilde{J}_1, \ldots, \tilde{J}_k}\left[ (v + \sum_{\ell=1}^{k} s_\ell)^{-n} h\left(v + \sum_{\ell=1}^{k} s_\ell\right) f_\rho(v) \prod_{\ell=1}^{k} s_\ell^{|c_\ell|} \rho(ds_\ell) \right].$$

$V$ is the surplus mass and the corresponding represented weights are $\left(\tilde{J}_\ell\right)_{\ell \in [k]}$. Thus, the posterior distribution for the partition is expressed as an expectation with respect to the auxiliary variables. After substituting the $\sigma$-Stable Lévy measure, the surplus mass can be sampled and the represented weights are the auxiliary variables that can be introduced. See Algorithm 17 for details.

The conditional probabilities have the following convenient form

$$\text{Pr}\left(\text{i joins cluster c' } | \Pi_{i-1}^p, \mathbf{y}_{1:i-1}, \left\{\tilde{J}_\ell \in ds_\ell\right\}_{\ell=1}^{|\Pi_{i-1}^p|}, T - \sum_{\ell \leq |\Pi_{i-1}^p|} \tilde{J}_\ell \in dv\right) \qquad (5.17)$$
$$\propto \begin{cases} s_{c'} F\left(Y_i \in dy_i\right) | \{y_j\}_{j \in c'}\right) & \text{if } c' \in \Pi_{i-1}^p \\ v F\left(Y_i \in dy_i\right) & \text{o.w.} \end{cases}$$

where

$$F\left(Y_i \in dy_i\right) = \int F(Y_i \in dy_i | x) H_0(dx)$$
$$F\left(Y_i \in dy_i\right) | \{y_j\}_{j \in c}\right) = \int F\left(Y_i \in dy_i | x\right) f\left(x | \{y_j\}_{j \in c}\right) dx.$$

**Algorithm 17** AuxiliarySMC

$\Pi_1^p = \{\{1\}\}, \forall p \in \{1, \ldots, L\}$
Sample **T=GenerallyTiltedStable**$(h_t, \sigma, L)$
$\tilde{\mathbf{J}}^1$=**ExactSampleNewTableSize**$(T, \sigma, L)$
**for** $i = 2 : n$ **do**
    **for** $p = 1 : L$ **do**
        Set $c'$ according to
$\Pr\left(\text{i joins cluster c'} \mid \Pi_{i-1}^p, \mathbf{y}_{1:i-1}, \left\{\tilde{J}_\ell \in ds_\ell\right\}_{\ell=1}^{|\Pi_{i-1}^p|}, T - \sum_{\ell \leq |\Pi_{i-1}^p|} \tilde{J}_\ell \in dv\right)$ ▷ from Equation (5.17)
        **if** $|c'| = 1$ **then**
            $\Pi_i^p = \Pi_{i-1}^p \cup \{\{i\}\}$
            $\tilde{J}_{|\Pi_i^p|}$=**ExactSampleNewTableSize**$\left(V := T - \sum_{\ell \leq |\Pi_{i-1}^p|} \tilde{J}_\ell, \sigma\right)$
            $V = V - \tilde{J}_{|\Pi_i^p|}$
        **else**
            $c' = c' \cup \{i\}, \; c' \in \Pi_{i-1}^p$
            $\Pi_i^p = \Pi_{i-1}^p$
        **end if**
        Compute weight
$w_i^p \propto w_{i-1}^p \times \Pr\left(Y_i \in dy_i \mid \Pi_i^p, \mathbf{y}_{1:i-1}, \left\{\tilde{J}_\ell \in ds_\ell\right\}_{\ell=1}^{|\Pi_{i-1}^p|}, T - \sum_{\ell \leq |\Pi_{i-1}^p|} \tilde{J}_\ell \in dv\right)$ ▷ from Equation (5.18)
    **end for**
    **if** ESS$<$ thresh $\times L$ **then**
        Resample $p' \sim$ Multinomial$\left(\tilde{w}_i^1, \ldots, \tilde{w}_i^L\right), \forall \ell \in \{1, \ldots, L\}, \Pi_i^p = \Pi_i^{p'}$ ▷ The $p$-th particle inherits the entire path of the $p'$-th particle, see Section 5.1.
    **end if**
**end for**

---

**Algorithm 18** ExactSampleNewTableSize$(V, \sigma)$

**if** $\sigma = 0.5$ **then**
    $G \sim$ Gamma$\left(\frac{3}{4}, 1\right)$
    $IG \sim$ Inverse Gamma$\left(\frac{1}{4}, \frac{1}{4^3}V^{-2}\right)$
    $V_{stick} = \frac{\sqrt{G}}{\sqrt{G}+\sqrt{IG}}$
    $\tilde{J}_{new} = V_{stick}V$
**else**
    **if** $\sigma < 0.5$ && $\sigma = \frac{u_\sigma}{v_\sigma}, u_\sigma, v_\sigma \in \mathbb{N}$ **then**
        $\lambda = u_\sigma^2 / v_\sigma^{\frac{v_\sigma}{u_\sigma}}$
        $IG \sim$ Inverse Gamma$\left(1 - \frac{u_\sigma}{v_\sigma}, \lambda\right)$
        $\frac{1}{G} \sim \mathcal{E}_\mathcal{T}\left(\lambda, L_{\frac{u_\sigma}{v_\sigma}}^{-1/u}\right)$ ▷ Samples an exponentially tilted random variable. See Chapter 3, Section 3.6 for details about these distributional identities (Favaro et al., 2014a)
        $V_{stick} = \frac{G}{G+IG}$
        $\tilde{J}_{new} = V_{stick}V$
    **end if**
**end if**

and

$$\Pr\left(Y_i \in dy_i \mid \Pi_i^p, \mathbf{y}_{1:i-1}, \left\{\tilde{J}_\ell \in ds_\ell\right\}_{\ell=1}^{|\Pi_{i-1}^p|}, T - \sum_{\ell \leqslant |\Pi_{i-1}^p|} \tilde{J}_\ell \in dv\right) \quad (5.18)$$
$$= \frac{v}{t} F(Y_i \in dy_i) + \sum_{c \in \Pi_i^p} \frac{s_c}{t} F(Y_i \in dy_i \mid \{y_j\}_{j \in c}).$$

If the prior for the allocation variables is used as a proposal, then

$$\Pr\left(\text{i joins cluster c'} \mid \Pi_{i-1}^p, \left\{\tilde{J}_\ell \in ds_\ell\right\}_{\ell=1}^{|\Pi_{i-1}^p|}, T - \sum_{\ell \leqslant |\Pi_{i-1}^p|} \tilde{J}_\ell \in dv\right) \quad (5.19)$$
$$\propto \left\{\begin{array}{ll} s_{c'} & \text{if } c' \in \Pi_{i-1}^p \\ v & \text{o.w.} \end{array}\right\}$$

and

$$\Pr(Y_i \in dy_i \mid \Pi_i^p, \mathbf{y}_{1:i-1}) = F\left(Y_i \in dy_i \mid \{y_j\}_{j \in c}\right), \quad i \in c. \quad (5.20)$$

A remarkable fact is that the conditional distribution of the new size-biased weight given the previous ones, the current partition and the total mass, depends only on the surplus mass, since, the sequence of surplus masses forms a Markov chain. This fact was mentioned in Chapter 2, when the generative process for size-biased sampling from a general Poisson–Kingman random probability measure is explained. The main difference is that, in the generative process, the sequence of surplus masses forms a Markov chain *a priori*, this is also true *conditionally on everything else*. Hence, we can use the identity of distribution given in Favaro et al. (2014a), for $\sigma \leqslant 0.5$, to produce an exact sample of the size-biased weight when the $i$-th observation is assigned to a new block, as in Algorithm 18. However, Algorithm 17 relies on being able to sample exactly the total mass from a generally tilted distribution. This forbids the use of all priors in the $\sigma$-Stable Poisson–Kingman class since it is only possible to sample from a polynomially or exponentially tilted stable distributions with the rejection samplers of Devroye (2009) and Hofert (2011) which correspond to the Pitman–Yor and Generalised Gamma processes, respectively. However, due to the relationship between the two processes introduced in Perman et al. (1992) and reviewed in Appendix A.4, we can effectively construct a sampler for all members of the $\mathcal{PG}(\sigma \leqslant 0.5)$ class (James, 2013).

Indeed, the generative process for sampling from an exponentially tilted Stable random variable with randomised tilting parameter is

$$G \sim f_G$$
$$X \mid G \sim \text{ExpTiltedStable}(G^{\frac{1}{\sigma}}). \tag{5.21}$$

For different choices of $f_G$ different members of the $\mathcal{PG}$ class can be recovered. If we choose $f_G := \text{Gamma}(1, k)$ then the Pitman-Yor process is obtained. In the next section, another SMC scheme is proposed for the entire class of Gibbs type priors with $\sigma \in (-\infty, 1)$, but it is no longer an exact inference scheme.

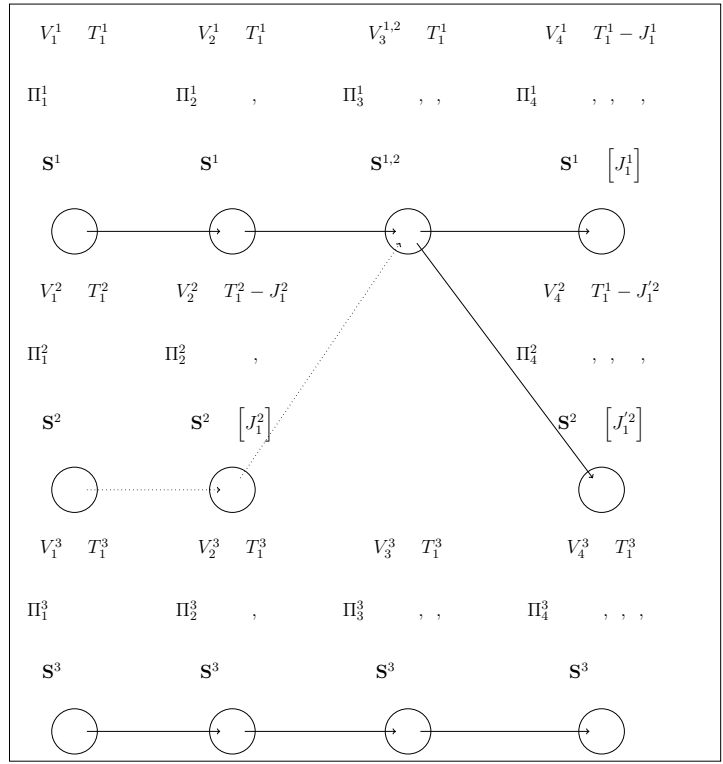

Figure 5.1: Ilustration of the auxiliary variable SMC for $L = 3$ particles. From left to right, top to bottom: each node represents adding observation $i$-th to the $p$-th particle path, with corresponding surplus mass $V_i^p$, partition $\Pi_i^p$ and vector of $k^p = |\Pi_i^p|$ size-biased weights $S^p = \left[\tilde{J}_1^p, \ldots, \tilde{J}_{k^p}\right]$. The dotted line represents the resampling step when the $p$-th particle inherits the $p'$-th particle's path.

## 5.4 An approximate SMC sampler for Gibbs type priors

In this section, an approximate SMC algorithm is presented for all Gibbs type priors with $\sigma \in (-\infty, 1)$. The sampler is no longer exact because, even though an unbiased and positive estimator for the $V_{n,k}$ can be obtained with the distributional identities of Equations (5.26) and (5.30), the ratio of unbiased estimators is no longer unbiased. See Jacob and Thiery (2015) for an interesting result about the difficulty in obtaining simultaneously non-negative and unbiased estimators.

---

**Algorithm 19** ApproxSMC

$\Pi_1^p = \{\{1\}\}, \forall p \in \{1, \ldots, L\}$
**for** $i = 2 : n$ **do**
    **for** $p = 1 : L$ **do**
        Let $\widehat{\mathbf{V}}_{(i-1,i,|\Pi_{i-1}^p|)} := \left[\hat{V}_{i-1,|\Pi_{i-1}^p|}, \hat{V}_{i,|\Pi_{i-1}^p|}, \hat{V}_{i,|\Pi_{i-1}^p|+1}\right]$

        $\widehat{\mathbf{V}}_{(i-1,i,|\Pi_{i-1}^p|)}=$MonteCarloEstimate$(i, |\Pi_{i-1}^p|, \sigma)$
            ▷ from Equation (5.30), if $\sigma < 0$ or from Equation (5.26), if $\sigma \in (0, 1)$
        Set $c'$ according to
$\Pr\left(\text{i joins cluster } c' \mid \Pi_{i-1}^p, \mathbf{y}_{1:i-1}, \widehat{\mathbf{V}}_{(i-1,i,|\Pi_{i-1}^p|)}\right)$   ▷ from Equation (5.22)
        **if** $|c'| = 1$ **then**
            $\Pi_i^p = \Pi_{i-1}^p \cup \{\{i\}\}$
        **else**
            $c' = c' \cup \{i\}, c' \in \Pi_{i-1}^p$
            $\Pi_i^p = \Pi_{i-1}^p$
        **end if**
        Compute weight
$w_i^p \propto \Pr\left(Y_i \in dy_i \mid \Pi_i^p, \mathbf{y}_{1:i-1}, \mathbf{V}_{(i-1,i,|\Pi_{i-1}^p|)}\right)$   ▷ from Equation (5.23)
    **end for**
    Normalise the weights $\tilde{w}_i^p = \frac{w_i^p}{\sum_{j=1}^L w_i^j}$
    ESS$= \frac{1}{\sum_{\ell=1}^L (\tilde{w}_i^p)^2}$
    **if** ESS$<$ thresh $\times L$ **then**
        Resample $p' \sim$ Multinomial $\left(\tilde{w}_i^1, \ldots, \tilde{w}_i^L\right), \forall p \in \{1, \ldots, L\}$
        $\Pi_i^p = \Pi_i^{p'}$ ▷ The $p$-th particle inherits the entire path of the $p'$-th particle, we are resampling at every step.
    **end if**
**end for**

---

The conditional probability for the allocation variables, for any member of the Gibbs type priors, if estimators of the $V_{n,k}$ are used, denoted by $\widehat{V}_{n,k}$, is

$$\Pr\left(i \text{ joins cluster } c' \mid \Pi_{i-1}^p, \mathbf{y}_{1:i-1}, \widehat{\mathbf{V}}_{(i-1,i,|\Pi_{i-1}^p|)}\right) \propto$$

$$\left\{ \begin{array}{ll} \dfrac{\widehat{V}_{i,|\Pi_{i-1}^p|}}{\widehat{V}_{i-1,|\Pi_{i-1}^p|}}(|c'|-\sigma)F(Y_i \in dy_i \mid \{y_j\}_{j \in c'}) & \text{if } c' \in \Pi_{i-1}^p \\ \dfrac{\widehat{V}_{i,|\Pi_{i-1}^p|+1}}{\widehat{V}_{i-1,|\Pi_{i-1}^p|}}F(Y_i \in dy_i) & \text{o.w.} \end{array} \right\} \quad (5.22)$$

$$p\left(y_i \mid \Pi_i^p, \mathbf{y}_{1:i-1}, \mathbf{V}_{(i-1,i,|\Pi_{i-1}^p|)}\right) = \dfrac{\widehat{V}_{i,|\Pi_{i-1}^p|+1}}{\widehat{V}_{i-1,|\Pi_{i-1}^p|}}F(Y_i \in dy_i)$$

$$+ \dfrac{\widehat{V}_{i,|\Pi_{i-1}^p|}}{\widehat{V}_{i-1,|\Pi_{i-1}^p|}} \sum_{c' \in \Pi_{i-1}^p} (|c'|-\sigma)F(Y_i \in dy_i \mid \{y_j\}_{j \in c'})$$

$$(5.23)$$

where

$$F(Y_i \in dy_i) = \int F(Y_i \in dy_i \mid x)H_0(dx)$$

$$F(Y_i \in dy_i) \mid \{y_j\}_{j \in c} = \int F(Y_i \in dy_i \mid x)f(x \mid \{y_j\}_{j \in c})dx.$$

If the prior for the allocation variables is used as a proposal

$$\Pr\left(i \text{ joins cluster } c' \mid \Pi_{i-1}^p, \mathbf{y}_{1:i-1}, \mathbf{V}_{(i-1,i,|\Pi_{i-1}^p|)}\right) \propto \left\{ \begin{array}{ll} \dfrac{\widehat{V}_{i,|\Pi_{i-1}^p|}}{\widehat{V}_{i-1,|\Pi_{i-1}^p|}}(|c'|-\sigma) & \text{if } c' \in \Pi_{i-1}^p \\ \dfrac{\widehat{V}_{i,|\Pi_{i-1}^p|+1}}{\widehat{V}_{i-1,|\Pi_{i-1}^p|}} & \text{o.w.} \end{array} \right\}$$

$$(5.24)$$

and

$$\Pr\left(Y_i \in dy_i \mid \Pi_i^p, \mathbf{y}_{1:i-1}\right) = F\left(Y_i \in dy_i \mid \{y_j\}_{j \in c}\right), \quad i \in c \quad (5.25)$$

where

$$F\left(Y_i \in dy_i \mid \{y_j\}_{j \in c}\right) = \int F\left(Y_i \in dy_i \mid x\right)f\left(x \mid \{y_j\}_{j \in c}\right)dx.$$

In the case $\sigma \in (0,1)$, which corresponds to the $\sigma$-Stable Poisson–Kingman priors,

the conditional distribution of an exchangeable random partition of $[n]$ is

$$\Pr{}_\sigma (\Pi_n = \pi \mid t) = \frac{\sigma^k}{\Gamma(n - \sigma k)} \left(t^{-\sigma}\right)^k [f_\sigma(t)]^{-1} \int_0^1 p^{n-\sigma k-1} f_\sigma ((1-p)t) \, dp \prod_{i=1}^k [1-\sigma]_{n_i-1}$$

where the first term can be expressed as

$$
\begin{aligned}
V_{n,k} &= \int_{\mathbb{R}^+} \frac{\sigma^k}{\Gamma(n-\sigma k)} \left(t^{-\sigma}\right)^k \int_0^1 p^{n-\sigma k-1} f_\sigma((1-p)t) \, dp \, h(t) dt \quad (5.26) \\
&= \frac{\sigma^{k-1} \Gamma(k)}{\Gamma(n)} \mathbb{E}\left(h(Z)\right).
\end{aligned}
$$

This double expectation can be computed as an expected value of a ratio of 2 random variables due to Theorem 2.1 of Ho et al. (2008). Indeed,

$$V_{n,k,t} = \frac{\Gamma(1-\sigma)}{\Gamma(n-k\sigma)} \left(\frac{\sigma}{t^\sigma}\right)^{k-1} \int_0^1 p^{n-1-k\sigma+\sigma} \tilde{f}_\sigma(p \mid t) \, dp$$

where

$$\tilde{f}_\sigma(p \mid t) = \frac{\sigma (pt)^{-\sigma}}{\Gamma(1-\sigma)} \frac{f_\sigma((1-p)t)}{f_\sigma(t)} I_{(0,1)}(p).$$

$\tilde{f}_\sigma$ is the density of a random variable $Z$, which is a ratio of independent random variables from which we can simulate

$$Z = \frac{S_{\sigma,k\sigma}}{\beta_{k\sigma,n-k\sigma}} \stackrel{d}{=} \frac{S_{\sigma,(1-k)\sigma}}{\beta_{(k-1)\sigma+1,n-1-(k-1)\sigma}}, \quad (5.27)$$

$\beta_{k\sigma,n-k\sigma}$ is a Beta random variable and $S_{\sigma,k\sigma}$ is a polynomially tilted Stable random variable with density

$$f_{S_{\sigma,k\sigma}}(t) = \frac{\Gamma(\sigma k+1)}{k+1} t^{-\sigma k} f_\sigma(t).$$

Hence, Equation (5.26) is the expectation of a function of the random variable from Equation (5.27). We could obtain an unbiased and positive estimator for the $\hat{V}_{n,k}$ by repeatedly sampling the random variables of the quotient in Equation (5.27) for all $V_{n,k}$'s that belong to the $\sigma$-Stable Poisson–Kingman family. We can use the samplers from Devroye (2009) and Hofert (2011) and the generative process for sampling from a polynomially tilted Stable random variable given in Equation (5.21). Furthermore, we could precompute and use the recursion from Equation (5.1) to diminish the computational cost to obtain the estimates $\hat{V}_{n,k}$. If we estimate them along the SMC sampler, without taking into account the resampling step, the computational cost is $\mathcal{O}(N_s \times n \times L)$ where $n$ is the sample size and the number of estimators in the last level of the triangular array, $N_s$ is the number of Monte Carlo samples per estimate and $L$ is the number of particles. If we instead precompute them, and exploit the recursion, the computational cost is $\mathcal{O}(N_s \times n + \frac{n(n-1)}{2})$. See Figure 5.2 for the graphical description of the triangular array.

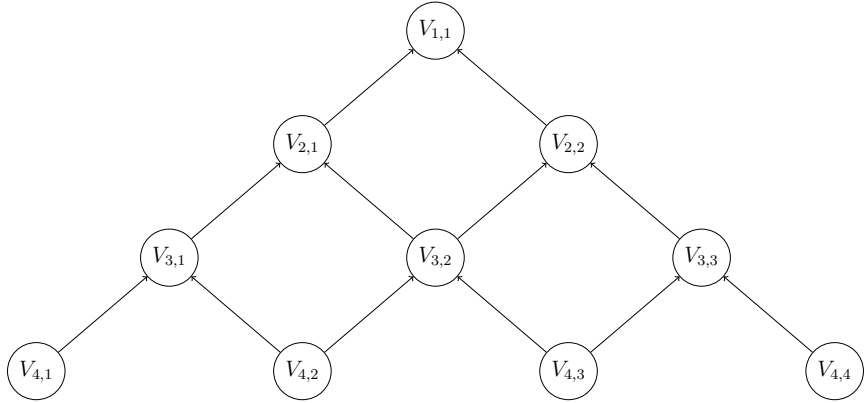

Figure 5.2: Infinite triangular array for the $V_{n,k}$ for $[n] = \{1, \ldots, 4\}$. The number of elements of the array is given by the triangular number, which is equal to $\frac{n(n+1)}{2}$.

In the case where $\sigma < 0$, different mixing distributions for the total number of components $m$ can be chosen, as reviewed in Chapter 2. The generative model for a mixture of finite mixtures (MFM) model is given by

$$M \sim \mathcal{Q} \tag{5.28}$$
$$\underline{\pi} \mid M = m \sim \text{Symmetric Dirichlet}(\gamma)$$
$$X_1, \ldots, X_n \mid \underline{\pi} \overset{\text{i.i.d.}}{\sim} \text{Categorical}(\underline{\pi}).$$

Integrating out $\underline{\pi}$ and separating the occupied and empty components, we obtain

$$p(X_1, \ldots, X_n \mid M = m) = \frac{\gamma^k \Gamma(m\gamma)}{\Gamma(\gamma m + n)} \prod_{\{\ell : n_\ell > 0\}} \frac{\Gamma(n_\ell + \gamma)}{\Gamma(\gamma + 1)}$$

$$= V_{n,k,m} \times \prod_{\{\ell : n_\ell > 0\}} W_{n_\ell}.$$

If the total total number of components is randomised, the combinatorial factor is added and let $\gamma = |\sigma|$

$$V_{n,k} = \sum_{m=1}^{\infty} |\sigma|^k \frac{m \Gamma(m)}{\Gamma(m - k + 1)} \frac{\Gamma(m|\sigma|)}{\Gamma(m|\sigma| + n)} \pi(m) \qquad (5.29)$$

where

$$|\sigma| \Gamma(|\sigma|) = \Gamma(|\sigma| + 1) = \Gamma(1 - \sigma).$$

In general, there is a simple way to obtain Monte Carlo estimators for the $V_{n,k}$: sample a total number of components from the distribution of Equation (5.28), then sample the rest of the partition and take the average. For this reason, Equation (5.29) can be expressed as an expectation with respect to the prior distribution over the total number of components, the corresponding estimator is

$$V_{n,k} = \sum_m V_{n,k,m} \pi(m)$$
$$= \mathbb{E}_M (V_{n,k,m})$$
$$\approx \frac{1}{N_m} \sum_{j=1}^{N_m} V_{n,k,m_j}. \qquad (5.30)$$

Hence, Equation (5.30) could be precomputed to obtain a triangular array, just as in the $\sigma \in (0, 1)$ case.

## 5.5 Some examples of distributions over the total number of components

In Chapter 2, a review about the set of extreme points for Gibbs type priors is provided. This result means that any distribution over the total number of components can be written as a mixture over these extreme elements (Gnedin and Pitman, 2006). In particular, when $\sigma < 0$, it corresponds to the two parameter Chinese restaurant family, with parameters $(\sigma, |\sigma|m)$, $m \in \mathbb{N}$. This result means that we can assign a different probability mass function for the total number of components $m$ and write its corresponding $V_{n,k}$ parameter as an expectation of the $V_{n,k}$ parameter for the two parameter Chinese with a DeFinetti mixing measure for the total number of components of our choice. A particular mixing distribution for $m$ is presented in Gnedin (2010) and reviewed in the first example herein.

i) **(Gnedin, 2010).** The author derives an EPPF for a model with a finite but random number of types. The probability mass function for the total number of components with $\sigma = -1$, and $\theta = m$ is

$$\mathcal{Q}(M = m) = \frac{\gamma(1-\gamma)_{m-1\uparrow}}{m!}, \quad m \in \mathbb{N}, \quad \gamma \in (0,1). \tag{5.31}$$

Substituting Equation (5.31) in Equation (5.28) and integrating out the total number of components $m$ we should obtain

$$\begin{aligned} V_{n,k} &= \frac{(k-1)!(1-\gamma)_{k-1}(\gamma)_{n-k}}{(n-1)!(1+\gamma)_{n-1}} \\ &= \frac{\gamma(k-1)!\Gamma(k-\gamma)\Gamma(\gamma+n-k)}{(n-1)!\Gamma(n+\gamma)\Gamma(1-\gamma)}. \end{aligned} \tag{5.32}$$

This is an interesting example, the parameter $\gamma$ is between $\gamma \in (0,1)$; as a consequence, there is an infinite expected number of occupied components. The main difference with a Gibbs type prior with $\sigma \in [0,1)$ is that the number of occupied components does not grow as a function of the number of observations. Furthermore, if we use this as the mixing distribution for the total number of components, the $V_{n,k}$ ratio can be computed analytically so an exact SMC scheme is obtained.

ii) **(Miller and Harrison, 2015).** $\mathbf{M - 1} \sim \mathbf{Poisson}(\lambda)$ and $|\sigma| = 1$ with corresponding probability mass function

$$\mathcal{Q}(M = m) = \exp(-\lambda)\frac{\lambda^{m-1}}{(m-1)!}, \quad m \in \mathbb{N}, \quad \lambda > 0.$$

iii) $\mathbf{M} \sim \mathbf{Poisson}(\gamma)$, restricted to the positive integers, with corresponding probability mass function

$$\mathcal{Q}(M = m) = \exp(-\gamma)\frac{\gamma^m}{(1-\exp-\gamma)m!}, \quad m \in \mathbb{N}, \quad \gamma > 0.$$

iv) $\mathbf{M} \sim \mathbf{Geometric}(\eta)$, with corresponding probability mass function

$$\mathcal{Q}(M = m) = (1-\eta)\eta^{m-1}, \quad m \in \mathbb{N}, \quad \eta \in (0,1).$$

Both algorithms presented in this section could include an MCMC step in order to rejuvenate the particles. For instance, Miller and Harrison (2015) incremental Gibbs sampler can be used for this rejuvenation step.

## 5.6 SMC and the Log-marginal likelihood

An advantage about using an SMC scheme is that the marginal likelihood can be directly estimated from the output. Indeed, the SMC marginal likelihood estimator is given by

$$\prod_{i=1}^{n} \frac{1}{L} \sum_{p=1}^{L} w_i^p. \tag{5.33}$$

This quantity is useful to construct the Bayes factor test to answer the statistical question of interest in the remaining sections. When the allocation variable is sampled from the prior, it is called the Bootstrap filter by Gordon et al. (1993). We could also sample it from the posterior given by Equation (5.22). Furthermore, we could choose only resample when the effective sample size (ESS) falls below some threshold, denoted by *thresh*, to lessen the path degeneracy problem. In all of our SMC samplers, each observation is "locked in" to the table they first sit in so we could do a Gibbs step to reassign tables and rejuvenate the particles. However, as the size of the dataset increases, the component-wise Gibbs move gets more expensive. Other clever strategies to diminish the path degeneracy are required, for instance, Bouchard-Côté et al. (2015) use particle Gibbs (Andrieu et al., 2010) with a split and merge proposal.

## 5.7 Bayes factors

The Bayes factor, first introduced in Jeffreys (1935) and successively reviewed in Kass and Raftery (1995), Robert (2001) allows us to compare the predictions made by two competing scientific theories represented by two statistical models. In general, the Bayes factor requires priors for the parameters representing each hypothesis or competing model. It is defined as the following ratio

$$\text{BF} = \frac{p(\mathbf{D} \mid \mathcal{M}_1)}{p(\mathbf{D} \mid \mathcal{M}_2)}$$

where

$$p(\mathbf{D} \mid \mathcal{M}_k) = \int p(\mathbf{D} \mid \mathcal{M}_k, \phi_k) f(\phi_k \mid \mathcal{M}_k) d\phi_k, \quad k = 1, 2. \tag{5.34}$$

where $\mathbf{D} = (y_1, \ldots, y_n)$ is our data, $\mathcal{M}_1$ is model one, $\mathcal{M}_2$, model two; $\phi_k$ is the parameter under the hypothesis or competing model $\mathcal{M}_k$, $k = 1, 2$ and $f(\phi_k \mid \mathcal{M}_k)$ is its corresponding prior density. The numerator and denominator are the marginal probabilities of the data also called predictive probabilities, marginal likelihood, integrated likelihood or model evidence. Geweke and Amisano (2010) comment how the predictive likelihood function is inherently Bayesian since it is a component of the likelihood function, integrated over the posterior distribution of the unobservables, latent variables and parameters, at the time the prediction is made. The logarithm of the Bayes factor can be written in the following way, with respect to the one step-ahead predictive probabilities

$$\begin{aligned}
\log \left[ \frac{p(y_1, \ldots, y_n \mid \mathcal{M}_1)}{p(y_1, \ldots, y_n \mid \mathcal{M}_2)} \right] &= \log p(y_1, \ldots, y_n \mid \mathcal{M}_1) - \log p(y_1, \ldots, y_n \mid \mathcal{M}_2) \\
&= \log \left[ p(y_1 \mid \mathcal{M}_1) p(y_2 \mid y_1, \mathcal{M}_1) \ldots p(y_n \mid y_{n-1}, \ldots, y_1, \mathcal{M}_1) \right] \\
&\quad - \log \left[ p(y_1 \mid \mathcal{M}_2) p(y_2 \mid y_1, \mathcal{M}_2) \ldots p(y_n \mid y_{n-1}, \ldots, y_1, \mathcal{M}_2) \right] \\
&= \log \left[ \frac{p(y_1 \mid \mathcal{M}_1)}{p(y_1 \mid \mathcal{M}_2)} \right] + \sum_{j=2}^{n} \log \left[ \frac{p(y_j \mid y_{j-1} \ldots, y_1 \mathcal{M}_2)}{p(y_j \mid y_{j-1} \ldots, y_1, \mathcal{M}_2)} \right].
\end{aligned}$$

Computing the marginal likelihood from Equation (5.34) requires integrating out the model's parameters. Numerical integration methods can be used for this purpose, but as the dimension of the parameter increases the numerical approximation error grows. Another possibility is to use the output from some posterior simulator, for instance, Geweke and Amisano (2010) suggest to use MCMC and compute the harmonic mean estimator (Newton and Raftery, 1994). This estimator has infinite variance and has been shown to perform badly in certain situations. For instance, when the underlying space of the random variable is infinite and unbounded. More precisely, the inverse of the likelihood function is not square integrable with respect to the posterior (Newton and Raftery, 1994). SMC allows to have an estimator of this quantity which requires neither of these approaches. Furthermore, the Bayes factor can be estimated sequentially to see the effect of the growing data size on the competing hypotheses.

## 5.8 Experiments

In this section, a Bayesian hypothesis test is constructed to show the empirical performance in terms of the marginal likelihood of an infinite mixture model versus a mixture of finite mixtures model. In the latter case, the distribution for the total number of components given by Example 1 (Gnedin, 2010) is chosen (vanillaSMCIII and SMCIII). This leads to a model that has an expected number of occupied components which is infinite, as all members of the $\sigma$-Stable Poisson–Kingman priors have when the sample size goes to infinity.

The MFM model will "saturate" if the corresponding predictive loglikelihood decreases to/near zero while the infinite mixture model will keep adapting. However, this is currently a claim since it is not clear if one can effectively witnesss the asymptotic properties of the occupied number of components random variable in terms of the proposed model criticism measure. Furthermore, it also depends on the dataset.

The galaxy dataset of Roeder (1990) is used for the experiments with a conjugate prior for the mean and known variance. This allows the model parameters to be integrated out, so the cluster assignment variables are the only ones that need to be sampled. See the Appendix A.6 for the corresponding expressions for the conditional distribution of the $i$-th observation given the first $i - 1$.

The implementation of the standard SMC (stdSMC) and standard vanilla SMC (vanillastdSMC) versions for the PY are based on Wood and Black (2008) which relies on the generalised Polýa urn representation. There is no Polýa urn analogue for the NGG process. The scheme I (SMCI and vanillaSMCI), is based on the auxiliary SMC scheme for the $\mathcal{PG}$ class presented here; scheme II, on the first SMC presented here for the PY and the NGG. The approximate SMC (approxSMC, vanillaapproxSMC) schemes use the last scheme presented in this chapter. All algorithms are implemented in Matlab.

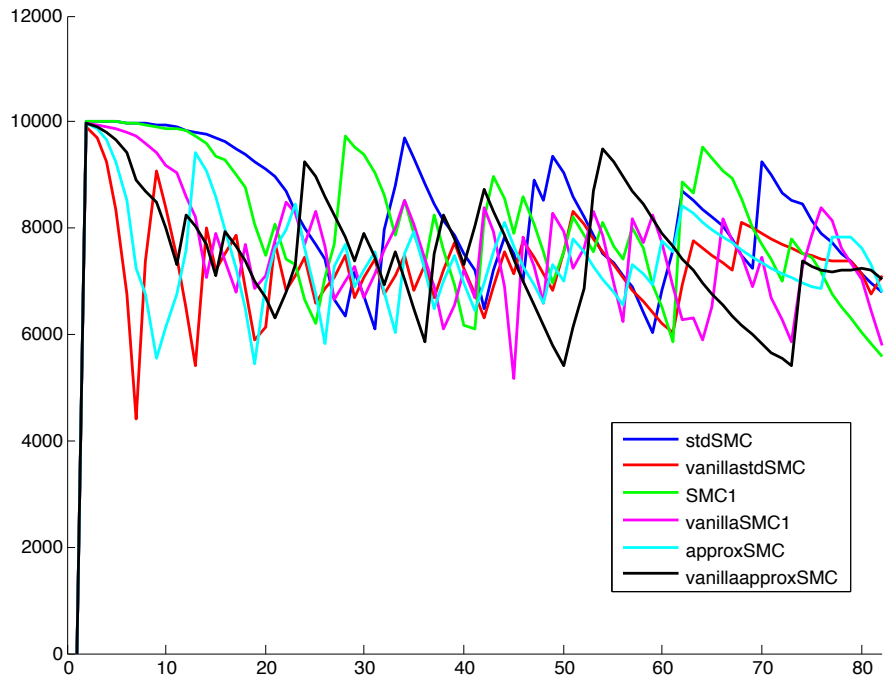

Figure 5.3: ESS for PY with different SMC algorithms. This plot shows how all algorithms have similar ESS, on average, some produce noisier quantities due to Monte Carlo approximations.

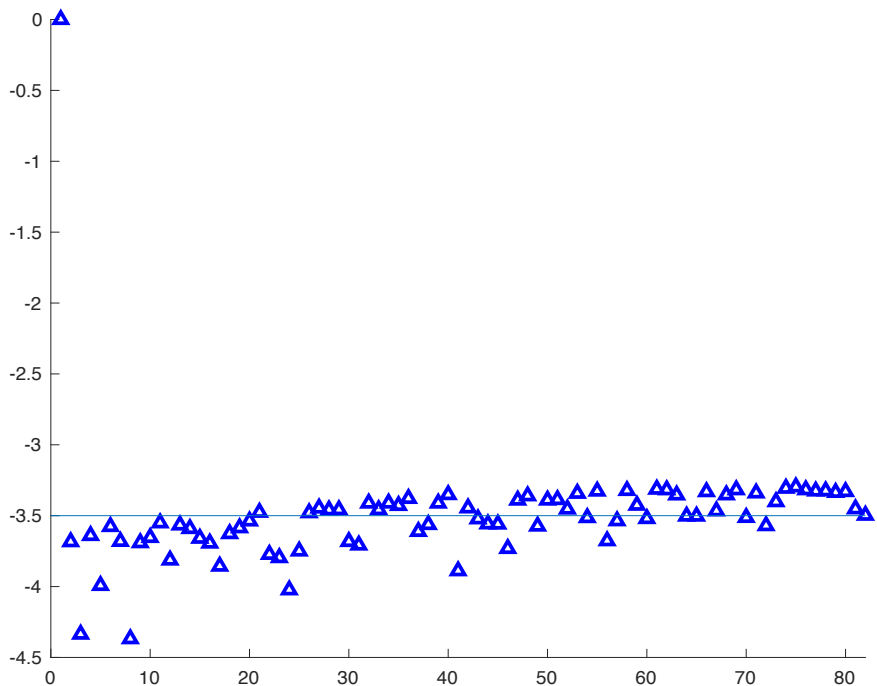

Figure 5.4: Log-Bayes factor $\log B_{1,2}$, model $\mathcal{M}_1$ is a PY infinite mixture and model $\mathcal{M}_2$ MFM with Gnedin's prior.

| Algorithm | Running time(±std) | log-Marginal likelihood(±std) |
|---|---|---|
| PY($\theta = 10, \sigma = 0.5$) | | |
| StandardVanillaSMC | 377.927 (35.29) | -294.622 (0.76) |
| StandardSMC | 445.839 (15.65) | -292.704 (0.65) |
| VanillaSMC I | 663.909 (39.36) | -297.865 (1.45) |
| SMC I | 649.042 (32.03) | -298.129 (0.86) |
| ApproxVanillaSMC | 543.429 (40.53) | -299.966 (0.50) |
| AproxSMC | 420.818 (23.38) | -295.093 (0.47) |
| NGG($\tau = 20, \sigma = 0.5$) | | |
| VanillaSMC I | 417.735 (13.60) | -286.591 (0.14) |
| SMC I | 429.590 (32.93) | -286.577 (0.35) |
| ApproxVanillaSMC | 568.531 (29.29) | -299.149 (0.02) |
| ApproxSMC | 511.341 (21.18) | -297.107 (0.13) |
| MFM($M \sim$ Gnedin($\gamma = 0.5$)) | | |
| VanillaSMC III | 433.536 (135.82) | -276.129 (0.77) |
| SMC III | 412.625 (116.99) | -276.427 (0.32) |

Table 5.1: Running times in seconds and log-marginal likelihood averaged over 5 runs, 10000 particles.

In Table 5.1, the algorithmic performance is reported in terms of running times per second and the marginal loglikelihood averaged over five runs with corresponding standard deviations. The number of particles, as well as the resampling threshold are $L = 10,000$ and $0.5L$, respectively. All vanilla versions of the algorithm choose the prior as a proposal distribution. The SMC I scheme for the Pitman–Yor mixture model takes longer than using an NGG due to the additional level of randomness introduced by randomising the exponentially tilted parameter for the initial values of the total mass. The SMC of Algorithm 16 is not reported because its running time is too slow compared to the other schemes. Further strategies are needed to reduce this running time.

Figure 5.3 shows the effective sample size for each SMC scheme. In general, it confirms that if an importance distribution that gets some information from the previous data points is used, then it leads to a higher ESS. The only exception is the approximate vanilla SMC scheme but this could be due to the noise introduced from the Monte carlo approximation.

Figure 5.4 shows the log-Bayes factor sequentially, as more data is added. A MFM model corresponds to the hypothesis or competing model $\mathcal{M}_1$ where the distribution for the total number of components is given by Gnedin (2010) versus a Pitman–Yor mixture model. In the beginning, the Log-Bayes factor estimate is noisy but, as the number of observations increases, its variance is reduced and stabilises around -3.5. There is positive evidence against the Pitman-Yor mixture model. However, since the log-Bayes factor estimates obtained from this procedure are biased further investigations need to be carried out to improve both marginal likelihood estimates and produce unbiased and possibly lower variance log-Bayes factors.

## 5.9  Summary of this Chapter

In this Chapter, we propose various SMC schemes for two model classes: an infinite mixture model and a finite mixture model where the total number of components is finite but unknown. We propose a sequential Bayesian hypothesis test based on the Bayes factor for each model class to formally test our assumptions about the data generating mechanism. This is possible because of the availability of a model evidence estimator from the SMC output but producing good marginal likelihood estimators is a difficult task on its own. Furthermore, if consistency with respect to the marginal likelihood holds for the BNP model, its corresponding model evidence will eventually be larger than the one obtained from a mixture of finite mixtures model. This scenario can only be observed if sequential inference is carried out and we provide the tools for making this task possible.

# Chapter 6

# Discussion and Conclusions

The main aim of this thesis was to introduce novel, general purpose, exact inference schemes using the Poisson-Kingman class of Bayesian nonparametric priors. Despite the infinite dimensional nature of Bayesian nonparametric models, we presented finite dimensional representations which allowed us to derive a variety of MCMC and SMC schemes. These inference schemes were then used for inference in an infinite mixture model with different Poisson-Kingman priors. The infinite mixture model is an example of a relatively simple compositional model but the inference schemes presented herein could be extended for more complicated compositional models, which could consist of more than one Bayesian nonparametric block. Hence, from a computational perspective, the contributions are mostly methodological. Nevertheless, we demonstrate that these novel inference schemes are more efficient and generic than existing MCMC and SMC methods for Bayesian nonparametric mixture models. Our methods, thus, allow the Poisson-Kingman priors to be used in practice to model real word data, considerably increasing the flexibility with which practitioners can model the world's inherent complexity.

In Chapter 3, a marginalised MCMC sampler is introduced for inference in $\sigma$-Stable Poisson–Kingman mixture models. This class of random probability measures is natural, because it is more general than the Dirichlet process but it is still mathematically tractable (De Blasi et al., 2015). The factorised Gibbs form makes it simple to deal with when constructing an inference scheme. However, certain intractabilities have hindered its use, as noted by Lijoi et al. (2008). For instance, the intractability associated to the $\sigma$-Stable density. We introduced an augmentation, based on an integral representation of the $\sigma$-Stable density, which makes this density no longer intractable. Furthermore, we showed that the marginal sampler is a competitive alternative to the conditional MCMC sampler (Favaro and Walker, 2012). The number of auxiliary variables that needs to be instantiated per iteration is smaller than for the conditional sampler, thus, it has smaller memory and storage requirements. Also, in our experiments, the marginal sampler has better ESS and running times. A possible future direction, is to incorporate split-merge proposals, for the cluster reassignment step, rather than using a component-wise Gibbs move to improve the mixing. Sometimes, Gibbs sampling might not explore all possible modes effectively. Another issue is that all the MCMC BNP methods give a flat clustering per iteration and it is not clear how to best summarise it. We used an adhoc agglomerative algorithm to post process the output and obtain a summarisation for our partitions. However, how to better summarise partitions is still an open question, an interesting possibility is given by Wade and Ghahramani (2015).

In Chapter 4, a novel hybrid MCMC sampler is introduced for inference in a Poisson–Kingman mixture model. A new compact but complete way of representing the infinite dimensional component was derived such that it is feasible to perform inference. A distinctive property of the hybrid MCMC scheme is that the auxiliary sequence of size-biased weights can be sampled retrospectively since a new size-biased weight needs to be sampled only when we observe a new cluster. In our experiments, this scheme outperforms both marginal and conditional samplers in running times in most cases and in ESS in all cases. However, there are still various challenges that remain. For instance, there are values of the parameter $\sigma > 0.5$ for which we are unable to perform inference with our hybrid MCMC sampler. One future direction, is to extend the distributional identities from Favaro et al. (2014a) to include all members of the $\mathcal{PG}$ class. Furthermore, it may be possible to improve the algorithmic efficiency by exploring alternative proposal distributions whenever a Metropolis–Hastings step is used.

In Chapter 5, various SMC samplers are presented for inference in infinite mixture models and for a mixture models where the total number of components is finite but unknown. The SMC samplers are shown to perform well in terms of running times and produce reasonable estimates of the marginal likelihood. Indeed, an immediate product of any SMC scheme is an estimate of the model evidence which we obtained for both model classes. This procedure enabled us to address an important question in Bayesian nonparametrics. Typically, the key insight behind Bayesian nonparametric models is that it is easier to perform inference in a mixture model with an infinite total number of components than with a finite but unknown total number of components. However, it is unclear whether these models are actually interchangeable: there might be datasets that are considerably better modelled by a mixture model with a finite number of components. Our experiments showed that there is positive evidence for preferring one model over the other. However, our conclusion is based on a biased point estimate of the log-Bayes factor and we are unable to quantify the amount of bias. Furthermore, in a severely misspecified scenario, one of the models could better encode the mechanisms by which a dataset is generated. For instance, the nonparametric model could better encode the case where the number of clusters is not stationary. A computational consideration is that the component-wise Gibbs step used as a rejuvenation step is inefficient when the number of data points is large. Bouchard-Côté et al. (2015) propose an efficient algorithm based on particle Gibbs (Andrieu et al., 2010) with a particular split merge proposal. This is a possible avenue of future research that could lead to improvements in the marginal likelihood estimates. Furthermore, improvements in the running time are possible if we modify the code to recursively update the sufficient statistics of each cluster component. We could also incorporate parameters per cluster, rather than integrating them out, in all SMC schemes for the use of non-conjugate base distributions. Finally, other model criticism tools could be proposed that can asses directly how the posterior over the number of components adapts, rather than using one based on the marginal likelihood such as the Bayes factor. A future methodological contribution is to formally show that the first two proposed schemes target the correct posterior distribution, called an exact approximation by Andrieu and Roberts (2009). The first scheme uses an unbiased and positive estimator for the proposal and the weights; the second scheme, an augmentation for both the proposal and the weights. It is not immediately clear whether these ideas render an exact approximation. The nested SMC (Naesseth et al., 2015) and the random weights SMC (Fearnhead et al., 2010) are examples of using unbiased and positive estimators for either the proposal or the weights. Griffin (2011) proposes an analogous auxiliary SMC sampler for the NRMI class of mixture models. These references could help set the path for how to proceed when proving our claims.

We have discussed how the problem of model selection with respect to the number of components in a mixture model is no longer an issue due to the Bayesian nonparametric framework. These ideas allowed us to bypass the step of selecting the number of components without overfitting. However, some assumptions are still made about the data generating mechanism when choosing a nonparametric building block for our models. In Chapter 5, we investigated whether a Bayesian nonparametric model is always the best alternative using some model selection tools in a sequential scenario. This research question falls under the umbrella of model criticism, an area which consists of assessing whether and in what ways should we should change our models. Even from a Bayesian perspective there are some assumptions that should be put under scrutiny: it is crucial to think about the model choices made, to be a critical Bayesian modeller. Furthermore, we should not only asses our models in terms of predictive performance but determine other relevant aspects of the model in question and some model criticism tools can be developed for this purpose.

Scalable Bayesian inference for BNP models is another interesting avenue for future research. Usually, Markov chain Monte Carlo and sequential Monte Carlo schemes scale poorly with respect to the dimensionality of the data and with respect to the number of data points. This problem is shared among all BNP MCMC and SMC methods because they deal with the likelihood term in the same way. For instance, when adding a new cluster, all methods sample its corresponding parameter from the prior distribution. In a high dimensional scenario, it is very difficult to sample parameter values which are close to the existing data points. It may be possible to construct some alternative strategies with good scalability properties such as: either an approximate MCMC scheme that only uses a subset of the data (Welling and Teh, 2011; Korattikara et al., 2014; Bardenet et al., 2014), a variational inference scheme (Blei and Jordan, 2006) or a combination thereof such as stochastic variational inference (Hoffman et al., 2013). A mean field variational inference is one possibility but other variational schemes could be considered, for instance, by keeping only some dependencies between the latent variables in the approximating distributions.

# Appendix A

## A.1 Infinite mixture model as a limit of a finite mixture model

$$M \sim \mathcal{Q}$$
$$\underline{\pi} \mid M = m \sim \text{Symmetric Dirichlet}(\alpha)$$
$$X_1, \ldots, X_n \mid \underline{\pi} \overset{\text{i.i.d.}}{\sim} \text{Categorical}(\underline{\pi}) \tag{A.1}$$

$$p(\underline{\pi} \mid M = m) = \frac{\Gamma(m\alpha)}{\Gamma(\alpha)^m} \prod_{\ell=1}^{m-1} \pi_\ell^{\alpha-1} (1 - \pi_1 - \ldots - \pi_{m-1})^{\alpha-1}$$

$$p(X_i \mid \underline{\pi}) = \prod_{\ell=1}^{m} \pi_\ell^{\mathbb{I}\{X_i = \ell\}}$$

$$p(\underline{\pi} \mid M = m) p(X_1, \ldots, X_n \mid \underline{\pi}) = \frac{\Gamma(m\alpha)}{\Gamma(\alpha)^m} \prod_{\ell=1}^{m-1} \pi_\ell^{\alpha + \sum_{i=1}^{n} \mathbb{I}\{X_i = \ell\} - 1}$$
$$\times (1 - \pi_1 - \ldots - \pi_{m-1})^{\alpha - 1 + n - \sum_{\ell < m} n_\ell}.$$

The weights are Dirichlet distributed with parameters $\alpha + \sum_{i=1}^{n} \mathbb{I}\{X_i = \ell\}$ for $\ell = 1, \ldots, m-1$ and $n - \sum_{\ell=1}^{m-1} \sum_{i=1}^{n} \mathbb{I}\{X_i = \ell\}$. Then, if the total number of components is split in $k$ occupied components and $k_0$ empty components

$$p(X_1, \ldots, X_n \mid M = m) = \frac{\alpha^k \Gamma(m\alpha)}{\Gamma(\alpha m + n)} \prod_{\{\ell : n_\ell > 0\}} \frac{\Gamma(n_\ell + \alpha)}{\Gamma(\alpha + 1)}$$

where $n_\ell = \sum_{i=1}^{n} \mathbb{I}\{X_i = \ell\}$. If we add the combinatorial factor, then

$$p(\Pi_n = \pi \mid M = m) = \frac{m!}{(m-k)!} \frac{\alpha^k \Gamma(m\alpha)}{\Gamma(\alpha m + n)} \prod_{\ell=1}^{k} \frac{\Gamma(n_\ell + \alpha)}{\Gamma(\alpha + 1)}.$$

Let $\alpha = \frac{\theta}{m}$, if the Stirling approximation for large m is used for each term in the quotient of Gamma functions, $m! \sim \sqrt{2\pi} m^{m+1/2} \exp(-m)$, and the fact that $\Gamma(m+1) = m!$,

$$\simeq \left(1 - \frac{k}{m}\right)^{k-m-1/2} \exp(-k) \frac{\theta^k \Gamma(\theta)}{\Gamma(\theta + n)} \prod_{\ell=1}^{k} \frac{\Gamma(n_\ell + \frac{\theta}{m})}{\Gamma(\frac{\theta}{m} + 1)}.$$

Let $m \to \infty$

$$p(\Pi_n = \pi) = \frac{\theta^k \Gamma(\theta)}{\Gamma(\theta + n)} \prod_{\ell=1}^{k} \Gamma(n_\ell).$$

Since,

$$\lim_{m \to \infty} \left(1 - \frac{k}{m}\right)^{k-m-1/2} = \exp\left[\lim_{m \to \infty} (k - m - 1/2) \ln\left(1 - \frac{k}{m}\right)\right]$$

Let $x = \frac{k}{m}$, $x \to 0$ when $m \to \infty$ and if L'Hopital rule is used, then,

$$\lim_{x \to 0} \left(k - \frac{k}{x} - 1/2\right) \ln(1 - x) = k.$$

Hence,

$$\lim_{m \to \infty} \left(1 - \frac{k}{m}\right)^{k-m-1/2} = \exp(k).$$

## A.2 Ascending and descending factorials notation

$$(x)_{n\uparrow\alpha} = x(x + \alpha) \ldots (x + (n-1)\alpha) = \prod_{\ell=1}^{n-1}(x + \ell\alpha) = \alpha^n \left(\frac{x}{\alpha}\right)_{n\uparrow}$$

$$(x)_{n\downarrow\alpha} = (x)_{n\uparrow-\alpha}$$

$$(x)_{n\uparrow 1} = \frac{\Gamma(x + n)}{\Gamma(x)}$$

$$(x)_{n\downarrow 1} = \frac{\Gamma(x)}{\Gamma(x + n)}.$$

## A.3 Subordinators and relationship between the Lévy measure $\rho$ and the corresponding density $f_\rho$

A *Lévy process* is a stochastic process $\{T(t), t \geqslant 0\}$ with independent and stationary increments. A *subordinator* $\{T(t), t \geqslant 0\}$ is a one dimensional Lévy process which is almost surely increasing, i.e. $T(t) \geqslant 0$ for each $t > 0$ and $T(t_1) > T(t_2)$ if $t_1 > t_2$. See Applebaum (2009) for an indepth study of Lévy processes. Its Laplace exponent takes the form

$$\psi(u) = \int_0^\infty (1 - \exp(uy))\rho(dy). \tag{A.2}$$

The Lévy measure $\rho$ satisfies

$$\rho(-\infty, 0) = 0 \text{ and } \int \min(y, 1)\rho(dy) < \infty.$$

The Laplace transform from Equation (A.2) is defined with respect to the Lévy measure $\rho$. The usual definition of the Laplace transform of a random variable is given with respect to its corresponding density, if it exists. Let $f_T(t)$ denote the density for each random variable $T(t)$ at time $t$, then

$$\int \exp(-uy) f_{T(t)}(y) dy = \int_0^\infty (1 - \exp(uy))\rho(dy).$$

We can recover the corresponding density from the Lévy measure and viceversa, the former will not always have an analytic form. Some examples are the following:

1. $\sigma$-Stable subordinator. Let $\sigma \in (0, 1)$ and $u \geqslant 0$

$$u^\sigma = \frac{\sigma}{\Gamma(1-\sigma)} \int_0^\infty (1 - \exp(-uy)) \frac{dy}{y^{1+\sigma}}.$$

Furthermore,

$$\frac{\sigma}{\Gamma(1-\sigma)} \int_0^\infty (1 - \exp(-uy)) \frac{dy}{y^{1+\sigma}} = \int \exp(-uy) f_{T(t)}(y) dy$$

2. Gamma subordinator. Let $\{T(t), t \geqslant 0\}$ be a gamma process with parameters

$a, b > 0$ so that each $T(t)$ has density

$$f_{T(t)}(y) = \frac{b^{at}}{\Gamma(at)} y^{at-1} \exp(-by)$$

and

$$\int \exp(-uy) f_{T(t)}(y) dy = \int_0^\infty (1 - \exp(-uy)) y^{a-1} \exp(-by) dy$$
$$= \left(1 + \frac{u}{b}\right)^{-at}.$$

The Stable subordinator is an example where the Laplace transform has an analytic form but not the underlying density whereas in the Gamma subordinator both objects have analytic expressions.

## A.4 Two parameter Chinese restaurant EPPF

$$\Pr\nolimits_{\rho, H_0}(T \in dt, \Pi_n = (c_\ell)_{\ell \in [k]}, X_\ell^* \in dx_\ell^*, \tilde{J}_\ell \in ds_\ell \text{ for } \ell \in [k])$$
$$= h(t) t^{-n} f_\rho(t - \sum_{\ell=1}^K s_\ell) dt \prod_{\ell=1}^k s_\ell^{|c_\ell|} \rho(ds_\ell) H_0(dx_\ell^*) \quad \text{(A.3)}$$

Let the $\sigma$-Stable Lévy measure and the corresponding $h(t)$ function for the Pitman–Yor process be chosen, given by

$$\rho_\sigma(s) = \frac{\sigma}{\Gamma(1-\sigma)} s^{-\sigma-1}$$
$$h(t) = \frac{\Gamma(\theta)}{\Gamma(\frac{\theta}{\sigma})} t^{-\theta}. \quad \text{(A.4)}$$

We want to recover, after marginalising everything out and a change of variables, the usual two parameter Chinese Restaurant distribution, namely

$$\Pr(\Pi_n = \pi) = \frac{\Gamma(\theta)}{\Gamma(\theta+n)} \frac{\sigma^k \Gamma(\frac{\theta}{\sigma}+k)}{\Gamma(\frac{\theta}{\sigma})} \prod_{\ell=1}^k \frac{\Gamma(|c_\ell|-\sigma)}{\Gamma(1-\sigma)}.$$

If we start with Equations (A.3), after Equations (A.4) are substituted, let $V =$

$T - \sum_{\ell=1}^{k} \tilde{J}_\ell$, then

$$\Pr{}_{\rho, H_0}(V \in dv, \Pi_n = (c_\ell)_{\ell \in [k]}, X_\ell^* \in dx_\ell^*, \tilde{J}_\ell \in ds_\ell \text{ for } \ell \in [k])$$
$$= h(v + \sum_{\ell=1}^{k} s_\ell)(v + \sum_{\ell=1}^{k} s_\ell)^{-n} f_\sigma(v) dv \prod_{\ell=1}^{k} \frac{\sigma}{\Gamma(1-\sigma)} s_\ell^{|c_\ell|-\sigma-1} H_0(dx_\ell^*)$$
$$= \frac{\Gamma(\theta)}{\Gamma(\frac{\theta}{\sigma})}(v + \sum_{\ell=1}^{k} s_\ell)^{-(n+\theta)} f_\sigma(v) \prod_{\ell=1}^{k} \frac{\sigma}{\Gamma(1-\sigma)} s_\ell^{|c_\ell|-\sigma-1} H_0(dx_\ell^*).$$

If we use the following disintegration

$$(v + \sum_{\ell=1}^{k} s_\ell)^{-\theta-n} = \frac{1}{\Gamma(\theta + n)} \int r^{\theta+n-1} \exp(-rv - r \sum_{\ell=1}^{k} s_\ell) dr$$

then,

$$= \frac{\Gamma(\theta)}{\Gamma(\frac{\theta}{\sigma})} \int \frac{r^{\theta+n-1} \exp(-rv)}{\Gamma(\theta+n)} f_\sigma(v)$$
$$\times \prod_{\ell=1}^{k} \exp(-rs_\ell) \frac{\sigma}{\Gamma(1-\sigma)} s_\ell^{|c_\ell|-\sigma-1} dr H_0(dx_\ell^*).$$

Since $s_\ell \sim \text{Gamma}(n_\ell - \sigma, r)$, for $\ell = 1, \ldots, k$, we can integrate out the size biased weights, then,

$$= \frac{\Gamma(\theta)}{\Gamma(\frac{\theta}{\sigma})} \int \frac{r^{\theta-\sigma k-1} \exp(-rv) dr}{\Gamma(\theta+n)} f_\sigma(v) \frac{\sigma^{k-1}}{\Gamma(1-\sigma)^k}$$
$$\times \prod_{\ell=1}^{k} \Gamma(n_\ell - \sigma) dr H_0(dx_\ell^*).$$

Finally, let us perform the following changes of variables $W = R^\sigma$, $R = W^{\frac{1}{\sigma}}$ with Jacobian $\frac{dR}{dW} = \frac{1}{\sigma} W^{\frac{1-\sigma}{\sigma}}$, then

$$= \frac{\Gamma(\theta)}{\Gamma(\frac{\theta}{\sigma})} \int \frac{w^{\frac{\theta}{\sigma}+k-1} \mathbb{E}\left[\exp\left(-w^{\frac{1}{\sigma}} v\right)\right] dw}{\Gamma(\theta+n)} \frac{\sigma^{k-1}}{\Gamma(1-\sigma)^k}$$
$$\times \prod_{\ell=1}^{k} \Gamma(n_\ell - \sigma) dr H_0(dx_\ell^*)$$

and, since the Laplace transform of $X \sim \text{Stable}(\sigma, 1, 0)$, is given by

$$\mathbb{E}(\exp(-\gamma X)) = \exp(-\gamma^\sigma)$$

then,

$$= \frac{\Gamma(\theta)}{\Gamma(\frac{\theta}{\sigma})} \int \frac{w^{\frac{\theta}{\sigma}+k-1}\exp(-w)dw}{\Gamma(\theta+n)} \frac{\sigma^{k-1}}{\Gamma(1-\sigma)^k} \prod_{\ell=1}^{k} \Gamma(n_\ell - \sigma) dr H_0(dx_\ell^*)$$

$$= \frac{\Gamma(\theta)}{\Gamma(\theta+n)} \frac{\sigma^k \Gamma(\frac{\theta}{\sigma}+k)}{\Gamma(\frac{\theta}{\sigma})} \prod_{\ell=1}^{k} \frac{\Gamma(|c_\ell|-\sigma)}{\Gamma(1-\sigma)} H_0(dx_\ell^*).$$

where the first term corresponds to the two-parameter EPPF times the base distribution evaluated at the unique cluster parameters.

## A.5 SMC for BNP in generic terms

Let us assume a conjugate normal Pitman–Yor mixture model. The goal is to approximate each posterior distribution for the partition when the first $i$ data points are observed, denoted by $\Pr(\Pi_i = \pi \mid y_1, \ldots, y_i)$, $\forall i \in \{1, \ldots, n\}$.

- If $i = 1$, then $\Pi_1^p = \{\{1\}\}$, $\Pr(\Pi_1^p = \pi \mid y_1) = \delta_{\{\{1\}\}}$, $\forall p = 1, \ldots, L$.

- For the $i$-th step, let $(\Pi_i^p)_{p=1}^L$ be the sequence of partitions for each particle, where $L$ is the number of particles. The $p$-th estimate is $\widehat{\Pr}(\Pi_i^p \mid y_1, \ldots, y_i)$. The exact propagation is

$$\Pr(\Pi_{i+1}^p = \pi' \mid y_1, \ldots, y_i) \propto \Pr(\Pi_{i+1}^p = \pi' \mid y_1, \ldots, y_i, \Pi_i^p = \pi)$$
$$\times \Pr(\Pi_i^p = \pi \mid y_1, \ldots, y_i).$$

If we substitute the Monte Carlo approximation, we obtain the following particle approximation

$$\Pr(\Pi_{i+1}^p = \pi' \mid y_1, \ldots, y_i, \Pi_i^p = \pi) \widehat{\Pr}(\Pi_i^p = \pi \mid y_1, \ldots, y_i) =$$
$$\frac{1}{L} \sum_{p=1}^{L} \Pr(\Pi_{i+1}^p = \pi' \mid y_1, \ldots, y_i, \pi^p). \quad (A.5)$$

We can sample from the Monte Carlo approximation $\forall p \in \{1 \ldots, L\}, \Pi_{i+1} \sim \Pr(\Pi_{i+1}^p = \pi' \mid y_1, \ldots, y_i, \Pi_i^p = \pi)$. Then, the Bayes update is

$$\Pr(\Pi_{i+1}^p = \pi' \mid y_1, \ldots, y_{i+1}) \propto \Pr(Y_{i+1} \in dy_{i+1} \mid \Pi_{i+1}^p = \pi' y_1, \ldots, y_i)$$
$$\times \Pr(\Pi_{i+1}^p = \pi' \mid y_1, \ldots, y_i).$$

Successively, we can substitute the particle approximation from Equation (A.5) to obtain the additional particle approximation

$$\Pr\left(Y_{i+1} \in dy_{i+1} \mid \Pi_{i+1}^p = \pi', y_1, \ldots, y_i\right) \widehat{\Pr}\left(\Pi_{i+1}^p = \pi' \mid y_1, \ldots, y_i\right)$$
$$= \sum_{p=1}^{L} \Pr\left(Y_{i+1} \in dy_{i+1} \mid \pi^p, y_1, \ldots, y_i\right).$$

We can now approximate expectations of a function of interest, denoted by $h$, with respect to the target distribution $\int h(x)f(x)dx$, using the particle approximation $\hat{f}$ in the following way

$$\int h(x)f(x)dx \longrightarrow \frac{1}{L}\sum_{p=1}^{L} h(x^p)$$

Analogously, for $\int h(x)g(x)f(x)dx$, we have

$$\int h(x)g(x)f(x)dx \simeq \frac{1}{L}\sum_{p=1}^{L} h(x^p)g(x^p) \simeq \frac{1}{L}\sum_{p=1}^{L} g(x^p)\delta_{x^p}(\cdot)$$

Since we have a particle approximation for $f$, denoted by $\hat{f}$, if $L \to \infty$, then,

$$\frac{\frac{1}{L}\sum_{p=1}^{L} h(x^p)g(x^p)}{\frac{1}{L}\sum_{p=1}^{L} g(x^p)} \longrightarrow \frac{\int h(x)g(x)f(x)dx}{\int g(x)f(x)dx}.$$

We can substitute the Monte Carlo approximation in the Bayes update

$$\frac{\Pr\left(Y_{i+1} \in dy_{i+1} \mid \Pi_{i+1}^p = \pi', y_1, \ldots, y_i\right) \widehat{\Pr}\left(\Pi_{i+1}^p = \pi' \mid y_1, \ldots, y_i\right)}{\int \Pr\left(Y_{i+1} \in dy_{i+1} \mid \Pi_{i+1}^p, y_1, \ldots, y_i\right) \widehat{\Pr}\left(\Pi_{i+1}^p \mid y_1, \ldots, y_i\right) d\Pi_{i+1}^p}$$
$$= \frac{\sum_{p=1}^{L} \Pr\left(Y_{i+1} \in dy_{i+1} \mid \pi, y_1, \ldots, y_i\right) \delta_\pi(\pi^p)}{\sum_{j=1}^{L} \Pr\left(Y_{i+1} \in dy_{i+1} \mid \pi^j, y_1, \ldots, y_i\right)}.$$

Let $w_{i+1}^p = \Pr\left(Y_{i+1} \in dy_{i+1} \mid \pi, y_1, \ldots, y_i\right)$, then,

$$\frac{\Pr\left(Y_{i+1} \in dy_{i+1} \mid \Pi_{i+1}^p = \pi', y_1, \ldots, y_i\right) \widehat{\Pr}\left(\Pi_{i+1}^p = \pi' \mid y_1, \ldots, y_i\right)}{\int \Pr\left(Y_{i+1} \in dy_{i+1} \mid \Pi_{i+1}^p, y_1, \ldots, y_i\right) \widehat{\Pr}\left(\Pi_{i+1}^p \mid y_1, \ldots, y_i\right) d\Pi_{i+1}^p}$$
$$= \frac{\sum_{p=1}^{L} w_{i+1}^p \delta_\pi(\pi^p)}{\sum_{j=1}^{L} w_{i+1}^j}.$$

## A.6 Posterior updates and prior predictive densities

Let $Y_1, \ldots, Y_n$ be our observations. The generative model is

$$P \sim \text{RPM}(param, H_0)$$
$$X_i \mid P \sim P$$
$$Y_i \mid X_i \sim F(\cdot \mid \phi_{X_i})$$

**Known precision parameter $\tau_1$ case.** The precision parameter a function of the standard deviation $\sigma$, namely $\tau = \frac{1}{\sigma^2}$. Let $Y \in \mathbb{R}$, $\{Y_j\}_{j \in c}$ are the observations currently assigned to cluster $c$, $F(Y_i \in dy_i \mid \phi_c) = \text{Normal}(y_i \mid \mu_c, \tau_1)$, $H_0 = \text{Normal}(\mu \mid \mu_0, \tau_0)$, in this case $\phi = \mu$.

$$\Pr\left(Y_i \in dy_i \mid \{Y_j \in dy_j\}_{j \in c}\right) = \int \Pr\left(Y_i \in dy_i \mid \phi\right) p\left(\phi \mid \{Y_j \in dy_j\}_{j \in c}\right) d\phi$$

where

$$p\left(\phi \mid \{Y_j \in dy_j\}_{j \in c}\right) \propto \exp\left\{-\frac{1}{2}(\tau_0 + |c|\tau_1)\left[\phi^2 - 2\phi \frac{(\mu_0 \tau_0 + \tau_1 \sum y_j)}{|c|\tau_1 + \tau_0}\right]\right\}$$

so then, the posterior distribution parameters are

$$\mu_* = \frac{\mu_0 \tau_0 + \tau_1 \sum_{j \in c} y_j}{\tau_0 + |c|\tau_1}$$
$$\tau_* = \tau_0 + |c|\tau_1$$

and the predictive posterior distribution parameters are

$$\mu_{\text{pred}} = \mu_*$$
$$\tau_{\text{pred}} = \frac{\tau_1 \tau_*}{\tau 1 + \tau_*}.$$

The predictive prior distribution is

$$\Pr(Y_i \in dy_i) = \int \Pr(Y_i \in dy_i \mid \phi) p(\phi) d\phi$$

with parameters given by

$$\begin{aligned}\mu_{\text{pred}} &= \mu_0 \\ \tau_{\text{pred}} &= \frac{\tau_1 \tau_0}{\tau_1 + \tau_0}.\end{aligned}$$

**Unknown precision parameter case.** Let $Y \in \mathbb{R}$, $\{Y_j\}_{j \in c}$ are the observations currently assigned to cluster $c$, $F(Y_i \in dy_i \mid \phi_c) = \text{Normal}(y_i \mid \mu_c, \tau_c)$, $H_0 = \text{Normal}(\mu \mid \mu_0, \tau_0 \tau)\text{Gamma}(\tau \mid a_0, b_0)$, in this case $\phi = [\mu, \tau]$.

$$\Pr\left(Y_i \in dy_i \mid \{y_j\}_{j \in c}\right) = \int \Pr\left(Y_i \in dy_i \mid \phi\right) p\left(\phi \mid \{x_\ell\}_{\ell=1}^{nn}\right) d\phi$$

$$\begin{aligned}\Pr(\{Y_j \in dy_j\}_{j \in c} \mid \mu, \tau) &= (2\pi)^{-\frac{|c|}{2}} \tau^{\frac{|c|}{2}} \exp\left\{-\frac{\tau}{2} \sum_{i=1}^{|c|} (y_i - \mu)^2\right\} \\ &= (2\pi)^{-\frac{|c|}{2}} \tau^{\frac{|c|}{2}} \exp\left\{-\frac{\tau}{2}\left[|c| * \mu^2 - 2\mu \sum_{j \in c} y_j + C\right]\right\} \\ p(\mu \mid \tau) p(\tau) &= (2\pi)^{-\frac{1}{2}} \tau_0^{\frac{1}{2}} \frac{b_0^{a_0}}{\Gamma(a_0)} \tau^{a_0 + \frac{1}{2} - 1} \exp\left\{-\frac{\tau}{2}\left[\tau_0(\mu - \mu_0)^2 + 2b_0\right]\right\} \\ &= \frac{1}{Z(\tau_0, a_0, b_0)} \tau^{a_0 + \frac{1}{2} - 1} \exp\left\{-\frac{\tau}{2}\left[\tau_0(\mu - \mu_0)^2 + 2b_0\right]\right\}.\end{aligned}$$

The posterior distribution is

$$\begin{aligned}p(\mu, \tau \mid \{y_j\}_{j \in c}) &\propto \Pr(\{Y_j \in dy_j\}_{j \in c} \mid \mu, \tau) p(\mu, \tau) \\ &= \frac{(2\pi)^{-\frac{|c|}{2}} \tau^{\frac{|c|+1}{2} + a_0 - 1} \tau^{\frac{|c|+1}{2} + a_0 - 1}}{p\left(\{y_j\}_{j \in c}\right) Z(\tau_0, a_0, b_0)} \exp\left\{-\frac{\tau}{2}\left[|c|\mu^2 - 2\mu \sum_{j \in c} y_j + C + \tau_0(\mu - \mu_0)^2 + 2b_0\right]\right\}.\end{aligned}$$

with posterior distribution parameters given by

$$\begin{aligned}C &= \sum_{j \in c} y_j^2 + |c| \bar{y} \bar{y}' \\ b_1 &= \frac{\tau_0 \mu_0^2}{2} + \frac{C}{2} + b_0 - \frac{\tau_1}{2} \mu_1^2 \\ a_1 &= a_0 + \frac{|c|}{2} \\ \tau_1 &= \tau_0 + |c| \\ \mu_1 &= \frac{\sum_{j \in c} y_j + \tau_0 \mu_0}{|c| + \tau_0}\end{aligned}$$

and

$$Z(\tau_0, a_0, b_0) = (2\pi)^{\frac{1}{2}} \tau_0^{-\frac{1}{2}} \Gamma(a_0) b_0^{-a_0}.$$

Since $\frac{Z(\tau_1, a_1, b_1)}{(2\pi)^{-\frac{|c|}{2}}} = Z(\tau_0, a_0, b_0) \Pr\left(\{Y_j \in \mathrm{d}y_j\}_{j \in c}\right)$, then,

The prior predictive is

$$\Pr\left(\{Y_j \in \mathrm{d}y_j\}_{j \in c}\right) = \frac{Z(\tau_1, a_1, b_1)(2\pi)^{-\frac{|c|}{2}}}{Z(\tau_0, a_0, b_0)}.$$

if $|c| = 1$

$$\begin{aligned}
\Pr(Y \in \mathrm{d}y) &= \frac{b_0^{a_0} \tau_0^{1/2} \Gamma(a_0 + 1/2)}{(1 + \tau_0^{1/2}) \Gamma(a_0) \sqrt{2\pi}} \left[ b_0 \left\{ 1 + \frac{\tau_0(y - \mu_0)^2}{2b_0(1 + \tau_0)} \right\} \right]^{-\frac{2a_0+1}{2}} \\
&= \text{Student-t}\,(y \mid \eta, \mu, \lambda)
\end{aligned}$$

with parameters given by

$$\begin{aligned}
\eta_{\text{pred}} &= 2a_0 \\
\mu_{\text{pred}} &= \mu_0 \\
\lambda_{\text{pred}} &= \frac{a_0 \tau_0}{b_0(1 + \tau_0)}.
\end{aligned}$$

The posterior predictive is

$$\begin{aligned}
p\left(\{Y_j \in \mathrm{d}y_j\}_{j \in c} \mid \{y_j\}_{j \in c}\right) &= \frac{\Pr\left(\{Y \in \mathrm{d}y\} \cap \{y_j\}_{j \in c}\right)}{\Pr\left(\{Y_j \in \mathrm{d}y_j\}_{j \in c}\right)} \\
&= \frac{Z(\tau_2, a_2, b_2)(2\pi)^{-\frac{nn+1}{2}}}{Z(\tau_0, a_0, b_0)} \frac{Z(\tau_0, a_0, b_0)}{Z(\tau_1, a_1, b_1)(2\pi)^{-\frac{|c|}{2}}} \\
&= \frac{Z(\tau_2, a_2, b_2)(2\pi)^{-\frac{1}{2}}}{Z(\tau_1, a_1, b_1)}
\end{aligned}$$

where

$$
\begin{aligned}
C &= \sum_{j=1}^{|c|+1} y_j^2 + (|c|+1)\bar{y}\bar{y}' \\
b_2 &= \frac{\tau_0 \mu_0^2}{2} + \frac{C}{2} + b_0 - \frac{\tau_1}{2}\mu_1^2 \\
a_2 &= a_0 + \frac{|c|+1}{2} \\
\tau_2 &= \tau_0 + |c| + 1 \\
\mu_2 &= \frac{\sum_{j=1}^{|c|+1} y_j + \tau_0 \mu_0}{|c|+1+\tau_0}.
\end{aligned}
$$

if $m = 1$

$$
\begin{aligned}
\Pr\left(\{Y_j \in \mathrm{d}y_j\}_{j=1}^m \mid \{y_j\}_{j\in c}\right) &= \frac{b_1^{a_1}(|c|+\tau_0)^{\frac{1}{2}} \Gamma(a_1 + 1/2)}{\sqrt{2\pi}\Gamma(a_1) b_1} \left[\frac{(nn+\tau_0)(y-\mu_1)^2}{2b_1(|c|+\tau_0+1)} + 1\right]^{-\frac{2a_1+1}{2}} \\
&= \text{Student-t}(y \mid \eta, \mu, \lambda)
\end{aligned}
$$

with parameters given by

$$
\begin{aligned}
\eta_{\text{pred}} &= 2a_1 \\
\mu_{\text{pred}} &= \mu_1 \\
\lambda_{\text{pred}} &= \frac{a_1(\tau_0 + |c|)}{b_1(|c|+1+\tau_0)}.
\end{aligned}
$$

## A.7  Additional pseudocode

The following equations give us a reassignment rule for the MFM model for a rejuvenation step in SMC in the conjugate case

$$
\begin{aligned}
\Pr(i \text{ joins cluster } c' \mid \text{rest}) &\propto (|c'| + \gamma) F(Y_i \in dy_i | \{y_j\}_{j \in c'}) \\
\Pr(i \text{ joins new cluster } c' \mid \text{rest}) &\propto \gamma \frac{V_{n,k+1}}{V_{n,k}} F(Y_i \in dy_i),
\end{aligned} \quad (A.6)
$$

**Algorithm 20** SliceSampler($x_j, f, E, L$)

    $N_1 = N_2 = 0$
    Sample $y \sim \mathrm{U}\,[0, f(x_j)]$     ▷ $f$ can be an unnormalized density
    Sample $l \sim \mathrm{U}(0, L)$     ▷ $L$ is the chosen length of the interval $(a, b)$
    Set $a = x_j - l$, $b = x_j - l + L$     ▷ $x_j$ is the previously accepted point
    **while** $f(a) < y \lor f(b) < y$ **do**     ▷ Samples $x_{j+1}$ uniformly from the set $f^{[-1]}[y, \infty)$
        **if** $f(a) < y$ **then**
            $a = a - E$     ▷ $E$ is the chosen size to enlarge the initial interval
            $N_1 = N_1 + 1$
        **end if**
        **if** $f(b) < y$ **then**
            $b = b + E$
            $N_2 = N_2 + 1$
        **end if**
        $L = b - a$
        Sample $l \sim \mathrm{U}(0, L)$
        Set $a = x_j - l$, $b = x_j - l + L$
    **end while**
    Set $c = x_j - l - N_1 L$, $d = x_j - l + N_2 L$
    Sample $w \sim \mathrm{U}(c, d)$
    **while** $f(w) < y$ **do**
        **if** $w < x_j < d$ **then**
            $w \sim \mathrm{U}(w, d)$
        **else**
            $w \sim \mathrm{U}(c, w)$
        **end if**
    **end while**
    Return $x_{j+1} = w$

---

**Algorithm 21** Conjugate-MixtureofFiniteMixtures($\{Y_i\}_{i=1}^n$)

    **for** $i = 1 : n$ **do**
        Let $c \in \Pi_n$ be such that $i \in c$
        $c \leftarrow c \setminus \{i\}$
        Set $c'$ according to $\Pr[i$ joins cluster $c' \mid \{Y_j\}_{j \in c}, \text{rest}]$     ▷ from Equation (A.6)
    **end for**

---

**Algorithm 22** CoClustering($C$)

    **for** $m = 1 \to M, i = 1 \to n, j = 1 \to n$ **do**     ▷ $M$ is the number of MCMC iterations
        **if** $c_{m,i} == c_{m,j}$ **then**     ▷ $n$ is the number of data points
            $A_{i,j,m} = 1$     ▷ $A$ is a $n \times n \times M$ array
        **else**
            $A_{i,j,m} = 0$
        **end if**
    **end for**
    $P = \mathrm{Sum}(A, 3)/M$
    **return**

## A.8 Agglomerative clustering algorithms

The coclustering probability matrix $P$ from Algorithm 22 can be used to obtain a distance matrix for the input of an agglomerative clustering algorithm. The $i, j$ th entry of the distance matrix is given by

$$d_{i,j} = 1 - P_{i,j} \tag{A.7}$$

---
**Algorithm 23** SingleLinkage($D$)
---
    for $i = 1 : n$ do
        $B_i = \{i\}$
    end for
    for $m = n - 1 : 1$ do
        $B_{\text{merged}} = B \cup B'$ with $i \in B, j \in B'$ such that $d_{i,j} \leq d_{k,\ell}$ for all $i, j, k, \ell \in \{1, \ldots, m\}$, $i \neq j$ and $k \neq \ell$.
        for $\ell = 1 : m$ do
            $d_{\ell,(i,j)} = \frac{1}{2}d_{\ell,i} + \frac{1}{2}d_{\ell,i}d_{\ell,j} - \frac{1}{2}|d_{\ell,i} - d_{\ell,j}|$
        end for
    end for
---

Algorithm 23 starts with all data points assigned to a separate cluster each. At each point, it merges clusters $i, j$ based on the following criteria: $\min_{i \in G, j \in H} d_{i,j}$. If $i, j$ are merged together, the $k$-th entry of the new distance matrix is calculated in the following way

$$d_{k,(ij)} = \frac{1}{2}d_{k,i} + \frac{1}{2}d_{k,i}d_{k,j} - \frac{1}{2}|d_{ki} - d_{k,j}| \quad \forall \ k = 1, \ldots, n-1. \tag{A.8}$$

The above procedure is repeated until all the data points are merged into the same cluster.

Other criteria for selecting how to merge clusters are:

Complete linkage: $\max_{i \in G, j \in H} d_{i,j}$.

Average linkage: $d_A = \frac{1}{N_G} \frac{1}{N_H} \sum_{i \in G} \sum_{j \in H} d_{i,j}$.

## A.9 Average leave-one-out and 5-fold predictive probabilities

The posterior predictive density for $Y_i$ given $\mathbf{X}^*$, $P$ and $\mathbf{Y}_{\setminus i} = \{Y_1, \ldots, Y_{i-1}, Y_{i+1}, \ldots, Y_n\}$ is

$$\Pr\left(Y_i \in \mathrm{d}y_i \mid R, W, \mathbf{Y}_{\setminus i}, \mathbf{X}^*, \Pi_n\right) = \eta_0(n, \sigma, k, w, r) \int \Pr(Y_i \in \mathrm{d}y_i \mid x) \, p(x) \mathrm{d}x$$
$$+ \eta_1(n, \sigma, |c|) \sum_{j \in c} (|c| - \sigma) f(y_i \mid x_c^*)$$

where

$$\eta_0(n, \sigma, |\Pi_n|, w, r) = \frac{e^{-w(1-\sigma)} r^{-\sigma}}{\Gamma(n - \sigma(|\Pi_n| + 1))}$$
$$\eta_1(n, \sigma, |\Pi_n|) = \frac{1}{\Gamma(n - \sigma|\Pi_n|)}.$$

The corresponding estimator, where $M$ is the size of the chain after burn in, is given by:

$$\hat{\Pr}\left(Y_i \in \mathrm{d}y_i \mid R, W, \mathbf{Y}_{\setminus i}, \mathbf{X}^*\right) = \frac{1}{M} \left[ \sum_{m=1}^M \eta_0(n, \sigma, \Pi_n^m, w_m, r_m) \int \Pr(Y_i \in \mathrm{d}y_i \mid x) \, p(x) \mathrm{d}x \right]$$
$$+ \left[ \eta_1(n, \sigma, |\Pi_n^m|) \sum_{j \in c_m} (|c| - \sigma) \Pr\left(Y_i \in \mathrm{d}y_i \mid x_{c_m}^*\right) \right].$$
(A.9)

To obtain the average leave-one-out predictive probability we do the following:

1. For $i = 1, \ldots, n$, remove the $i$-th observation from the dataset and run the MCMC with the rest of the data points.

2. At each MCMC iteration evaluate the predictive probability of the $i$-th datapoint.

3. Average over the $M$ MCMC iterations to get the average predictive probability for the $i$-th observation given by Equation (A.9).

After we do this for all observations we take the average of the leave-one-out predictive probabilities to obtain the estimates reported in Table 2.

Similarly, to obtain the average 5-fold predictive probability we do the following:

1. For $j = 1, \ldots, 5$, randomly split the dataset into training data (4/5) and test data (1/5) making sure each observation belongs to the test data only once, in other words, sample it without replacement.

2. At each MCMC iteration evaluate the predictive probability of each test data point and take the average.

3. Average over the $M$ MCMC iterations to get the average predictive probability for the $j$-th batch of test data.

After we do this for all test data batches we take the average of their corresponding 5-fold predictive probabilities to obtain the estimates reported in Table 3.

## A.10  Geweke's Getting it Right

Geweke (2004) has a formal test for testing the correctness of an MCMC algorithm using a two sample hypothesis test or a Q-Q plot. We compare the output of the Marginal-conditional simulator from Equation (A.10), where the updates follow the generative model, versus the output from the Succesive-conditional simulator from Equation (A.11), where the posterior simulator, which correctness we want to check, is used.

$$
\begin{aligned}
&\text{for } t \geq 1 \\
&1. \text{ Sample } x^{(t)} \sim f(x) \\
&2. \text{ Sample } \phi^{(t)} \mid x^{(t)} \sim f\left(\phi \mid x^{(t)}\right) \\
&3. \text{ Sample } y^{(t)} \mid \phi^{(t)}_{x^{(t)}} \sim \text{Normal}\left(y \mid \phi_{x^{(t)}}\right)
\end{aligned}
\tag{A.10}
$$

$$
\begin{aligned}
&\text{for } t = 0 \\
&1. \text{ Sample } x^{(0)} \sim f(x) \\
&2. \text{ Sample } \phi^{(0)} \mid x^{(0)} \sim f\left(\phi \mid x^{(0)}\right) \\
&\text{for } t \geq 1 \\
&1. \text{ Sample } y^{(t)} \mid \phi^{(t-1)}_{x^{(t-1)}} \sim \text{Normal}\left(y \mid \phi^{(t-1)}_{x^{(t-1)}}\right) \\
&2. \text{ Sample } \phi^{(t)} \mid x^{(t-1)}, y^{(t-1)} \sim f\left(\phi \mid x^{(t-1)}, y^{(t-1)}\right) \\
&3. \text{ Sample } x^{(t)} \mid \phi^{(t)}, y^{(t)} \sim f(x \mid y^{(t)})
\end{aligned}
\tag{A.11}
$$